\documentclass[pra,aps]{revtex4-1}
\usepackage{amssymb}
\usepackage{amssymb}
\usepackage{graphicx}
\usepackage{amsmath}
\usepackage{times}
\usepackage{color}
\usepackage{subfigure}
\usepackage{setspace}
\usepackage{bm}

\begin{document}

\title{Probing many-body interactions  in an optical lattice clock}

\author{A.~M. Rey$^{1 *}$, A.~V. Gorshkov$^{2}$, C.~V. Kraus$^{3,4}$, M.~J.~Martin$^{1,5}$, M.~Bishof$^1$, M.~D.~Swallows$^1$, X.~Zhang$^1$, C.~Benko$^1$,  J. Ye$^1$,  N.~D.  Lemke$^6$ and A.~D. Ludlow$^6$ }

\affiliation{$^{1}$JILA, NIST and University of Colorado, Department of Physics, Boulder, CO  80309, USA}
\affiliation{$^\ast$ E-mail: arey@jilau1.colorado.edu}
\affiliation{$^{2}$ Joint Quantum Institute, NIST and University of Maryland, Department of Physics, College Park, MD 20742, USA}
\affiliation{$^{3}$Institute for Quantum Optics and Quantum Information of the Austrian Academy of Sciences, A-6020 Innsbruck, Austria}
\affiliation{$^{4}$Institute for Theoretical Physics, University of Innsbruck, A-6020 Innsbruck, Austria}
\affiliation{$^{5}$Institute for Quantum Information and Matter, California Institute of Technology, Pasadena, CA 91125, USA}
\affiliation{$^{6}$National Institute of Standards and Technology,  Boulder, CO  80305, USA}

\begin{abstract}
We present a unifying theoretical framework that describes recently observed
many-body effects during the interrogation of  an optical lattice clock operated with thousands of fermionic alkaline earth atoms.
The framework is based on  a many-body master equation that accounts for the interplay between elastic and inelastic $p$-wave  and $s$-wave interactions, finite temperature effects and  excitation inhomogeneity during the quantum dynamics of the interrogated atoms. Solutions of the master equation in different parameter regimes are presented and compared. It is shown that  a general solution can be  obtained by using the so called Truncated Wigner Approximation which is applied in our case in the context of an open quantum system.
We use the developed framework  to model the density shift and decay of the  fringes observed during  Ramsey spectroscopy in the JILA ${}^{87}$Sr  and  NIST  ${}^{171}$Yb   optical lattice clocks. The developed framework opens a suitable path for dealing with a variety of strongly-correlated and  driven open-quantum spin systems.
\end{abstract}

\date{\today }
\maketitle

\section{Introduction}

One of the ultimate goals of modern physics is  to
 understand and fully control quantum mechanical systems and to exploit
them both, at the level of basic research and for numerous
technological applications  including  navigation, communications, network management, etc. To accomplish these objectives, we aim at developing the most advanced and novel measurement techniques capable of probing quantum matter at the fundamental
level.

Some  years ago, the second -- the international unit of time -- was defined by the Earth's rotation.
However, with the discovery of quantum mechanics and the quantized nature of the atomic energy levels, it became clear that atomic clocks could be more accurate and more precise than any mechanical
or celestial reference previously known to man.  Thus, in 1967 the second was redefined
as the duration of 9,192,631,770 periods of the radiation corresponding to the transition between the two hyperfine energy levels of a caesium atom. Since then, the accuracy of atomic clocks has improved
dramatically, by a factor of 10 or so every decade. The characterization  of the unit of time plays a central role within the International System of Units (SI) because of its unprecedented high accuracy and because it is also used in the definitions of other units such as meter, volt and ampere.

Thanks to the development of laser trapping and cooling techniques \cite{Wieman1999,Phillips1998}, the best caesium standards have reached an accuracy of one part in $10^{16}$. However, caesium  clocks are limited by the fact that they are based on atomic transitions in the microwave domain. Because the quality factor of the clock is proportional to the frequency, optical clocks with  frequencies
that can be  $10^6$ times higher than microwaves, offer an impressive potential gain over their microwave counterparts. Optical frequencies on the other hand
are very difficult to measure, as the oscillations are  orders of magnitude faster than
what electronics can measure. The implementation  of frequency comb technology~\cite{Hall2006} has provided a coherent link between the optical and
microwave regions of the electromagnetic spectrum, greatly simplifying
 optical frequency measurements of high accuracy.
  After the  development of frequency combs,   the interest in optical clocks has grown rapidly.  Now, optical clocks  based on  single trapped ions and neutral atoms are  the new generation of frequency standards  with a sensitivity and accuracy as high as one part in $10^{18}$ \cite{Chou2010,Rosenband2008,Bloom2013}.

Optical clocks operated with fermionic neutral alkaline earth
atoms (AEA), such as $^{87}$Sr or ${}^{171}$Yb, have matured considerably. Those employ an optical lattice to tightly confine the atoms  so that Doppler and photon-recoil related effects on the transition
frequency are eliminated.  State-of-the-art neutral-atom-optical clocks have  surpassed the accuracy  of the Cs standard \cite{Ludlow2008} and just recently, thanks to  advances in modern precision laser spectroscopy,  are   reaching  and even surpassing the accuracy of  single ion standards \cite{Bloom2013}.   The most stable of these clocks now operate near the quantum noise limit \cite{Nicholson2012,Ludlow2013}. The stability arises from the intrinsic atomic physics of  two-valence-electron
atoms that possess extremely long lived singlet and triplet states (clock states),
with intercombination lines nine orders of magnitude narrower
than a typical dipole-allowed electronic transition.

The potential advantage of neutral-atom clocks over  single trapped ion clocks is that, in the former,
 a large number of atoms is simultaneously interrogated. This could lead to  a large signal-to-noise improvement; however, high atom numbers combined with tight confinement also lead to high atomic
densities and the potential for non-zero collisional frequency shifts via contact atom-
atom interactions.  With atom-light coherence times reaching several seconds, even very weak interactions (\textit{e.g.}, fractional energy level shifts of order $\geq 1\times 10^{-16}$) can dominate the dynamics of these systems.

To suppress these interactions, the use of ultracold, spin-polarized fermions was proposed. The idea was to  exploit the Fermi suppression of $s$-wave
 contact interactions while freezing out $p$-wave  and higher wave collisions at ultracold atomic
temperatures. Indeed, at precision level of $10^{-15}$, the JILA Sr  clock did not exhibit a density-dependent frequency shift \cite{Boyd2007}, however as the measurement precision progressed, density-dependent frequency shifts were measured in spin polarized fermions, at the  JILA  Sr clock \cite{Campbell2009,Blatt2009,Swallows2011} and  at the NIST Yb clock \cite{Lemke2009}.

When those density-dependent frequency shifts were first observed, they were attributed to $s$-wave collisions allowed by inhomogeneous excitation \cite{Takamoto2006,Campbell2009,Rey2009,Gibble2009,Yu2010}, under the assumption that $p$-wave interactions were suppressed at the operating temperatures ($T\sim \mu$K) \cite{Campbell2009}. The basic initial understanding, obtained from a mean-field treatment, was that excitation inhomogeneities induced by the optical probing laser made the initially indistinguishable fermionic  atoms  distinguishable, and thus allowed them to interact via $s$-wave collisions.

However, studies of the cold collision shift in the NIST Yb optical lattice clock using Ramsey spectroscopy revealed that $p$-wave interactions were the dominant elastic interactions in that system \cite{Lemke2011}.  Furthermore,
evidence of inelastic $p$-wave interactions was reported in both Yb and Sr atomic clocks \cite{Ludlow2011,Bishof2011b}. Although the importance of many-body interactions in optical clocks has been recognized theoretically \cite{Rey2009, Gibble2009,Yu2010}, only recent measurements have revealed their many-body nature \cite{Martin2012,Martinthesis}. In those measurements, the role of $s$-wave collisions was further suppressed by operating the Sr clock with  highly homogeneous atom-laser coupling.
 This results in dominant $p$-wave interactions with a collective character, as we will explain below.

At this point it is important to emphasize that recent advances in modern precision laser spectroscopy,
with record levels of stability and residual laser drift less than  mHz/s \cite{Nicholson2012,Matt2011,Martinthesis} are the crucial developments that  are allowing us to deal with
clocks operated at a very different  conditions than those ones dealt with just few years ago.
 The level of precision  spectroscopy achievable in current atomic clocks is now providing  the required spectral resolution to systematically
resolve and study the complex excitation spectrum of an interacting many-body  system. This was certainly  not the case in prior
clock experiments where interaction effects were subdominant and  where a mean-field  treatment was more than
 enough to describe the clock behavior. For example in 2006, a 2 Hz spectral resolution has achieved for the  Sr atomic transition and no interaction effects were observable  at the time \cite{Boyd2006}.

In this paper we present a unifying theoretical framework that goes beyond a simple mean-field treatment and  that is capable of describing the full many-body dynamics of nuclear spin-polarized alkaline earth atoms  during clock interrogation. The two clock states are treated as an effective spin degree of freedom. Both, elastic and inelastic two-body collisions and single-particle losses are present during the dynamics, and thus a pure Hamiltonian formulation is not sufficient. Instead,  we develop a master equation formulation which is capable of treating  the quantum evolution of an open spin system. We provide analytic/exact solutions of the master equation dynamics in parameter regimes where  exact treatments are possible.  For the more generic situations we solve the  dynamics relying on the so called   Truncated Wigner Approximation (TWA) \cite{Blakie2008,Polkovnikov2010}. In contrast to previous theoretical treatments of the clock dynamics, which were limited to treating two-particles or many-particles but at the mean-field level or under the  all-to-all approximation \cite{Rey2009,Gibble2009,Rey2009,Gibble2009,Yu2010,Swallows2011}, the TWA method allows us  to   include both  elastic and inelastic collisions beyond the mean-field level, finite temperature effects and inhomogeneities generated by either the laser during the pulse interrogation or by many-body interactions. Those are shown to be crucial for properly modeling observed many-body dynamics, especially at $T\gtrsim 10\mu$K, at which excitation inhomogeneities can not be neglected. To our knowledge this is the first time that the TWA is applied to describe an open quantum system in the presence of inelastic losses.

Although this paper focuses on optical lattice clocks, the developed theoretical framework is generic for   driven open-quantum systems and  should be a useful platform for dealing with a variety of current experimentally relevant systems including trapped ions \cite{Britton2012,Kim2010}, polar molecules \cite{Neyenhuis2012,Hazzard2012,Gorshkov2011a,Gorshkov2011b,Hazzard2011b,Bo2013}, nitrogen vacancy centers \cite{Prawer2008},  and atoms in optical cavities \cite{Baumann2010,Gopalakrishnan2011} among others.

 \begin{figure*}
  \begin{center}
    \includegraphics[width=100mm]{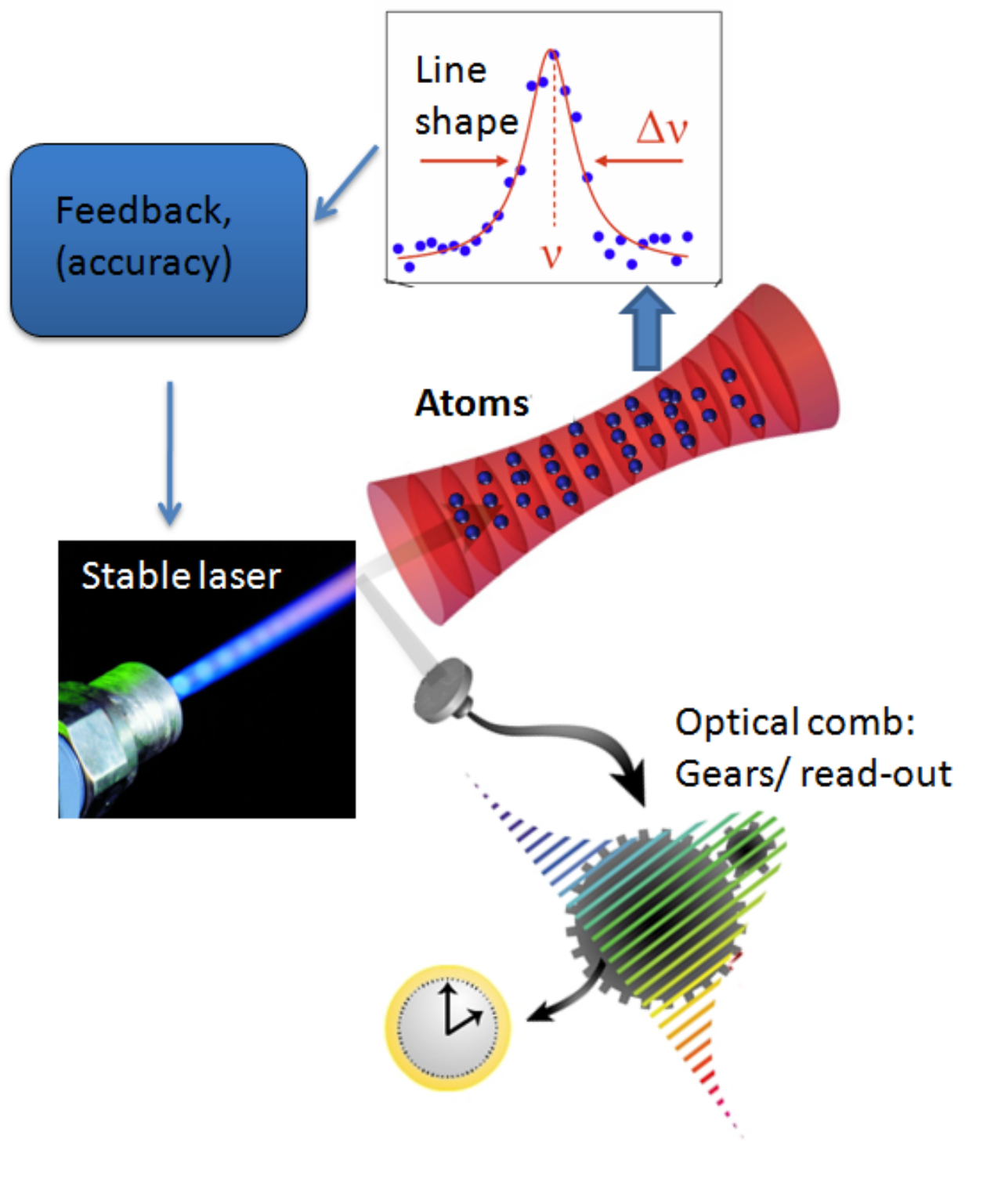}
  \end{center}
 \caption{ (Color online) General design of an optical lattice  clock}\label{clock}
\end{figure*}

The remainder of the paper is organized as follows. In section II, we  introduce the reader to  the basic operation of an  atomic clock and  derive the many-body Hamiltonian that describes the dynamics of nuclear spin-polarized fermionic atoms during clock interrogation. We then proceed to derive a simplified effective spin model which assumes frozen motional-excitations during the dynamics. In section III, we solve for the dynamics under the assumption  of collective spin interactions (all-to-all interactions) in a closed system. In Section IV, we show how to treat the observed two-body losses and introduce a master equation, which we solve under the collective-interactions approximation.  We also show how to  use the TWA to deal with  the  open quantum system dynamics. In section V, we relax the frozen-motional-degrees-of-freedom approximation and  derive an improved spin Hamiltonian with cubic spin-spin interactions which account for the virtual occupation of excited motional modes. In section VI, we go beyond the all-to-all interactions approximation and present a more general prescription that can address  both non-collective interactions and losses  and single-particle inhomogeneities. In section VII, we apply  the developed theoretical framework to model the dynamics during Ramsey spectroscopy observed in the JILA${}^{87}$Sr and the NIST  ${}^{171}$Yb optical lattice clocks, and finally, in section VIII, we present the conclusions.
In Appendixes 1-5, we present some details omitted in the main text.

\section{ Many-body physics during clock operation}

\subsection{A simple  overview of an optical lattice clock }

The general design of an optical lattice clock is shown in Fig.~\ref{clock}.
It consists of two components, a laboratory  radiation source  and an atomic system  with a
natural reference frequency determined by quantum mechanics
to which the laboratory radiation source can be compared.
 Here, the laboratory  radiation source
is an ultra-stable continuous-wave  laser.  It  acts as the local oscillator (or pendulum) for the clock and is used to probe an electromagnetic resonance in an atom. The atomic signal can then be used to determine the
difference between the laser frequency and that of the reference atom, allowing
laser frequency to be monitored and stabilized to the preferred value.  A frequency comb  \cite{Hall2006}
provides the gears of the clock, allowing measurements of the laser frequency
relative to other high accuracy clocks in either the optical or microwave domains.

The two main quantities that characterize the performance of a clock are
the accuracy and the precision. The accuracy  is determined
by how well the measured frequency matches that of the atom's natural frequency.
In general, the accuracy will depend on the atomic species used and how well
it can be isolated from environmental effects during spectroscopy. The precision
of the clock is more commonly referred to as the stability, which
represents the repeatability of the measured clock frequency over a given averaging time $\tau$.
For quantum projection noise limited measurements, it is typically expressed as  \cite{Bauch2003}
{\small \begin{equation}
\sigma(\tau)\approx \frac{\Delta \nu}{\nu} \sqrt{\frac{t_c}{\tau N}}.\label{allan}\end{equation}}
Here, $\frac{\nu}{\Delta  \nu}=Q$ is the line quality factor of the clock transition for a linewidth
$\Delta  \nu$ and  $\sqrt{N}$ is associated with the signal-to noise-ratio achieved for interrogating $N$ atoms  in the measurement cycle time $t_c$.

Fermionic AEA such as ${}^{87}$Sr and ${}^{171}$Yb have unique
properties that make them ideal candidates for the realization of atomic clocks \cite{Katori2011}.  The clock states are  the  ground singlet state, $^1S_0$, and a long lived triplet state $^3 P_0$, with intercombination lines both electric and magnetic dipole
forbidden and as narrow as a few mHz, see Fig. \ref{schma}. In the  ground state ($^1S_0$), and  to leading order in   the excited  state ($^3 P_0$),    the electronic degrees
of freedom have neither spin nor orbital angular momentum \cite{Boyd2006}. This means that the atoms in the clock states only interact with external magnetic fields through the nuclear spin degrees of freedom which have a
$g$-factor  1000 times smaller than the electronic orbital one. Therefore, AEA are much less sensitive to magnetic field fluctuations and/or to intensity and phase noise on the optical fields than conventional alkali atoms.

The clock transition is only allowed (i.e. laser light   weakly couples  $^1S_0$ to $^3 P_0$) because in the excited state the hyperfine
interaction leads to a small admixture of the higher-lying $P$ states \cite{Boyd2007}. This small admixture strongly affects
the magnetic moment and causes the $g$-factor of the excited state to be significantly different  from that of the ground
state ($\sim 60\%$ for Sr). The different $g$-factor allows for the addressability of the various Zeeman levels  in the presence of a
bias magnetic field as  demonstrated in Ref. \cite{Boyd2006}.

Optical spectroscopy in atomic clocks
is sensitive to atomic motion  due to   the Doppler effect. To overcome this limitation,
 atoms are confined in a tight  optical lattice formed by a standing wave light pattern to eliminate broadening
and frequency shifts due to atomic motion. Within a lattice site the atoms are tightly trapped in the so-called Lamb-Dicke regime
where the length scale associated with their motion is much smaller than the wavelength of the
laser probing the atoms. Moreover the optical lattice can be carefully designed to operate at the so-called magic wave length at which the light shifts on the clock states are equal and the clock frequency is not perturbed \cite{Ye2008}.

In optical lattice clocks, one can simultaneously probe large samples $\sim 10^{3-5}$ of
laser-cooled atoms which can potentially lead to  high frequency stability.
Nevertheless,  this precision may come at the cost of systematic inaccuracy due to atomic interactions. As mentioned before, the use of  ultracold spin-polarized fermions was thought to be the key to avoid interaction effects. However, it has been  recently shown that this is not the case  \cite{Campbell2009,Swallows2011,Bishof2011,Bishof2011b,Martin2012, Lemke2011,Ludlow2011}.

\subsection{Many-body Hamiltonian for spin  polarized fermionic atoms}

 \begin{figure*}
  \begin{center}
 \includegraphics[width=120mm]{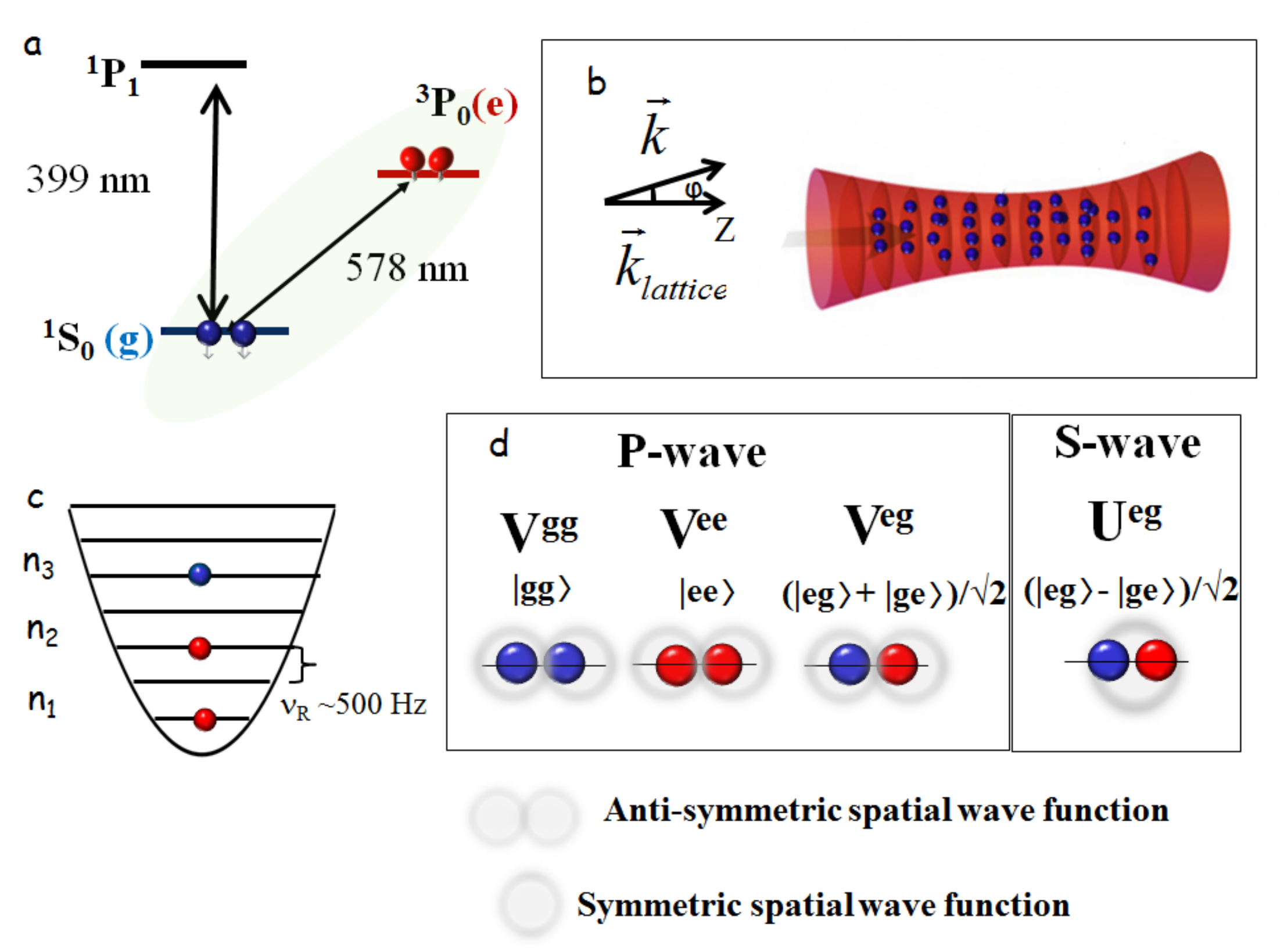}
 \end{center}
  \caption{(Color online) Schematics: a. Energy levels of alkaline earth atoms.  The $^1S_0$-$^3P_0$ electronic levels, which we denote as $g$ and $e$ respectively,   define the spin-$1/2$ system. Here the red-blue colors indicate electronic levels. The arrows show the nuclear spins which are fully polarized. The indicated energy level spacings are specific of   ${}^{171}$ Yb atoms.
  The atoms are confined in a deep one-dimensional optical lattice (generated by a laser with wavevector $\vec{k}_{\rm lattice}$ along the axial dimension), which creates an array of disk-like potentials (pancakes). In each pancake, the atoms occupy many transverse  vibrational levels at current $\sim\mu$K temperatures, as shown in c. The probing laser with wavevector $\vec{k}$, can be slightly misaligned from the lattice beam, and this gives rise to excitation inhomogeneities.  The relevant  interactions are $p$-wave and $s$-wave. The corresponding interaction parameters are indicated in panel d. To derive the spin model, one assumes that the atoms are frozen in the vibrational modes and that the only relevant dynamics, therefore,  happen in the electronic  degree of freedom. }
\label{schma}
\end{figure*}

In order to model the dynamics during clock interrogation, we will consider a nuclear spin-polarized ensemble of fermionic atoms   with two accessible   electronic degrees of freedom associated with the $^1S_0$-${}^3P_0$ electronic levels (see Fig.~\ref{schma}) which we denote as $g$ and $e$ respectively. $g$ stands for the ground state and $e$ for a excited state. We focus on the case where the atoms  are  trapped in an external potential $V_{ext}(\mathbf{R})$ which is the same for $g$ and $e$ ({\it i.e.}~at the ``magic wavelength'' \cite{Ye2008}). If the atoms are  illuminated by a linearly polarized laser beam with bare Rabi frequency $\Omega_0$, they are governed by the following
many-body Hamiltonian \cite{Gorshkov2009,Gorshkov2010,Rey2009,Yu2010,Hazzard2011b,Martin2012}
{\small
\begin{eqnarray}
&&\hat H =\hat H_0+ \hat H_1, \label{ham0}  \\&& \hat H_0= \sum_{\alpha }  \! \int  \!\! d^3  \mathbf{R} \hat \Psi^\dagger_{\alpha }(\mathbf{R})   \left(- \frac{\hbar^2}{2 m} \nabla^2 + V_{ext}(\mathbf{R})\right) \hat \Psi_{\alpha }(\mathbf{R})   +\frac{4 \pi \hbar^2 a_{eg}^-}{m}   \! \int  \!\! d^3 \mathbf{R} \hat \Psi^\dagger_{e }(\mathbf{R}) \hat \Psi_{e }(\mathbf{R}) \hat \Psi^\dagger_{g }(\mathbf{R}) \hat \Psi_{g }(\mathbf{R}) \nonumber \\&&
+\frac{ 3 \pi \hbar^2}{m} \sum_{\alpha,\beta} b_{\alpha \beta}^3\! \int  \!\! d^3 \mathbf{R} \Bigg[ \Big(\vec{\nabla} \hat \Psi_{\alpha }^\dagger(\mathbf{R})\Big)  \hat \Psi_{\beta }^\dagger (\mathbf{R}) -\hat \Psi_{\alpha }^\dagger (\mathbf{R}) \Big(\vec{\nabla} \hat \Psi_{\beta }^\dagger (\mathbf{R}) \Big) \Bigg]\cdot \Bigg[ \hat \Psi_{\beta }(\mathbf{R}) \Big(\vec{\nabla} \hat \Psi_{\alpha }(\mathbf{R})\Big)-\Big(\vec{\nabla} \hat \Psi_{\beta }(\mathbf{R})\Big) \hat \Psi_{\alpha }(\mathbf{R}) \Bigg]\nonumber \\&& +\frac{1}{2}\hbar \omega_0  \! \int  \!\! d^3 \mathbf{R} \Bigg[\hat \rho_e(\mathbf{R})  - \hat \rho_g(\mathbf{R})  \Bigg],\nonumber \\&&
\hat H_1=- \frac{ \hbar \Omega_0}{2 }  \! \int  \!\!  d^3 \mathbf{R} \Bigg[\hat \Psi^\dagger_{e }(\mathbf{R})  e^{-i (\omega_L t- \bm{k} \cdot \bf{ R})} \hat \Psi_{g }(\mathbf{R})  + {\rm h.c.}\Bigg] \nonumber.
\end{eqnarray}}
   Here $\hat \Psi_{\alpha }(\mathbf{R})$ is a fermionic field operator at position $\mathbf{R}$ for atoms with mass $m$  in electronic  state  $\alpha = g$ ($^1S_0$) or $e$ ($^3P_0$) while $\hat \rho_{\alpha }(\mathbf{R}) =
\hat \Psi^\dagger_{\alpha }(\mathbf{R}) \hat \Psi_{\alpha }(\mathbf{R})$ is the corresponding density operator. We consider two possible interaction channels:  $s$-wave and  $p$-wave (see Fig.~\ref{schma}).
 Since polarized fermions are in a symmetric nuclear state, their  $s$-wave interactions are characterized by only one   scattering
length $a_{eg}^-$,  describing collisions between two atoms in the antisymmetric electronic state, $\frac{1}{\sqrt{2}}(|ge\rangle-|eg\rangle)$. The $p$-wave interactions can have three different scattering volumes $b_{gg}^3$, $b_{ee}^3$, and $b_{eg}^3$ associated to the three possible electronic symmetric states ($|gg\rangle$, $|ee\rangle$, and $\frac{1}{\sqrt{2}}(|ge\rangle+|eg\rangle)$ respectively.
$\hat{H}_1$ takes into account the interrogation of the atoms by a laser that has frequency $\omega_L$ and wavevector $\bm{k}$  and is detuned from the atomic transition frequency $\omega_0$ by $\delta=\omega_L-\omega_0$.

\subsection{Effective spin model }\label{spinmod}
We  consider the situation of a  deep 1D lattice, $V_{ext}(\mathbf{R})$,  along $Z$, which creates an array of two-dimensional discs or ``pancakes'' and induces a weak harmonic radial (transverse) confinement
with an angular   frequency $\omega_R = 2 \pi \nu_R$. The lattice confines the atoms to the lowest axial vibrational mode.
This analysis can be straightforwardly generalized to the case of a 2D lattice in which two directions are frozen and only one is thermally populated.

We expand the field operator in a non-interacting atom  basis,
 $\hat{\Psi}_{\alpha}(\mathbf{R})=\phi^Z_0(Z) \sum_{\bf{ n}} \hat{c}_{\alpha \bf{n}} \phi_{n_X} (X) \phi_{ n_Y} (Y) $, where
$\phi_0^{Z}$ is  the ground longitudinal mode in  a lattice site and $\phi_n$ are transverse  harmonic oscillator eigenmodes.  $\hat{c}^\dagger_{\alpha  \bf{n}}$ creates a fermion in mode  ${\bf{n}}=(n_X,n_Y)$  and  electronic state $\alpha$.  In this basis and in the rotating frame of the laser, $\hat H$ can be rewritten as \cite{Swallows2011,Lemke2011,Martin2012}:
{\small
\begin{eqnarray}
&&\hat H_0=- \hbar \delta \sum_{\bf {n}} \hat{n}_{e \bf {n}}+\sum_{\alpha,\bf {n}} E_{\bf {n}} \hat{n}_{\alpha \bf {n}}  + \sum_{\alpha,\beta,\bf {n},\bf {n'},\bf {n''},\bf {n'''}
} \frac{\hbar}{4}\Bigg ((1-\delta_{\alpha,\beta} ) u S_{\bf {n} \bf {n'}\bf {n''} \bf {n'''}}+ v^{\alpha, \beta}
P_{\bf {n} \bf {n'}\bf {n''} \bf {n'''}} \Bigg)\hat{c}^\dag_{\alpha \bf {n}}\hat{c}^\dag_{\beta \bf {n'}}\hat{c}_{\beta \bf {n''}}\hat{c}_{\alpha \bf {n'''}} , \notag \\
&&u=4\sqrt{2\pi}   \sqrt{\omega_{Z} \omega_{R}} \frac{ a_{eg}^-}{a_{ho}^R},\quad \quad
v^{\alpha, \beta}= \frac{12}{ \sqrt{2\pi}} \sqrt{\omega_{Z} \omega_{R}} \frac{ b_{\alpha,\beta}^3}{{a_{ho}^R}^3}.\label{Habe}
\end{eqnarray}} Here $\hat{n}_{\alpha \bf {n}}=\hat{c}_{\alpha \bf{n}}^\dagger \hat{c}_{\alpha \bf{n}}$ is the atom number operator in mode $ {\bf n}$ and state $\alpha$, $\delta_{\alpha,\beta}$ is a Kronecker delta function, $a_{ho}^R=\sqrt{\hbar/(m\omega_R)}$ is the radial harmonic oscillator length, and $E_{{\bf n}}$ are single-particle energies in the trap. We have used a Gaussian approximation  for $\phi_0^{Z}$, which is excellent for the deep lattice used in experiments. $\hbar\omega_Z=2\sqrt{E_r V_Z}$ with $E_r$ the recoil energy, $\hbar^2 {k}_{\rm lattice}^2/(2m)$, ${k}_{\rm lattice}$ is the lattice beams' wave-number and $V_Z$ is the lattice depth. The coefficients  $S_{\bf {n} \bf {n'}\bf {n''} \bf {n'''}}$ and $P_{\bf {n} \bf {n'}\bf {n''} \bf {n'''}}$ characterize $s$- and $p$-wave matrix elements,  respectively, which depend on the harmonic oscillator modes and satisfy   $S_{\bf {n} \bf {n'}\bf {n''} \bf {n'''}} = S_{\bf {n} \bf {n'}\bf {n'''} \bf {n''}} = S_{\bf {n'} \bf {n}\bf {n''} \bf {n'''}} = S_{\bf {n'} \bf {n}\bf {n'''} \bf {n''}}$ and  $P_{\bf {n} \bf {n'}\bf {n''} \bf {n'''}} = -P_{\bf {n} \bf {n'}\bf {n'''} \bf {n''}} = -P_{\bf {n'} \bf {n}\bf {n''} \bf {n'''}} = P_{\bf {n'} \bf {n}\bf {n'''} \bf {n''}}$. Explicitly,
{\small
\begin{eqnarray}
&& S_{\bf {n} \bf {n'}\bf {n''} \bf {n'''}} = s (n_X ,n_X',n_X'',n_X ''') s( n_Y ,n_Y',n_Y'',n_Y '''), \\
&& P_{\bf {n} \bf {n'}\bf {n''} \bf {n'''}} =s( n_X ,n_X',n_X'',n_X ''') p(n_Y ,n_Y',n_Y'',n_Y ''' )+ p( n_X ,n_X',n_X'',n_X ''') s(n_Y ,n_Y',n_Y'',n_Y ''' ),\label{phis} \\
&& s( n ,n',n'',n''')\equiv \frac{\int d\xi e^{-2\xi^2 }H_{n}(\xi) H_{n'}(\xi)  H_{n''}(\xi) H_{n'''}(\xi)d\xi}{\pi \sqrt{2^{n+n'+n''+n'''} n! n'!n''! n'''!}} , \notag \\
&& p(n, n', n'', n''')=\frac{\int d\xi e^{-2\xi^2 }\left[\Big(\frac{d}{d\xi}H_{n}(\xi)\Big)  H_{n'}(\xi)-H_{n}(\xi) \Big(\frac{ d}{d\xi}H_{n'}(\xi)\Big) \right]  \left[\Big(\frac{d}{d\xi} H_{n''}(\xi)\Big)H_{n'''}(\xi)- H_{n''}(\xi) \Big(\frac{d}{d\xi}H_{n'''}(\xi) \Big)\right]}{\pi\sqrt{ 2^{n+n'+n''+n'''} n! n'!n''! n'''!}}. \notag\end{eqnarray} } Here  $H_n(x)$ are  Hermite polynomials.

\begin{figure}
  \begin{center}
 \includegraphics[width=70mm]{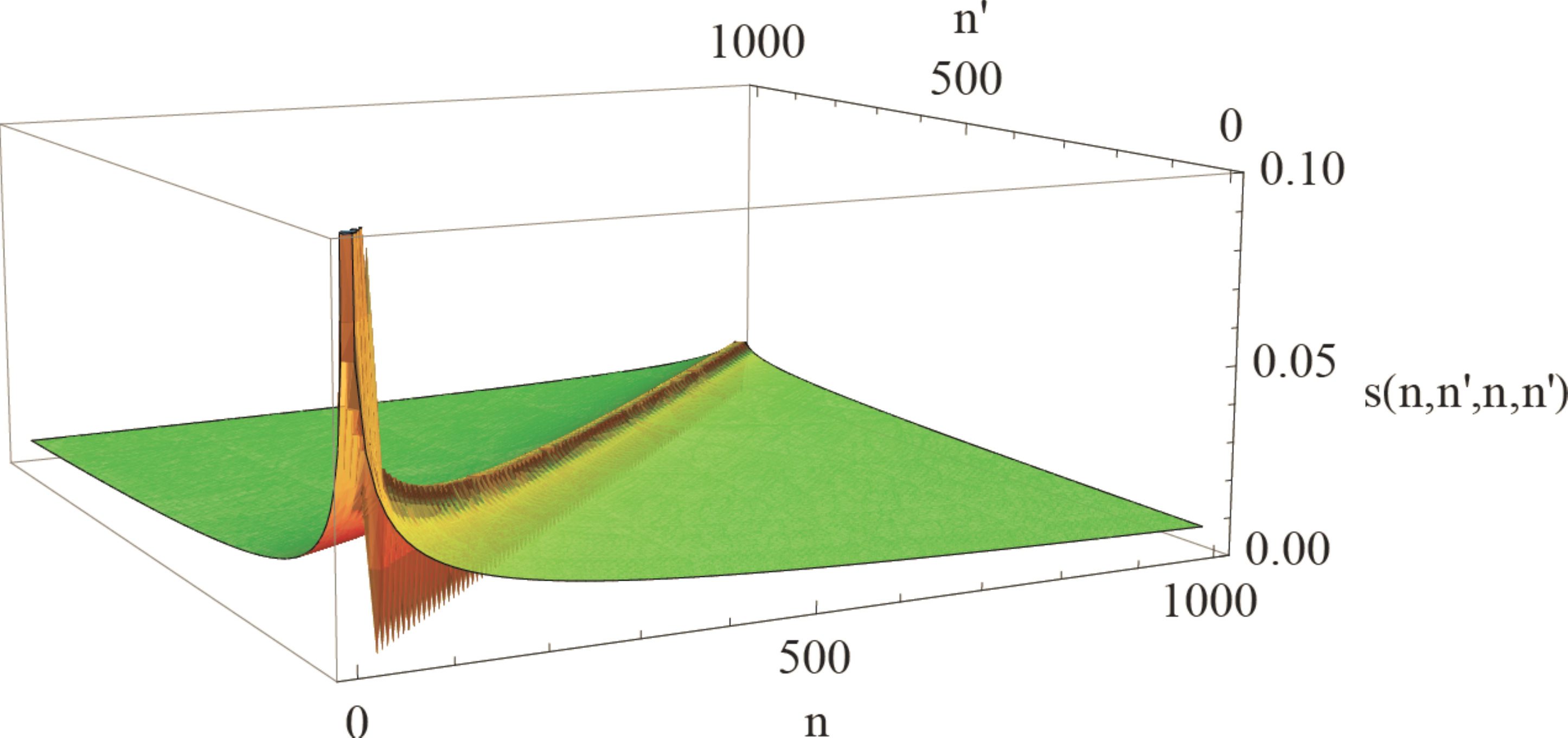}\\
 \includegraphics[width=70mm]{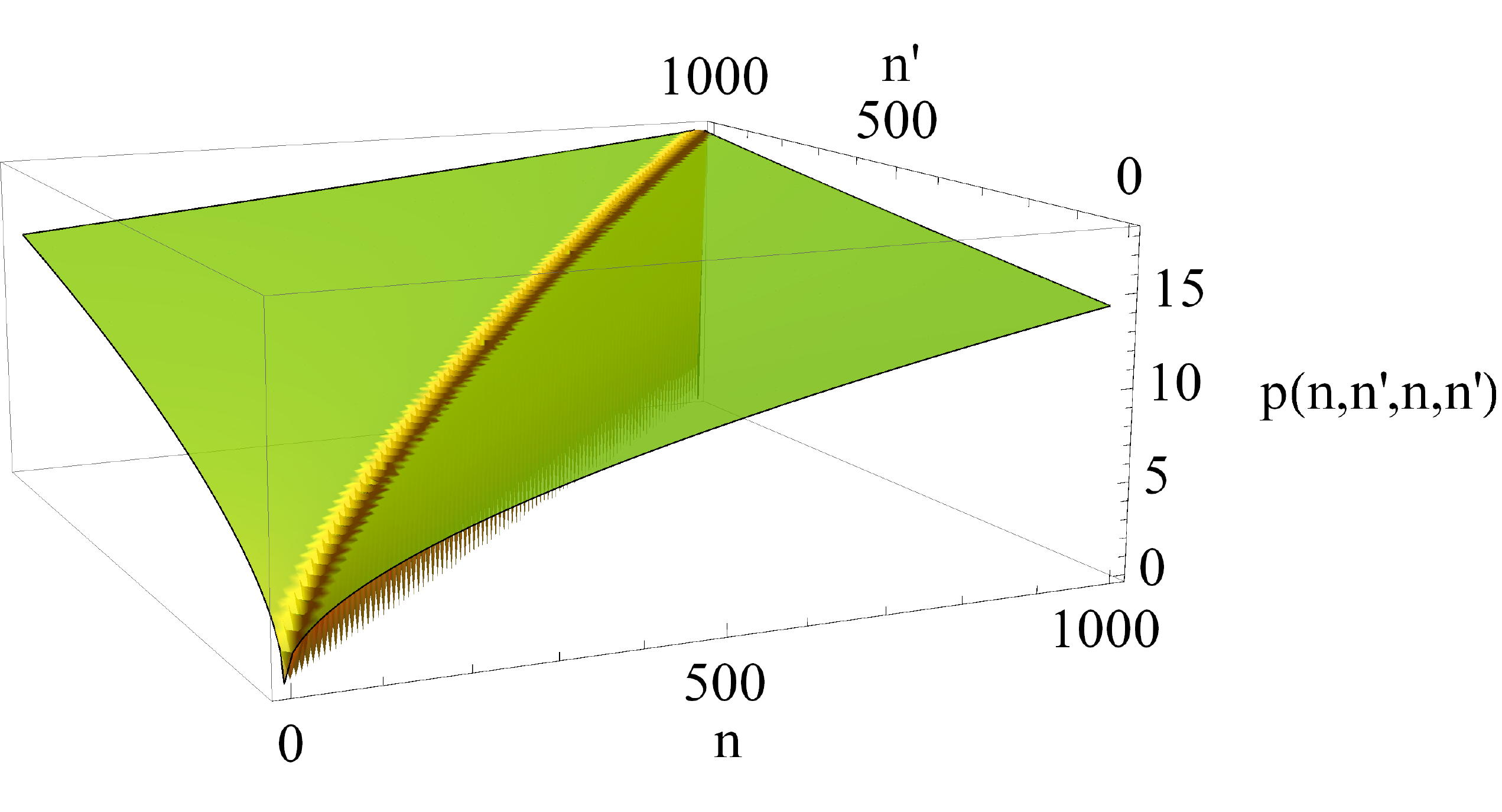}
 \end{center}
  \caption{(Color online)  Mode dependence of the functions $s( n ,n',n,n')$ and $p( n ,n',n,n')$.}
\label{inte}
\end{figure}

\begin{figure}
  \begin{center}
  \includegraphics[width=70mm]{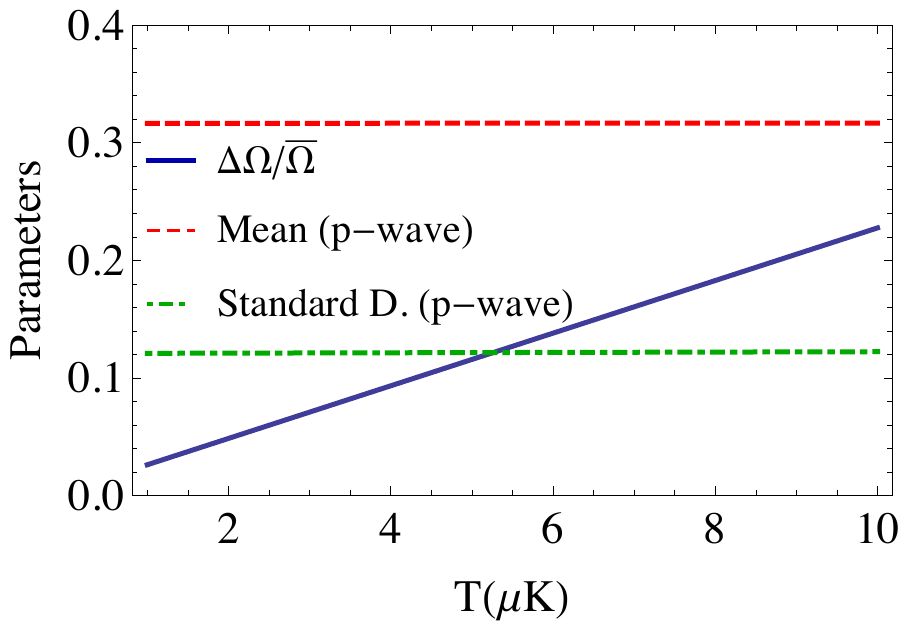}\\
    \includegraphics[width=70mm]{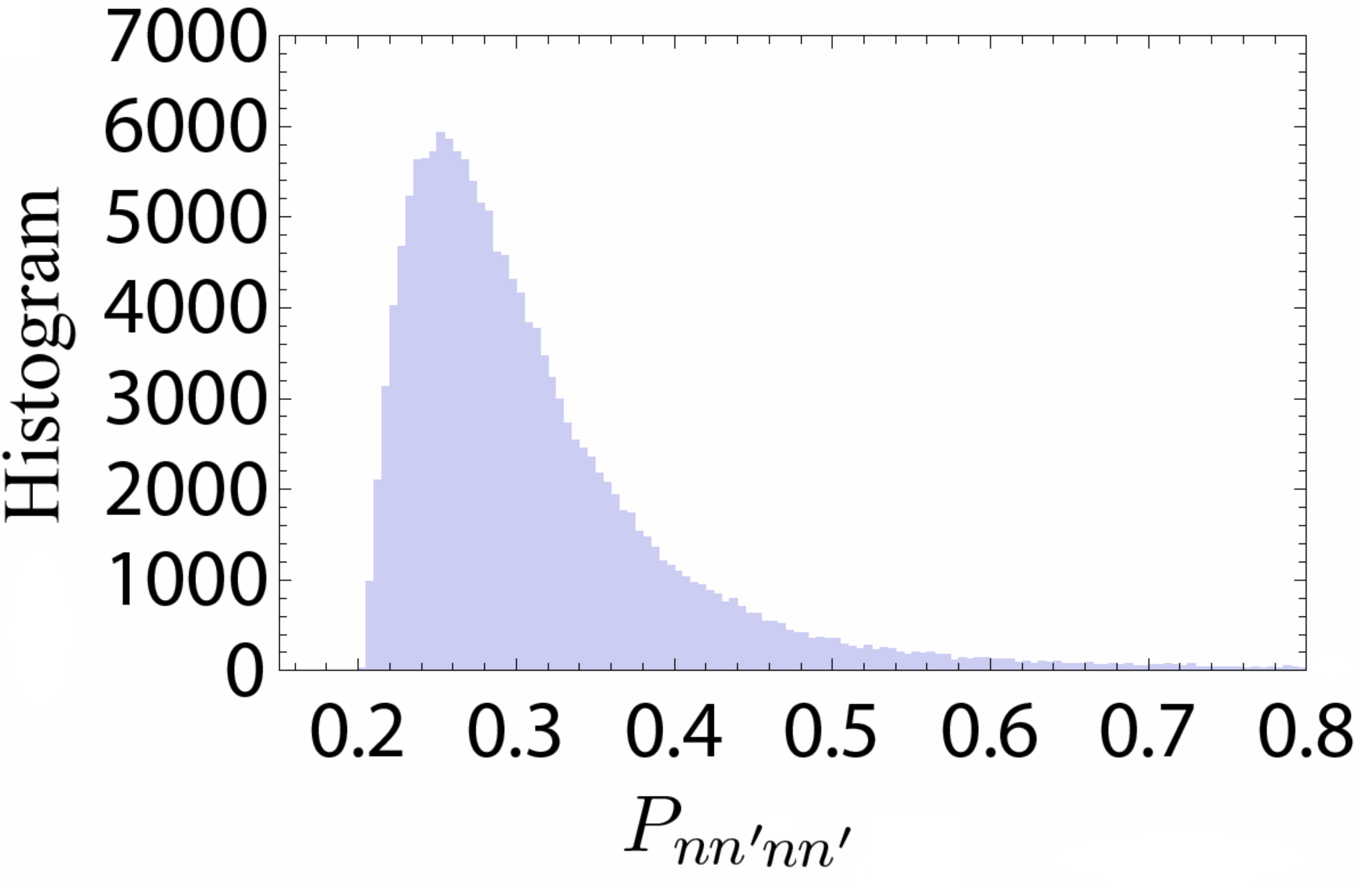}\\
  \end{center}
  \caption{(Color online)  The top panel shows  the mean (red-dashed line) and standard deviation (green-dot-dashed line) of the $p$-wave interactions vs temperature.  It also shows $  \Delta \Omega/\bar{  \Omega}$ (blue-solid line) vs temperature. We used the Yb clock parameters with a misalignment of $\varphi=5$ mrad. The bottom panel is a histogram of the $ P_{\bf {n} \bf {n'}\bf {n} \bf {n'}}$ at 5 $\mu$K. See Eq.~(\ref{phis}). }
\label{inte2}
\end{figure}

 In Fig. \ref{inte} we show the mode dependence of the functions $p( n ,n',n,n')$ and $s( n ,n',n,n')$.
Since those are computed in the harmonic oscillator mode basis, they are long-range in mode-space. While the $p( n ,n',n,n')$ function scales (for $|n-n'|\gg 1$) as $\sqrt{n+n'}$,  and  grows with increasing energy, as expected from $p$-wave interactions, the $s( n ,n',n,n')$ function scales (for $|n-n'|\gg 1$) as $1/\sqrt{|n-n'|}$,  and thus decreases with increasing energy. In Fig. \ref{inte2}, we also show the dependence of the mean and standard deviation of the  $p$-wave interaction parameters $P_{\bf {n} \bf {n'}\bf {n} \bf {n'}} $ as a function of temperature (T). There one can see that  $P_{\bf {n} \bf {n'}\bf {n} \bf {n'}} $ is  almost $T$ independent in this quasi-2D geometry. This is expected because while the actual $p$-wave interactions for fixed density should increase linearly with $T$ \cite{Kanjilal2004}, the latter growth is compensated by the linear decrease with $T$  of the density in a 2D harmonic trap. Fig. \ref{inte2}~(bottom)  shows a histogram of $P_{\bf {n} \bf {n'}\bf {n} \bf {n'}} $ which is peaked  about its average value. The histogram was computed at $T=5\mu$K but it is almost $T$ independent.

 For the laser-atom interaction Hamiltonian we follow Refs.~\cite{Campbell2009,Blatt2009,Rey2009,Gibble2009,Yu2010} and assume that the probe is slightly misaligned with a small component along the  $X$-direction:
 \begin{eqnarray}
\bm{k} = k (\sin\varphi \hat X + \cos\varphi \hat Z),  \label{inhoeq}
\end{eqnarray} with $\varphi \ll 1$  the misalignment angle (see Fig \ref{schma}). We also assume we are in a regime in which laser induced  sideband  transitions can be neglected,  and  define a mode-dependent effective Rabi frequency given by
  \begin{eqnarray}
  \Omega_{\bf{n}} = \Omega_0 L_{n_X}(\eta_X^2) L_0(\eta_Z^2)  e^{-(\eta_Z^2 + \eta_X^2)/2}, \label{laser}
     \end{eqnarray} where $\eta_{X,Z}=k_{X,Z}a_{ho}^{R,Z}$ are the  Lamb-Dicke parameters, $a_{ho}^{\alpha}=\sqrt{\frac{\hbar}{\omega_\alpha m}}$ and $L_n$ are Laguerre polynomials \cite{Wineland1979}.  In Fig. \ref{inte2} we show the ratio between the mean  Rabi frequency, $\bar{  \Omega}$,  and the standard deviation, $  \Delta \Omega$, as a function of temperature. $  \Delta \Omega$ increases with T since the atomic cloud spreads  as it heats up.
In the Lamb-Dicke  regime, $\eta_{X,Z}\ll 1$,  $\hat H_1$ becomes:
\begin{eqnarray}
&&\hat H_1=\sum_{\bf {n}}{\frac{\hbar \Omega_{\bf{n}}}{2 i }(\hat{c}^\dag_{g \bf{n}}\hat{c}_{e \bf{n}}- \rm{h.c})}, \label{many3}
\end{eqnarray}

Under typical operating conditions, $\nu_R\sim 450$~Hz, and  as  will be discussed below, the system is in the regime where its mean  interaction energy per particle is about two orders of magnitude weaker than the energy splitting between neighboring single-particle transverse vibrational modes. We refer to this regime as the vibrational-weakly-interacting regime.  Thus, to leading order, only collision events that conserve the total single-particle energy need to be considered.  Under these conditions the atom population is frozen in the initially populated modes and only the electronic degree of freedom $g,e$ vary during the clock interrogation. For an initial state with at most one atom per mode ($|g\rangle$-polarized state),  it is possible under these conditions to reduce  $\hat H$  to a spin-$1/2$ model with the spin encoded in the $g,e$ states. For  $N$ atoms and labeling the thermally populated harmonic oscillator modes as ${\bf n}_j=(n_{X j},n_{Yj})$ with $j \in\{1,2,\dots N\}$, the spin model can be written as:
\begin{eqnarray}
&&\hat H^{S}_{\vec{\bf n}}/\hbar =-{\delta}\sum_{j=1}^N {\hat  S}^z_{{\bf n}_j}-\sum_{j=1}^N \Omega_{\bf{n}_j} {\hat  S}^y_{{\bf n}_j}+ \sum_{j\neq j'}^N \Big[ J^\perp_{{\bf n}_j ,{\bf n}_{j'}} (\vec{{ S}}_{{\bf n}_j}\cdot \vec{{ S}}_{ {\bf n}_{j'}}) +\chi_{{\bf n}_j ,{\bf n}_{j'}}{ \hat S}^z_{{\bf n}_j}{\hat  S}^z_{ {\bf n}_{j'}}\Big]+ \label{manyspin}\\\notag
 && \sum_{j\neq j'}^N \Big[ \frac{C_{{\bf n}_j ,{\bf n}_{j'}}}{2} ({\hat  S}^z_{{\bf n}_j} I_{{\bf n}_{j'}}+{\hat  S}^z_{{\bf n}_{j'}} I_{{\bf n}_{j}})+\frac{K_{{\bf n}_j ,{\bf n}_{j'}}}{4} I_{{\bf n}_j}I_{{\bf n}_{j'}} \Big].
\end{eqnarray}

Here  $\vec{{S}}_{{\bf n}_j} = \frac{1}{2}\sum_{\alpha,\beta}\hat{c}^\dag_{\alpha {\bf n}_j}\vec{\sigma}_{\alpha \beta}\hat{c}_{\beta {\bf n}_j}$, with $\sigma^{x,y,z}_{\alpha \beta}$ Pauli matrices and   $I_{{\bf n}_j}=\sum_{\alpha,\beta}\hat{c}^\dag_{\alpha {\bf n}_j}\hat{c}_{\beta {\bf n}_j}$ the identity matrix.
\begin{eqnarray}
 J^\perp_{{\bf n}_j ,{\bf n}_{j'}}&=&\frac{V^{eg}_{{\bf n}_j ,{\bf n}_{j'}}-U^{eg}_{{\bf n}_j ,{\bf n}_{j'}}}{2},\\
\chi_{{\bf n}_j ,{\bf n}_{j'}}&=&\frac{V^{ee}_{{\bf n}_j ,{\bf n}_{j'}}+ V^{gg}_{{\bf n}_j ,{\bf n}_{j'}}-2V^{eg}_{{\bf n}_j ,{\bf n}_{j'}}}{2},\\
C_{{\bf n}_j ,{\bf n}_{j'}}&=& \frac{(V^{ee}_{{\bf n}_j ,{\bf n}_{j'}}-V^{gg}_{{\bf n}_j ,{\bf n}_{j'}})}{2}\\
K_{{\bf n}_j ,{\bf n}_{j'}}&=& \frac{(V^{ee}_{{\bf n}_j ,{\bf n}_{j'}}+V^{gg}_{{\bf n}_j ,{\bf n}_{j'}}+V^{eg}_{{\bf n}_j ,{\bf n}_{j'}}+U^{eg}_{{\bf n}_j ,{\bf n}_{j'}})}{2}
\end{eqnarray} The quantities
 \begin{eqnarray}
  &&V^{\alpha\beta}_{{\bf n}_j ,{\bf n}_{j'}}=v^{\alpha, \beta}P_{{\bf n}_j ,{\bf n}_{j'},{\bf n}_{j'},{\bf n}_{j}}\equiv v^{\alpha, \beta} P_{{\bf n}_j ,{\bf n}_{j'}},  \\ && U^{eg}_{{\bf n}_j ,{\bf n}_{j'}}=u S_{{\bf n}_j ,{\bf n}_{j'},{\bf n}_{j'} ,{\bf n}_{j}}\equiv u S_{{\bf n}_j ,{\bf n}_{j'}}, \label{qua}
 \end{eqnarray} encapsulate the temperature dependence of the interactions.

 Note, however, that in the case of  a pure harmonic spectrum, mode changing collisions are energetically allowed even under  weak interactions due to (i) the  linearity of the  harmonic oscillator spectrum and (ii) the separability of the harmonic oscillator potential along the $X$ and $Y$ directions. Condition (i) allows two particles in modes   $(n_X,n_Y)$ and $(m_X,m_Y)$  to collide and scatter into modes $(n_X+k,n_Y+k')$ and $(m_X-k,m_Y-k')$ without violating the energy conservation constraint. Condition (ii) allows the same two particles  to scatter into  modes  $(n_X,m_Y)$ and $(m_X,n_Y)$. Those issues, in principle, can impose important limitations on the validity of the spin model  in a harmonic trap. In practice, however, the trapping potential is not fully harmonic. It comes from the Gaussian beam profile of the lasers and  is given by $V_R\approx -A e^{-\frac{2 R^2}{w_0^2}}$ with $w_0$ the beam waist.  To leading order,  the trapping potential is harmonic $V_R\sim \frac{m\omega_R^2}{2} R^2$ but for an atom in a mode $\{n_X,n_Y\}$ there are higher order corrections of the energy beyond leading order: $E_{\bf n}=\hbar \omega_R(n_X+n_Y+1)+\Delta E_{\bf n}$, with $\Delta E_{\bf n}\sim  \hbar \omega_R\Big (2 \frac{a_{ho}^{R}}{w_0}\Big )^2 \Big(3(n_X^2+n_Y^2) +4{n}_X{n}_Y +5({n}_X +{n}_Y+1)  \Big)$.  At typical operating conditions of the Yb and Sr experiments: $\omega_R\sim 2\pi \times 450$Hz, $w_0\sim 30- 100 \mu$m, $T> 1 \mu$K, and a mean occupation mode numbers  ${\bar n}_{X,Y}> 50 $, the difference of $\Delta E_{\bf n}$ for nearby modes  is larger than $2\pi \times 10$ Hz which is not negligible compared to typical interaction energy scales $\sim$Hz.  The first term in  $\Delta E_{\bf n}$ thus prevents processes (i), while  the second term, which breaks the separability of the potential,  prevents processes (ii). Based on this argument we first restrict our analysis to only processes that  conserve the number of particles per mode.

  A further  simplification of Eq.~(\ref{manyspin}) can be made when atoms are initially prepared in  the totally symmetric Dicke manifold  with $S=N/2$ \cite{Arecchi1972} at time $t=0$.  Here, $S(S+1)$ is the eigenvalue of the collective operator  $\vec{S} \cdot \vec{S}$ and  $\hat S^{\tau=x,y,z} = \sum_{j=1}^N {\hat S}^{\tau}_{\mathbf{n}_j}$.

  In this case there are two important physical mechanisms that prevent leakage of the population   outside the symmetric Dicke manifold. (i) The
  weak dependence of the interaction matrix elements on the thermally populated modes so that the mode-dependent coupling constants $J^{\perp}_{{\bf n}_j,{\bf n}_j'}\vec{{S}}_{{\bf n}_j}$, $\chi_{{\bf n}_j,{\bf n}_j'}$, and $C_{{\bf n}_j,{\bf n}_j'}$ are  peaked at their averages  $J^{\perp}_{\vec{\bf n}}=\frac{\sum_{j\neq j'} J^{\perp}_{{\bf n}_j, {\bf n}_{j'}}}{N(N-1)}$,  $\chi_{\vec{\bf n}}=\frac{\sum_{j\neq j'} \chi_{{\bf n}_j, {\bf n}_{j'}}}{N(N-1)}$
and  $C_{\vec{\bf n}}=\frac{\sum_{j\neq j'} C_{{\bf n}_j,{\bf n}_{j'}}}{N(N-1)}$ (See Figs. \ref{inte}-\ref{inte2}). (ii) The fact that  $J^{\perp}_{\vec{\bf n}}\gg \Delta \chi_{\vec{\bf n}},\Delta C_{\vec{\bf n}}$. Here $\Delta \chi_{\vec{\bf n}},\Delta C_{\vec{\bf n}}$ are the corresponding standard deviations.  The latter is satisfied in part because $J^{\perp}_{\vec{\bf n}}$ is the only interaction term that has a contribution arising from $s$-wave interactions. In general, those are expected to dominate over $p$-wave interactions  since $p$-wave collisions  are suppressed by the centrifugal barrier which  is estimated to be  greater than $\sim 25\mu$K \cite{Campbell2009,Lemke2011}. It must be said, nevertheless, that the actual values of the $s$-wave and $p$-wave scattering parameters between two $e$ atoms and one $e$ and one $g$  are not known.  (i) and (ii)  impose  a large energy gap in the Hamiltonian which suppresses transitions between manifolds with different total collective spin  $S$, caused by the inhomogeneities $\Delta \chi$ and $\Delta C$ \cite{Rey2008a}. Consequently, to a very good approximation it is expected that the dynamics can be projected into the $S=N/2$ manifold with an effective Hamiltonian given  in terms of collective operators:
 \begin{eqnarray}
 &&\hat H^{S}_T = \hat H^{S}+  \hat H^{S}_\Omega\\
&&\hat H^{S}/ \hbar =  - \delta \hat S^{z}  + J^\perp_{\vec{\bf n}}\vec{S}\cdot\vec{S}+ \chi_{\vec{\bf n}} \left(\hat S^{z}\right)^{2}+ C_{\vec{\bf n}} \left(N-1\right) \hat S^{z}  \notag ,\\
&&\hat H^{S}_\Omega = - \hbar \bar{\Omega} \hat S^{y},
\label{Hamiltonian}
\end{eqnarray} The term $\vec{S}\cdot\vec{S}$ is a constant of motion and does not play any role in the collective dynamics.

In addition to the above conditions on the interactions, staying in the $S=N/2$ manifold requires that the probing laser generates negligible excitation inhomogeneity, {\it i.e.}~$\Delta\Omega/\bar{\Omega} \ll 1$. In this case we can write the  atom-light Hamiltonian  in terms of collective operators. The validity of this condition depends on the misalignment angle, $ \varphi$ [Eq.~(\ref{inhoeq})], and the average vibrational mode population \cite{Blatt2009}. At the current operating conditions, the JILA Sr clock can achieve  $\Delta\Omega/{\bar \Omega} < 0.1$ \cite{Martin2012},  and the collective mode approximation for the atom-light Hamiltonian is well satisfied. The latter is not necessarily the case for the Yb optical clock experiment at NIST due to the higher temperatures, as we will elaborate later. The restriction of the dynamics to the  Dicke manifold is relaxed in Sec.~\ref{noncoll} where we also investigate the parameter regime in which it is valid.

\section{ Ramsey interrogation: Collective case}

\begin{figure}
  \begin{center}
 \includegraphics[width=70mm]{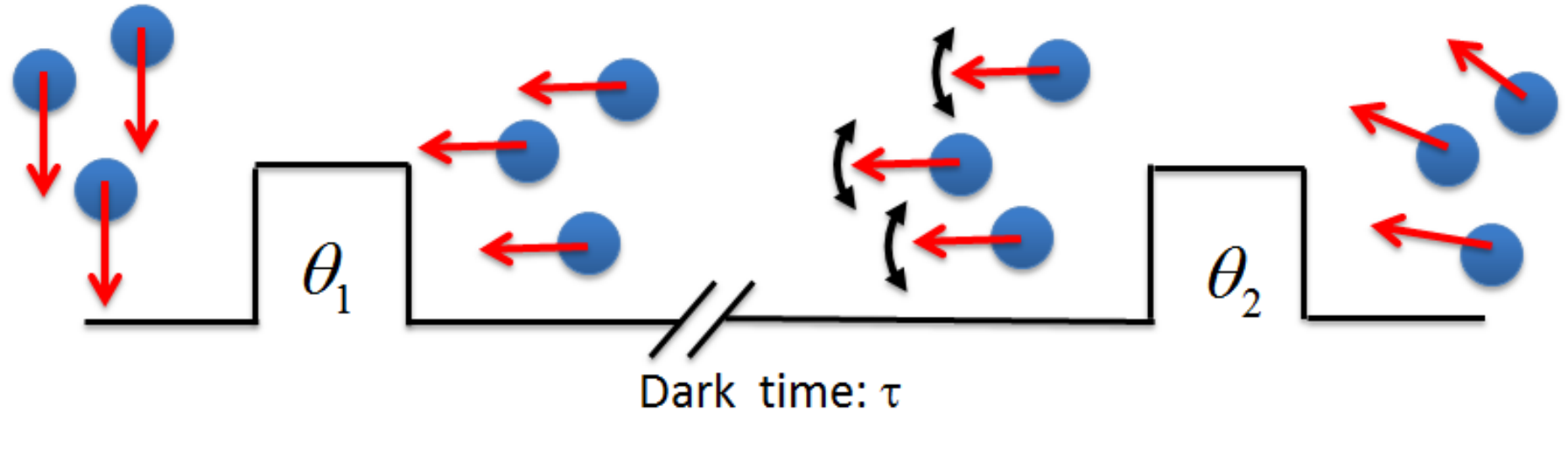}
 \end{center}
  \caption{(Color online) In  Ramsey spectroscopy two short pulses of area $\theta_1$ and $\theta_2$ are applied,  separated
by a dark time $\tau$.}
\label{ram}
\end{figure}

In Ramsey spectroscopy (see Fig.~\ref{ram}), a well established tool in atomic physics, atoms are typically
prepared in the same internal state, say $g$ (e.g. via optical pumping). Next,
one applies  a strong resonant linearly polarized light pulse for time $t_1$.
 By strong, we mean that the Rabi frequency must be much larger than the atomic interaction energy scales but weaker than  the harmonic oscillator frequency, $\nu_R$, to avoid laser induced mode changing processes.

This first pulse rotates the spin state of the atom at mode ${\bf n}$ to $\cos(\theta_1^{\bf n})  |g\rangle+\sin(\theta_1^{\bf n}) |e\rangle$, here $\theta_1^{\bf n}=\Omega_{\bf n} t_1$ is the pulse area. Subsequently, the  atoms are allowed to freely evolve for a dark time $\tau$. Finally a second pulse   is applied for a time $t_2$ and the population of  the $e,g$  states  measured.

In the following, we will discuss the case of a fixed number of atoms, $N$, prepared in the modes $\vec{\bf{n}}=\{\bf{n}_1,\dots,{\bf n}_N \}$, and only later we will worry about spatial and thermal averages.
Note that under  the collective-mode approximation $\theta_{1,2}=\bar{\Omega} t_{1,2}$.

\subsection{Analytic solution}

Let us  first start dealing with the situation in which there is no excitation inhomogeneity and the initial (second) pulses are just  rotations of the collective Bloch vector by an angle $\theta_{1} (\theta_{2})$, respectively.  In this case,  the state of the system before the free dynamics is

\begin{small}
  \begin{eqnarray}
|\psi (0^-)\rangle=\sum_{k=0}^{N} \sqrt{
                                              N \choose
                                              k
} ( \cos\frac {\theta_1}{2})^{N-k} (\sin \frac {\theta_1}{2})^k|N/2,k-N/2\rangle, \notag \label{state}
\end{eqnarray} \end{small}with  $|N/2,M\rangle$ being  collective Dicke states. During the free evolution, the Hamiltonian reduces to  $  \chi_{\vec{\bf n}}  ({\hat S}^{z})^2 + C_{\vec{\bf n}}  (N-1) { S}^z$  which introduces just  a phase, $e^{-i \tau[ \chi_{\vec{\bf n}} (M)^2 + C_{\vec{\bf n}}  (N-1) { M}]} $ to each of the $|N/2,M\rangle$  states. Here $M$ is an eigenvalue of ${ S}^z$ and takes integer or half-integer values (depending on whether $N$ is even or odd) satisfying  $-N\leq 2M \leq N$. Expressions for the evolution of the  spin operators after a dark-time evolution $\tau$ can be exactly computed:

{\small
  \begin{eqnarray}
  &&\langle {\hat S}^{x}\rangle=-\frac{N \sin (\theta_1)  }{2} Z_{\vec{\bf n}}^{N-1} \cos\left[ \tau \left(\delta -2 \pi \Delta\nu_{\vec{\bf n}}\right)\right],  \label{shipcx}\\
   &&\langle {\hat S}^{y}\rangle=-\frac{N \sin (\theta_1)}{2}Z_{\vec{\bf n}}^{N-1} \sin\left[ \tau \left(\delta -2 \pi \Delta\nu_{\vec{\bf n}}\right)\right], \label{shipcy} \\
 &&\langle {\hat S}^{z}\rangle=-\frac{N \cos\theta_1}{2}, \label{shipcz}
    \end{eqnarray}} with $Z_{\vec{\bf n}}$ and $\zeta_{\vec{\bf n}}$ and $\Delta\nu_{\vec{\bf n}}$ given  by

  \begin{eqnarray}
&& Z_{\vec{\bf n}}^2\equiv 1-\sin^2(\theta_1) \sin^2 \left (\chi_{\vec{\bf n}} \tau\right),\label{Zn} \\
&& \tan (\zeta_{\vec{\bf n}}\tau) \equiv \tan \left (\chi_{\vec{\bf n}} \tau\right)\cos(\theta_1 ), \label{zetan}\\
&&2 \pi \Delta\nu_{\vec{\bf n}}=(N-1)(C_{\vec{\bf n}}-\zeta_{\vec{\bf n}} ). \label{spn}
\end{eqnarray}


The normalized contrast, which is a measure of the amplitude  of the Ramsey fringes, is defined   as the magnitude of the projection of the collective Bloch vector on the x-y plane  normalized by half of the total number of atoms:
 \begin{eqnarray}
   {\mathcal C}&\equiv& \frac{2 \sqrt{\langle {\hat S}^{x}\rangle^2+\langle {\hat S}^{y}\rangle^2}}{N}, \label{defco}
 \end{eqnarray} from the above expression one obtains that
  \begin{eqnarray}
   {\mathcal C}&=&\Big |\sin (\theta_1) Z_{\vec{\bf n}}^{N-1}\Big|.
 \end{eqnarray}

 The contrast is  extracted   from measurements of the  fraction  of excited atoms, $N_{e, {\bf \vec n}}(t_1,t_2)/N$ by varying the laser  detunning, or by varying the phase of the second pulse along the x-y plane.  For the former case:

 \small{ \begin{eqnarray}
&&{\mathcal C}=\frac{1}{N \sin (\theta_2)} \sqrt{\Big(N_{e,\vec{\bf n}}\Big|_{\delta=0}-N_{e,\vec{\bf n}}\Big|_{\tau\delta=\pi}\Big)^2+
\Big(N_{e,\vec{\bf n}}\Big|_{\delta=0}+ N_{e,\vec{\bf n}}\Big|_{\tau\delta=\pi}-2 N_{e,\vec{\bf n}}\Big|_{\tau\delta=\pi/2} \Big)^2}\\
&&N_{e,\vec{\bf n}}=  \frac{N}{2}+\cos(\theta_2) \langle {\hat S}^{z}\rangle-\langle {\hat S}^{x}\rangle \sin(\theta_2). \label{shipcs}
    \end{eqnarray}}

Let's now discuss the physics encapsulated in Eqs.~(\ref{shipcx}-\ref{shipcz}). The term $\Delta\nu_{\vec{\bf n}}$ is the so called density-dependent frequency shift,  which gives rise to a density-dependent measurement of the atomic transition frequency. The quantity  $Z_{\vec{\bf n}}^{N-1}$
determines the contrast of the Ramsey fringes.  In the weakly interacting regime, $\chi_{\vec{\bf n}} \tau \ll 1$,  and for  $N\gg 1$,   $2 \pi \Delta\nu_{\vec{\bf n}}\approx ( N C_{\vec{\bf n}}+ 2  \chi_{\vec{\bf n}} \langle {\hat S}^{z}\rangle)$. This means that interactions  act as an effective magnetic field along the quantization axis with magnitude  $B_{\rm{eff}}=  ( N C_{\vec{\bf n}}+ 2  \chi_{\vec{\bf n}} \langle {\hat S}^{z}\rangle)$, which depends both on the total atom number and the population difference between excited and ground state. This is consistent with just a simple interpretation of the frequency shift as being    the average energy difference experienced by  an atom in state $e$ with respect to an atom in state $g$
 due to the presence of other atoms. As we explain in Sec.~\ref{mef}, this can be derived from a  mean field analysis which factorizes the interaction term as  $ \langle ({\hat S}^{z})^2\rangle \propto {\hat S}^{z}\langle {\hat S}^{z}\rangle$.  In this regime,
 the condition $\cos(\theta_1^{\rm {opt}})= C_{\vec{\bf n}}/\chi_{\vec{\bf n}}$ determines the
 pulse area $\theta_1^{\rm {opt}}$ at which the shift is canceled. This is the ideal operating pulse area for a clock \cite{Ludlow2011}.  Note that if $C_{\vec{\bf n}}$ is equal to zero, no density shift is expected at $\langle {\hat S}^{z}\rangle=0$. This is consistent  with the intuition that  due to the equal population of both $e$ and $g$ states, the mean energy  shift experienced  by an atom in state $e$ due to others is exactly  canceled by the opposite energy shift experienced by  an atom in $g$. In this weakly interacting regime, $Z_{\vec{\bf n}}\approx 1$ as expected in the case  that interactions act just as a mere effective magnetic field, which causes   the Bloch vector just to precess with  no Ramsey fringe-contrast decay.

 Outside the weakly interacting regime,  two important corrections to this picture arise. One is the fact that the shift is no longer linear in $\langle {\hat S}^{z}\rangle$ [from Eq. (\ref{zetan})]. The second one is that the Ramsey fringe-contrast  collapses and revives. The collapse is well approximated by a Gaussian decay,  $Z_{\vec{\bf n}}^{N}\sim e^{-N /2 \sin^2(\theta_1) \chi_{\vec{\bf n}} ^2\tau^2}$. The revivals take place  at times  $ \chi_{\vec{\bf n}} T_n= n \pi $ with $n$ an integer. This behavior of the contrast  is closely linked to the decay of coherence in a matter-wave  due to the nonlinearities arising from the atom-atom
interactions and subsequent revival due to the discreteness of the spectrum of the many-body system. The  observation of collapses and revivals of a Bose-Einstein condensate (BEC) loaded in an optical lattice  was first reported in Ref.~\cite{Greiner2002b}.

\subsection{Mean-field solution}\label{mef}
Even though the all-to-all interactions  allows for an exact solution of the many-body dynamics  in Ramsey spectroscopy, it is convenient to introduce an approximate mean-field treatment. The mean-field treatment will be very helpful for dealing with inelastic collisions,  which are   experimentally relevant.
A simple and enlightening way to carry out a mean-field treatment is to use the Schwinger-boson representation, which maps spin operators to two-mode  bosons subject to a constraint \cite{Auerbach1994}. It represents the spin operators as
\begin{figure}
  \begin{center}
  \includegraphics[width=80mm]{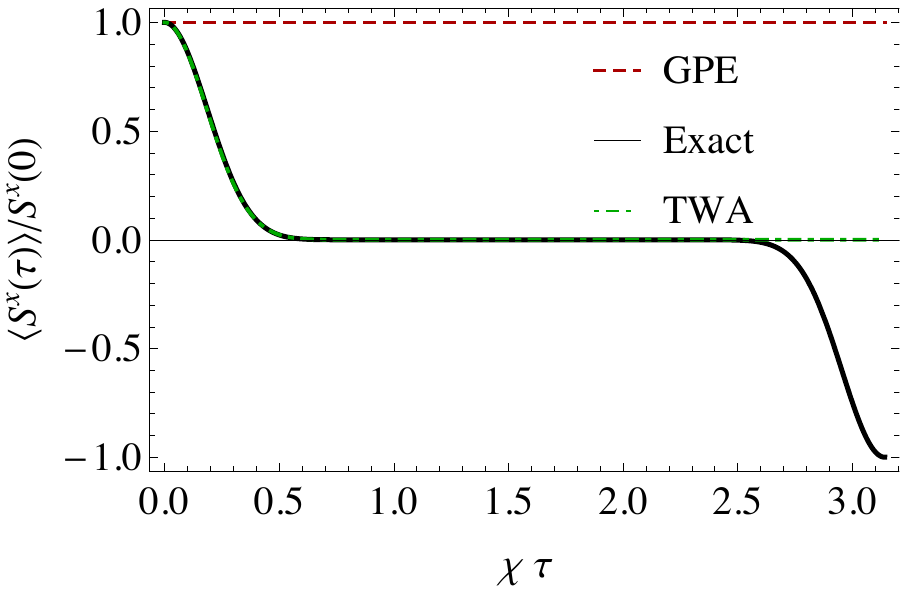}\\
\includegraphics[width=80mm]{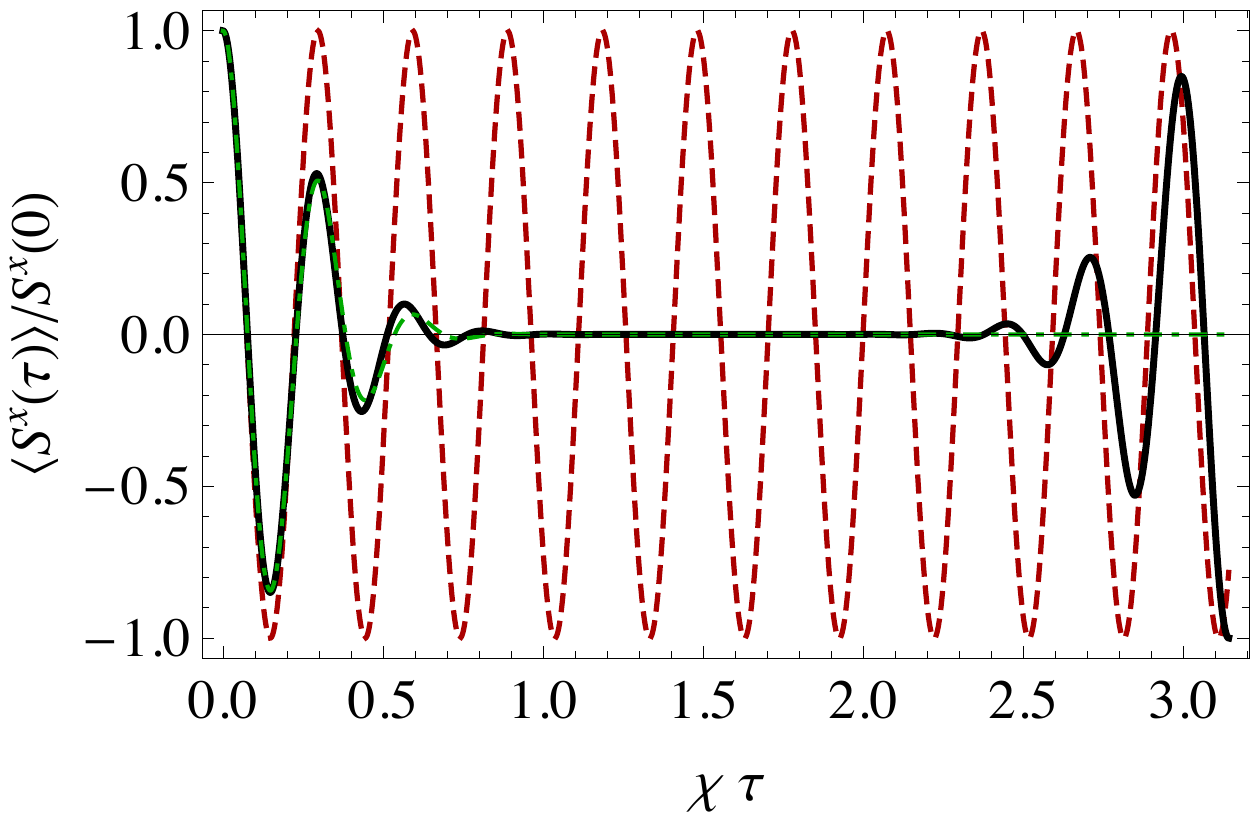}
  \end{center}
  \caption{(Color online) {Comparisons between the  GPE (mean-field, red-dashed  line), the  TWA  ( green-dot-dashed line) and the exact solution (black-solid line)}. Here we assumed $N=30$ and $C_{\vec{\bf n}} =\delta=0$.
  The top plot is for $\theta_1=\pi/2$, the lower plot for $\theta_1=\pi/4$.}
\label{comp}
\end{figure}
  \begin{eqnarray}
2{\hat S}^z&=&\hat{\Psi}_e^\dagger \hat{\Psi}_e-\hat{\Psi}_g^\dagger \hat{\Psi}_g, \label{sch}\\
{\hat S}^+&=&\hat{\Psi}_e^\dagger \hat{\Psi}_g, \\
{\hat S}^-&=&\hat{\Psi}_g^\dagger \hat{\Psi}_e, \\
\end{eqnarray} with the constraint
\begin{eqnarray}
N&=&\hat{\Psi}_e^\dagger \hat{\Psi}_e+\hat{\Psi}_g^\dagger \hat{\Psi}_g, \label{schc}
\end{eqnarray} with $\hat{\Psi}_\alpha$ a bosonic annihilation  operator of mode $\alpha=e,g$.
The mean-field treatment,  which gives rise to the so-called Gross-Pitaevskii Equation (GPE) \cite{pethick02}, replaces the field operators $\hat{\Psi}_{e,g}$ by c-numbers, $\hat{\Psi}_{e,g}\to \Phi_{e,g}$. The latter approximation is justified in the weakly interacting regime when
 there is a macroscopic population of those modes and the state of the system can be regarded  as a simple product state with no significant entanglement.  Translated to the spin language, those conditions imply that  the system can be well described as a spin coherent state.
For  purposes that will become clearer later, it is convenient to introduce the density matrix $\rho_{\alpha\alpha}=N_\alpha$ and $\rho_{\alpha\beta}={\Phi}_{\alpha}^*{\Phi}_{\beta}$. The latter satisfies the following equations of motion:

\begin{eqnarray}
&& \frac{\partial}{\partial t} \langle{{\hat S}^{z}}\rangle^{\rm{GPE}}= \bar{\Omega} \langle{{\hat S}^{x}}\rangle^{\rm{GPE}},\label{gpe3}\\
&&\frac{\partial}{\partial t}  {\rho}_{eg}= \frac{\partial}{\partial t} {\rho}_{ge}^* \equiv M_{eg} \label{egg}, \\
&& M_{eg}\equiv -\bar{\Omega} \langle {\hat S}^{z}\rangle^{\rm GPE}-i \left [\delta-N C_{\vec{\bf n}}-2\chi_{\vec{\bf n}}\langle {\hat S}^{z}\rangle^{\rm GPE}\right ] {\rho}_{eg}. \label{geg}
\end{eqnarray} Note ${\rho}_{ge}={\rho}_{eg}^*$ and
 \begin{eqnarray}
&& \langle{{\hat S}^{z}}\rangle^{\rm{GPE}}=\frac{{\rho}_{ee}-{\rho}_{gg}}{2},\\
&&\langle{{\hat S}^{x}}\rangle^{\rm{GPE}}=\frac{{\rho}_{eg}+{\rho}_{ge}}{2},\\ &&\langle{{\hat S}^{y}}\rangle^{\rm{GPE}}=\frac{{\rho}_{eg}-{\rho}_{ge}}{2 { i}},
\end{eqnarray} are  the components of the Bloch vector.

For the Ramsey dynamics, the mean-field treatment is almost trivial, and  time evolution  corresponds to a precession of the Bloch vector induced by an effective magnetic field
 \begin{eqnarray}
B^{\rm eff}_{\vec{\bf n}}=2\pi \Delta \nu_{\vec{\bf n}}^{\rm GPE}=N[C_{\vec{\bf  n}}-\chi_{\vec{\bf n}}\cos(\theta_1 )]. \label{beff}
\end{eqnarray} This behavior can be clearly seen in Eq. (\ref{egg}) by noticing that during the dark time $\bar{\Omega}=0$,  thus $\langle {\hat S}^{z}\rangle^{\rm GPE}$ is a constant of motion, and ${\rho}_{eg}$ precesses at a rate $\delta- 2\pi \Delta \nu_{\vec{\bf n}}^{\rm GPE}$. Note  that in striking disagreement  to the exact many-body solution, at mean-field level, the Ramsey fringe contrast never decays \cite{FossFeig2013b}.
This implies that it agrees with the exact solution only when $Z_{\vec{\bf n}}^N-1\ll 1$, which can be satisfied when $\sqrt{N}\sin(\theta_1) \chi_{\vec{\bf n}}\tau\ll 1$ , {\it i.e.}  short times, weak interactions, or small pulse areas.

\subsection{  Truncated Wigner Approximation  }
The truncated Wigner Approximation (TWA)  has proven to be a successful approach to incorporate the leading quantum corrections to
the mean-field dynamics. See, for example, Refs. \cite{Blakie2008,Polkovnikov2010} and references therein. To implement  the TWA, one needs to solve the mean-field equations of motion
supplemented by random initial conditions distributed according to the Wigner function. For a spin coherent state   with $S=N/2$, pointing initially in the x-z plane,  $\hat{n}_{\theta_1}=(\sin\theta_1, 0,-\cos\theta_1)$, the Wigner function is given by:

   \begin{eqnarray}
  \wp(S^x_0,S^y_0,S^z_0)&=&\left(\frac{1}{\pi S}\right)\delta(S^z_0 \cos\theta_1-S^x_0 \sin\theta_1+S) e^{-\frac{ (S^y_0)^2}{S}} e^{-\frac{ (S^z_0 \sin\theta_1+S^x_0 \cos\theta_1)^2}{S}}.\label{distr}
   \end{eqnarray} This Wigner function has a transparent interpretation. For example, if the Bloch vector points  along the direction $-\hat{z}$, {\it i.e.} $\theta_1=0$, because of the uncertainty principle, the transverse  components still fluctuate  so that  $\langle \hat{S}^y \hat{S}^y\rangle=\langle \hat{S}^x \hat{S}^x\rangle=S/2$. Quantum mechanical expectation values of the spin  operators relevant for this work can be computed  as $\overline{\langle {\mathcal O} ({\tau})\rangle }=\int dS^x_0 dS^y_0 dS^z_0 \langle {\mathcal O} ({\tau})\rangle \wp(S^x_0,S^y_0,S^z_0)$ with $\langle {\mathcal O} ({\tau})\rangle$ the classical evolution of the observable calculated using  the mean-field equations.

Using this prescription we obtain:

\begin{eqnarray}
 \overline{\langle {\hat S}^{x}(\tau)\rangle}&=&\rm{Re}[\overline{\langle {\hat S}^{+}(\tau)\rangle}] \label{twax},\\
\overline{\langle {\hat S}^{y}(\tau)\rangle}&=&\rm{Im}[\overline{\langle {\hat S}^{+}(\tau)\rangle}] \label{tway},\\
 \overline{\langle {\hat S}^{z}(\tau)\rangle}&=&-\frac{N \cos\theta_1}{2} \label{noisegp},\\
 \overline{\langle \hat S^+(\tau)\rangle}&=& -\frac{N \sin (\theta_1 )}{2}(1+\rm{i}\chi_{\vec{\bf n}} \tau \cos[\theta_1])  e^{ -\frac{N}{2} \chi_{\vec{\bf n}}^2
   \tau^2 \sin ^2(\theta_1)}  e^{\rm{i}\tau(\delta-2\pi \Delta \nu_{\vec{\bf n}}^{\rm GPE})} \label{twa+}.
   \end{eqnarray} Here $\rm{Re}$ and $\rm{Im}$ correspond to the real and imaginary parts, respectivelly.
      The essential physics is encapsulated  in the  term that exhibits a Gaussian decay, proportional to $e^{ -\frac{N}{2} \chi_{\vec{\bf n}}^2
   \tau^2 \sin ^2(\theta_1)}$,  and in the   phase shift, proportional to  $2\pi \Delta \nu_{\vec{\bf n}}^{\rm GPE}$. The exponential term leads to a decay of the Ramsey fringe contrast  missing at the mean-field level. Note that in addition to the exponential decay of the contrast and the mean-field phase shift, there is an extra  term, $\rm{i}\chi_{\vec{\bf n}} \tau \cos[\theta_1]$ in  Eq.(\ref{twa+}). For $|\chi_{\vec{\bf n}} \tau|<1$, this term just introduces  a finite $N$ correction ({\it i.e.} replaces $N\to N-1$) in the phase shift, irrelevant in the large-$N$ limit.
    The  TWA approximation, nevertheless, fails to  reproduce the periodic revivals of the coherence,  which take place in the exact solution.   However, for current experimental probing times,  which are much shorter than the revival time, the TWA is  an excellent alternative for capturing the quantum dynamics. Fig. \ref{comp} shows comparisons between the GPE, the  TWA and the  exact solution for  different pulse areas and for $N=30$.
   The power of the  TWA is demonstrated in its capability to fully capture the decay of the Ramsey fringe contrast. We note that other approximation methods, typically used to account for quantum fluctuations beyond mean-field, such as the time-dependent Bogoliubov approximation, generally fail to reproduce the decay of quantum coherences \cite{Rey2004}.  Moreover, as we will show below in Sec.~\ref{loss}, the  TWA can be straightforwardly generalized to deal with the many-body dynamics of an open quantum system.

\section{ Ramsey interrogation: Including losses}\label{loss}

\subsection{Master equation}

The Hamiltonian formulation described above is valid only for a closed system.  To account for losses due to inelastic   $e$-$e$ or $e$-$g$ collisions, recently observed in experiments \cite{Bishof2011b,Ludlow2011}, one needs to use instead a master equation:
\begin{equation}
\hbar \frac{d}{dt} {\hat \rho}= -{\rm i}[ {\hat H_{\vec{\bf n}}^S} , {\hat \rho} ] +  {\mathcal L}{\hat\rho}. \label{master}
\end{equation} Here  $\hat \rho$ is the  reduced density matrix operator of the many-body system. $\hat H_{\vec{\bf n}}^S$ is the  Hamiltonian given by Eq. (\ref{manyspin}),  and $ {\mathcal L}$ is a is a Lindbladian superoperator that accounts for inelastic processes. Considering $p$-wave $e$-$e$ and $e$-$g$ losses ($s$-wave losses do not occur if one restricts the dynamics  to the fully symmetric manifold \cite{FossFeig2012} as we will assume below),  $ {\mathcal L}$ is given by
{\small
    \begin{eqnarray}&& \frac{\mathcal L}{\hbar}=\sum_{j\neq j',\alpha=ee,eg}  \Gamma^\alpha_{{\bf n}_j,{\bf n}_{j'}} \Big[\hat{A}^\alpha_{{\bf n}_j,{\bf n}_{j'}} {\hat \rho} (\hat{A}^\alpha_{{\bf n}_j,{\bf n}_{j'}} )^\dagger \Big] \label{master}\\&& \notag- \sum_{j\neq j',\alpha=ee,eg}  \frac{\Gamma^\alpha_{{\bf n}_j,{\bf n}_{j'}}}{2}\Big[\big(\hat{A}^\alpha _{{\bf n}_j,{\bf n}_{j'}} )^\dagger \hat{A}_{{\bf n}_j,{\bf n}_{j'}}  {\hat \rho} + {\hat \rho}
(\hat{A}^\alpha_{{\bf n}_j,{\bf n}_{j'}} )^\dagger \hat{A}^{\alpha}_{{\bf n}_j,{\bf n}_{j'}} \Big ].
\end{eqnarray} }Here the jump operators are $\hat{A}^{ee}_{{\bf n}_j,{\bf n}_{j'}}=\hat{c}_{e {\bf n}_j}\hat{c}_{e {\bf n}_{j'}}$ and
 $\hat{A}^{eg}_{{\bf n}_j,{\bf n}_{j'}}=(\hat{c}_{e {\bf n}_j}\hat{c}_{g {\bf n}_{j'}}+\hat{c}_{g {\bf n}_j}\hat{c}_{e{\bf n}_{j'}})/\sqrt{2}$, while $\Gamma^{ee,eg}_{{\bf n}_j,{\bf n}_{j'}}=\gamma^{ee,eg} P_{{\bf n}_j {\bf n}_{j'} }$. The expression for $\gamma^{ee,eg}$ is identical to $v^{e,e}$ and $v^{e,g}$ respectively (see Eq.~(\ref{Habe})) up to the replacement of the $p$-wave elastic scattering volume by the inelastic one.

The last two terms in Eq.~(\ref{master}) can be thought of as generating an effective Hamiltonian, in which   the terms $\{V^{ee}_{{\bf n}_j,{\bf n}_{j'}} ,V^{ eg}_{{\bf n}_j,{\bf n}_{j'}} \}$ are replaced by $\{(V^{ee}_{{\bf n}_j,{\bf n}_{j'}}-\frac{i}{2} \Gamma^{ee}_{{\bf n}_j,{\bf n}_{j'}}),(V^{eg}_{{\bf n}_j,{\bf n}_{j'}}-\frac{i}{2} \Gamma^{eg}_{{\bf n}_j,{\bf n}_{j'}})\}$ respectively.  On the other hand, the first term in Eq.~(\ref{master}), the so-called recycling term, requires a full density matrix formulation and thus  the dimension of the Hilbert space  grows much faster with  $N$ than the corresponding dimension of the pure Hamiltonian case. For example just for 4 particles the required Hilbert space is $\sim 10^3$ states. Nevertheless, keeping the first term is crucial to conserve the trace of the density matrix.

 An important point to keep in mind is that the recycling term connects sectors of the Hilbert space with different atom number. However, since the environment knows how many particles left the system,   there are no coherences between sectors of different atom numbers and the master equation can be solved in a ``block-diagonal way'' \cite{Gorshkov2012}.

 The master equation can be further simplified under the collective mode approximation. Let $\rho_\mathcal{N}$ be the density matrix for a single sector of $\mathcal{N}$ particles, and let's assume that, at time $t=0$, we start in the sector with $\mathcal{N}=N$ particles.  As explicitly shown  in Appendix 1,  to solve for the master-equation dynamics, we need to solve a series of differential equations for each of the subspaces with cascading atom numbers.
  The sector $\mathcal{N}=N$  does not have driving terms and can be solved by merely solving the effective Hamiltonian dynamics. We  then use the
    integrated solutions in the   $\mathcal{N}=N$  sector  as a driving terms  for $\mathcal{N} = N-2$, the latter for $\mathcal{N}= N-4$, etc.
    After the dynamics of  all the density matrix  sectors are known, one can compute any observable. For $N \leq 50$,  the above procedure can be efficiently performed numerically \footnote{ At the highest operating densities of current optical clock experiments a cut-off of 50 atoms per lattice site  is sufficient}.

\begin{figure*}
  \begin{center}
  \includegraphics[width=55mm]{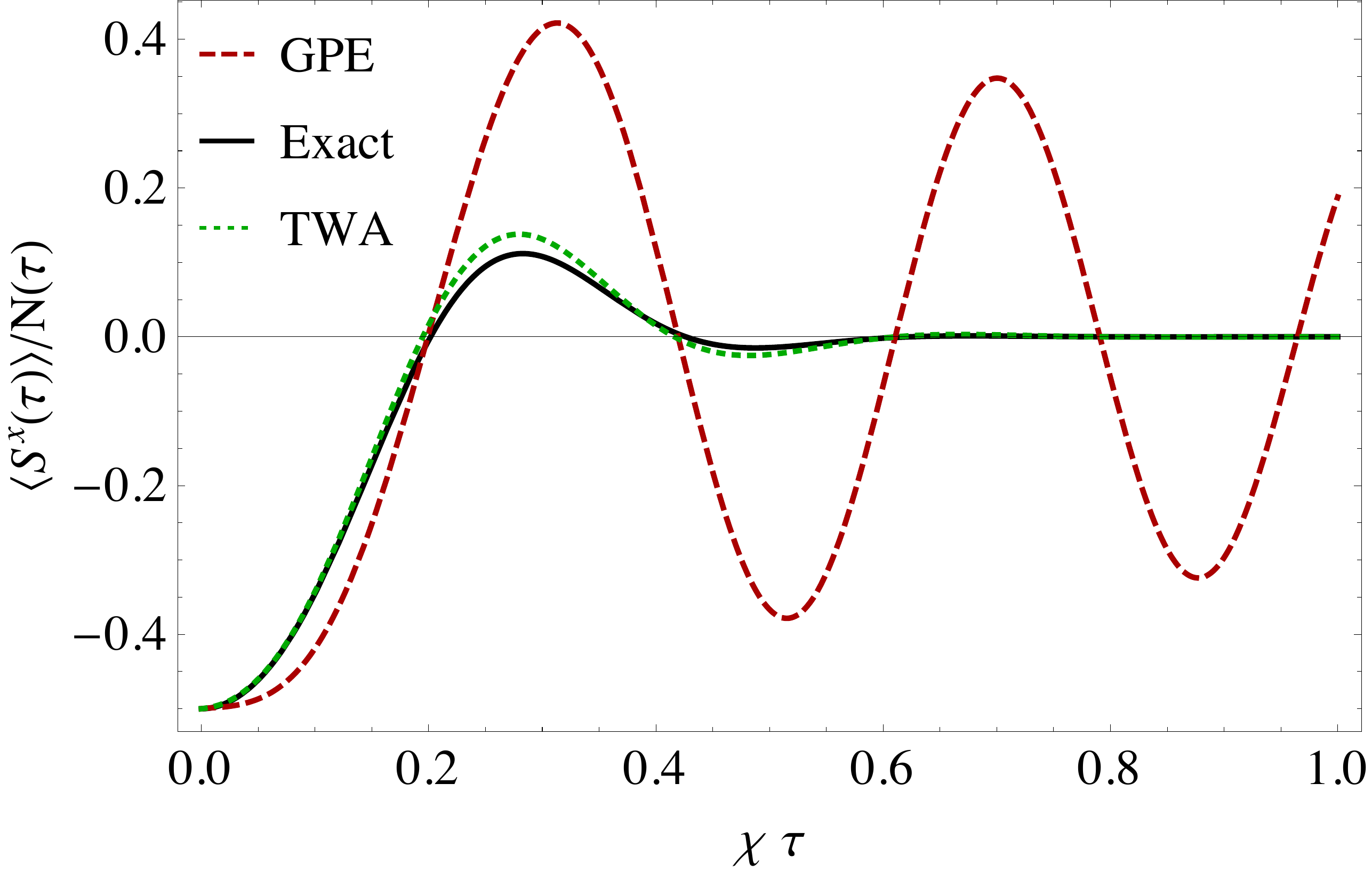}\quad
\includegraphics[width=55mm]{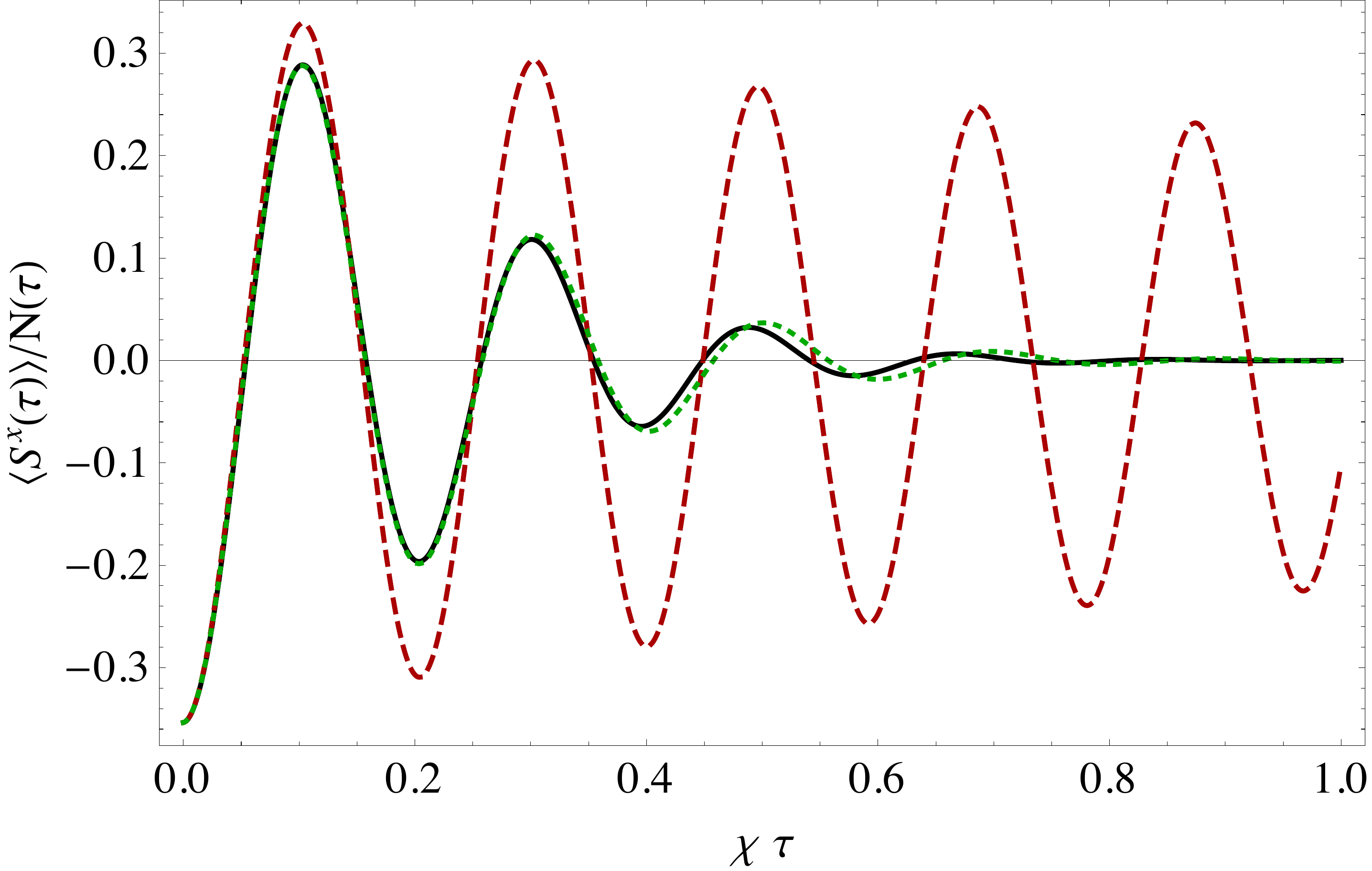}\quad
\includegraphics[width=55mm]{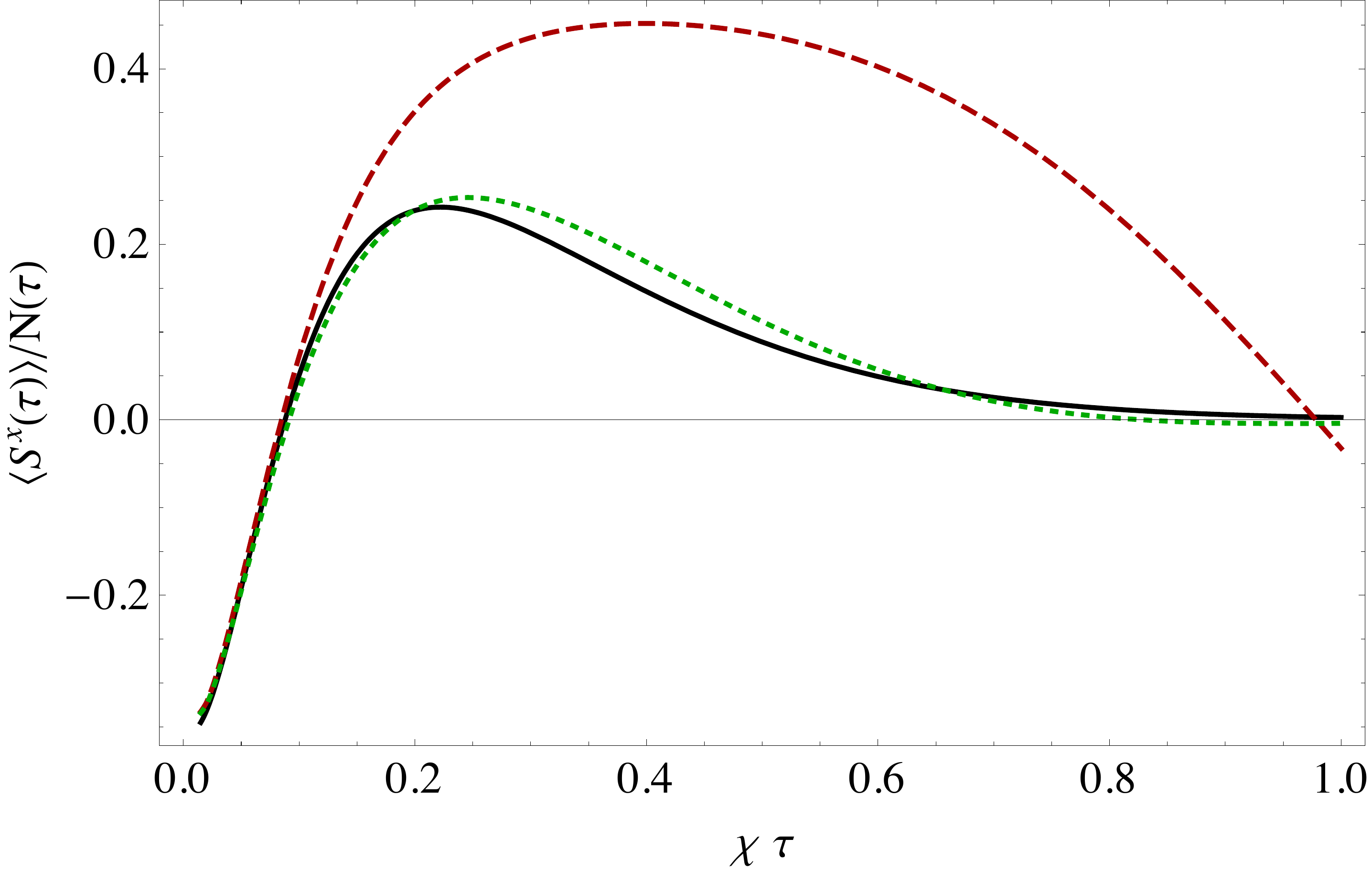}
  \end{center}
  \caption{(Color online) {Comparisons between the  GPE (red-dashed line), the  TWA (green-dotted line) and exact solution (black-solid line)}. Here we assumed $N=30$, ${C}_{\vec{\bf n}} =\delta=0$, ${\Gamma}^{e,e}_{\vec{\bf n}}=\chi_{\vec{\bf n}}/3$,  and ${\Gamma}^{e,g}_{\vec{\bf n}}=0$.
  The  plots are for $\theta_1=\pi/2$ (left),  for $\theta_1=  \pi/4 $ (middle) and $\theta_1=3 \pi/4$ (right). }
\label{comde}
\end{figure*}

\subsection{Mean-field Treatment}

At the mean-field level, accounting for losses is significantly simpler.
 We assume that the  reduced density matrix of the many-body system can be  factorized  as:
\begin{eqnarray}
\hat \rho&=&\bigotimes_j \hat{\rho}_{\vec{\bf n}_j} \quad\quad\quad \hat{\rho}_{\vec{\bf n}_j} \equiv \sum_{\alpha,\beta=e,g,0} \tilde{\rho}^j_{\alpha,\beta} |\alpha\rangle \langle \beta|  \label{fac} \end{eqnarray}
Here $\hat{\rho}_{\vec{\bf n}_j}$ is the reduced density matrix of  atom in mode $\vec{\bf n}_j$. Here  $g,e$ label the two possible spin states of the atom and 0 is the vacuum.

 By using this ansatz in Eq.~(\ref{master}), one can write closed equations of motion for $\tilde{\rho}^j_{\alpha,\beta}$. At the mean-field level, the equations of motion for
  $\tilde{\rho}^j_{\alpha,\beta}$ for $\alpha, \beta \in (e,g)$ are fully decoupled from the ones for  $\tilde{\rho}^j_{\alpha,0}$ and  $\tilde{\rho}^j_{0,\alpha}$, meaning that if there are not  particle-hole coherences initially in the system, they do not develop in the mean-field dynamics. Thus the only role the recycling terms play in the dynamics is  to populate the vacuum state, which can be accounted for by the constraint:  $\sum_{\alpha=e,g}\tilde{\rho}^j_{\alpha,\alpha}=1-\tilde{\rho}^j_{0,0}$.

 When  the recycling terms do not contribute to the equations of motion of $\tilde{\rho}^j_{\alpha,\beta}$, the
effect of the losses can be incorporated   by adding  to  the Hamiltonian a non-hermitian effective part.
 For the case of all-to-all interactions, the total net Hamiltonian that determines the mean-field dynamics,  including single-particle terms and both  elastic and inelastic collisions, can be written in terms of the collective variables ${\rho}_{\alpha,\beta}=\sum_{j=1}^N\tilde{\rho}^j_{\alpha,\beta}$ as:

\begin{eqnarray}
&&H^{MF}_{\vec{\bf n}}/\hbar =-\frac{\delta}{2}( {\rho}_{ee}-{\rho}_{gg})-\frac{\bar\Omega}{2 i} ( {\rho}_{eg}-{\rho}_{ge})
-\frac{\chi_{\vec{\bf n}}}{4}( {\rho}_{ee}-{\rho}_{gg})^2+\frac{C_{\vec{\bf n}}}{2} ( {\rho}_{ee}^2-{\rho}_{gg}^2)\notag\\
&& -i\frac{\Gamma^{e,e}_{\vec{\bf n}}}{4} {\rho}_{ee}^2-i\frac{\Gamma^{e,g}_{\vec{\bf n}}}{2} {\rho}_{ee}{\rho}_{gg},  \label{manyspinme}
\end{eqnarray} with $\Gamma^{e,g}_{\vec{\bf n}}=\frac{\sum_{j\neq j'} \Gamma^{e,g}_{{\bf n}_j, {\bf n}_{j'}}}{N(N-1)}$ and $\Gamma^{e,e}_{\vec{\bf n}}=\frac{\sum_{j\neq j'} \Gamma^{e,e}_{{\bf n}_j, {\bf n}_{j'}}}{N(N-1)}$.

The collective density matrix equations of motion obtained from that Hamiltonian, Eq~(\ref {manyspinme}) are:

\begin{eqnarray}
 &&\frac{\partial}{\partial t} {\rho}_{gg}= -\frac{\bar{\Omega}}{2} ({\rho}_{eg}+{\rho}_{ge})- \Gamma^{e,g}_{\vec{\bf n}} {\rho}_{ee}{\rho}_{gg}, \label{eqgp0}\\
 &&\frac{\partial}{\partial t} {\rho}_{ee}=  \frac{\bar{\Omega}}{2} ({\rho}_{eg}+{\rho}_{ge})- \Gamma^{e,e}_{\vec{\bf n}} {\rho}_{ee}^2- \Gamma^{e,g}_{\vec{\bf n}} {\rho}_{ee}{\rho}_{gg},\\
&& \frac{\partial}{\partial t} {\rho}_{eg}= M_{eg}- \left [ \frac{\Gamma^{e,e}_{\vec{\bf n}} {\rho}_{ee}}{2}+   \frac{\Gamma^{e,g}_{\vec{\bf n}} ({\rho}_{ee}+{\rho}_{gg})}{2} \right ] {\rho}_{eg}.
\label{eqgp}
\end{eqnarray} Note the term  $ M_{eg}$ gives the mean-field  dynamics of ${\rho}_{eg}$ in the absence of losses and was defined in Eq.~(\ref{geg}). The terms proportional to $\Gamma^{e,g}_{\vec{\bf n}}$ and $\Gamma^{e,e}_{\vec{\bf n}}$ give rise to classical rate equations for the population in the presence of two-body losses.

\subsection{ Truncated Wigner Approximation }
 Following a similar procedure to the one explained for the Hamiltonian  case without decay, one can include quantum correlations by first solving the mean-field equations of motion determined by  Eqs.~(\ref{eqgp0}-\ref{eqgp}) supplemented by random initial conditions distributed according to the Wigner distribution, Eq.~(\ref{distr}). In Fig.~\ref{comde}, we compare the solution of the exact many-body master equation, the mean-field, and the TWA with losses. We can see  some  difference between the TWA and the master equation for $N=30$, but overall the TWA does an excellent job reproducing the full dynamics.

\section{Improved Spin Model: Including virtual motional excitations}\label{effes}

\subsection{Resonant terms}
As discussed in Sec.~\ref{spinmod}, in a pure harmonic spectrum, mode changing collisions are energetically allowed and  impose important limitations on the validity of the spin model. However, when one takes into account the Gaussian shape of the actual potential,  corrections to the harmonic spectrum    are  at the level of the interaction
energy at current experimental conditions and are enough to prevent  mode-changing collisions. Consequently,
 resonant transitions arising from the perfect linearity of an harmonic oscillator spectrum can be treated as off-resonant. Off-resonant terms, nevertheless, can provide corrections to the dynamics predicted by the spin model, and, in this section, we explain a way to  incorporate those in the dynamics.

 \subsection{Off-Resonant terms}

The spin Hamiltonian neglects collision processes that do not preserve the single-particle energy. However,  off-resonant collisions can still take place virtually and will introduce corrections to the spin model. To account for those, we split  the N-particle Hilbert space into the resonant, $\Sigma$, and off-resonant, $\Upsilon$,  manifolds respectively, spanned by the states:
\begin{eqnarray}
|\Phi^\Sigma_{\vec{\bf\sigma}_{\vec{\bf n}}} \rangle&=&|\sigma_{{\mathbf n}_1},\sigma_{{\mathbf n}_2},\dots,\sigma_{{\mathbf n}_N}\rangle,\quad E_0^{tot}\equiv\sum_{j=1}^N E_{{\bf n}_j},\\
|\Psi^\Upsilon_{\vec{ \bf \sigma}_{\vec{\bf k}}}\rangle&=&|\sigma_{{\mathbf k}_1},\sigma_{{\mathbf k}_2},\dots,\sigma_{{\mathbf k}_N}\rangle, \quad E_{ {\vec {\bf k}}}^{tot}\equiv\sum_{j=1}^N E_{{\bf k}_j}\neq E_0^{tot}. \notag
\end{eqnarray} These states are  written in the occupation basis and $\sigma \in  \{g,e\}$. Since the same mode ${\mathbf k}_j$ can be occupied simultaneously by a $g$ atom and an $e$ atom, $g_{{\mathbf k}_j}$ and $e_{{\mathbf k}_j}$ can occur simultaneously in $|\Psi^\Upsilon _{\vec{ \bf \sigma}_{\vec{\bf k}}}\rangle$.

The spin Hamiltonian was obtained  by directly projecting the interaction part of the many-body Hamiltonian on $\Sigma$.  An effective Hamiltonian that accounts for the leading order corrections generated by   virtual occupation of off-resonant states   can be derived by using the Schrieffer-Wolff transformation \cite{Duan2003}:
\begin{eqnarray}
\hat H^{\rm eff}_{\vec{\bf  n}}={\hat H}^{S}_{\vec{\bf  n}}+ { \hat  H}^{S_2}_{\vec{\bf  n}}.
\end{eqnarray} Here, $\hat H^{S}_{\vec{\bf  n}}$ is the spin model given by Eq.~(\ref{manyspin}). $ {\hat H}^{S_2}_{\vec{\bf  n}}$ is obtained via  second-order perturbation theory as follows:
{\small \begin{equation}
\langle \Phi^\Sigma_{\vec{\sigma}_{\vec{\bf n}}} |{\hat H}^{S_2}|\Phi^\Sigma_{\vec{\sigma'}_{\vec{\bf n}}} \rangle=-\sum_{{\bf \sigma}_{\vec{\bf k}}}\frac{\langle \Phi^\Sigma_{\vec{\sigma}_{\vec{\bf n}}}|{\hat H}|\Psi^\Upsilon _{\vec{ \bf \sigma}_{\vec{\bf k}}}\rangle\langle \Psi^\Upsilon _{\vec{ \bf \sigma}_{\vec{\bf k}}}|{\hat H}|\Phi^\Sigma_{\vec{\sigma'}_{\vec{\bf n}}}\rangle}{ E_ {\vec {\bf k}}^{tot}-E_0^{tot}},
\end{equation}} Here ${\hat H}$ is the interaction part of Eq. (\ref{Habe}).

\begin{figure}[htb]
   \begin{center}
\includegraphics[width=0.55 \columnwidth]{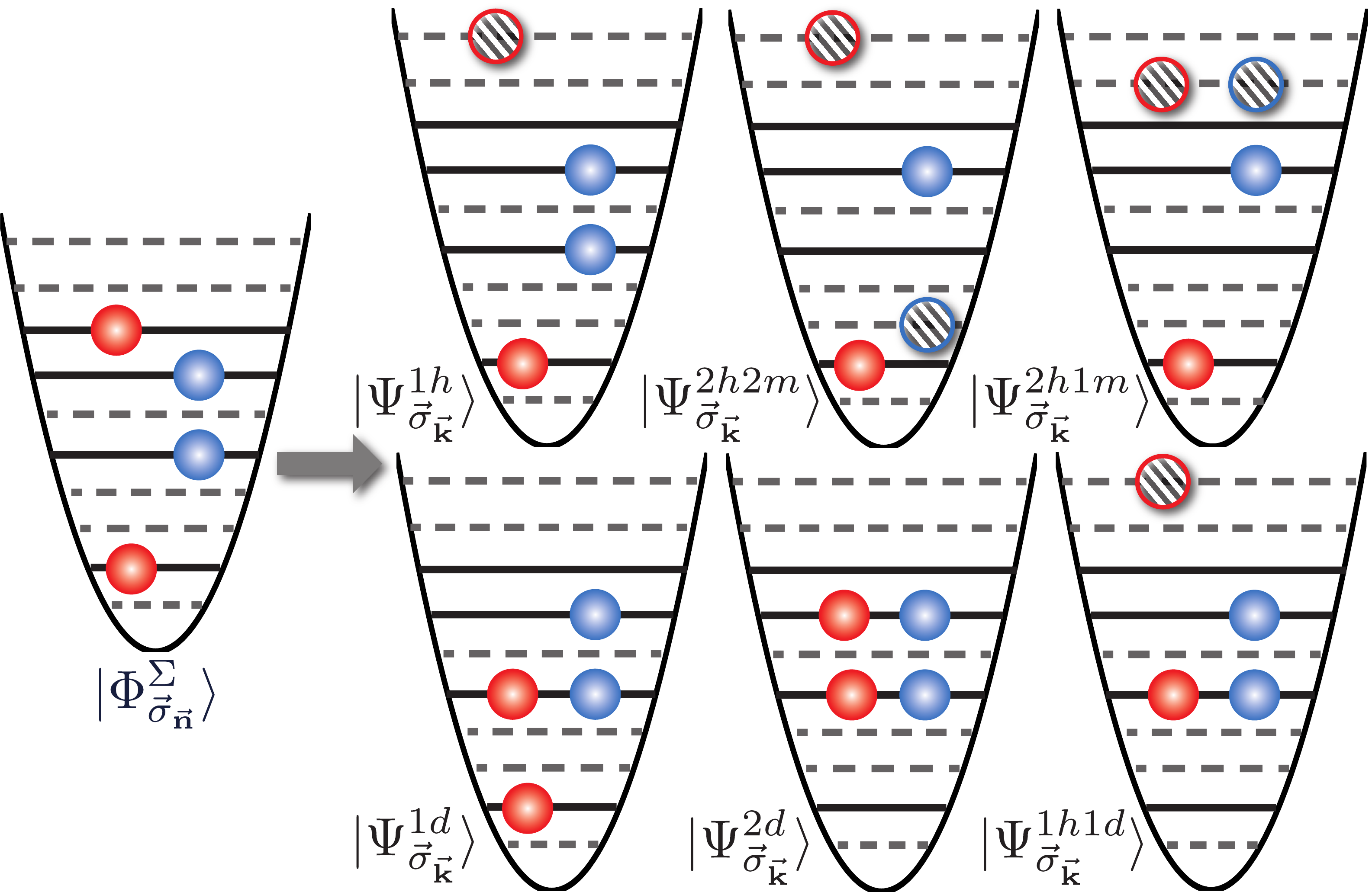}
  \end{center}
\caption{(Color Online) Schematic representation of the various virtual processes. Here the filled circles denote $g$ (blue) and $e$ (red) atoms in the initially populated levels (also shown with a solid line). The striped circles denote atoms virtually populating off-resonant levels,  shown with a dashed line. Note that the classification is made in terms of the number of doublons or populated states outside the initially populated manifold. Processes in which, in addition to the formation of a hole or a doublon   within the occupied manifold, the  spin is switched between two atoms  are also  included  in each classification but are not necessary shown in the schematics. Also, since this is a schematic plot, for simplicity, we show two $e$ atoms and two $g$ atoms only and  the color  was selected randomly. For example, for the cases (1h), (1d), (2d) and (1h1d)   corresponding processes exist where  g (blue) atoms are the ones that get excited.} \label{modes}
\end{figure}

 Since  ${\hat H}^{S_2}_{\vec{\bf  n}}$ acts on the $\Sigma $ subspace, it is convenient to explicitly write it in terms of spin operators. To accomplish that, we  divide the states $|\Psi^\Upsilon _{\vec{ \bf \sigma}_{\vec{\bf k}}}\rangle$ into 6 different categories, $|\Psi^{1h} _{\vec{ \bf \sigma}_{\vec{k}}}\rangle,  |\Psi^{2h2m}_{\vec{ \bf \sigma}_{\vec{\bf k}}}\rangle,|\Psi^{2h1m}_{\vec{ \bf \sigma}_{\vec{\bf k}}}\rangle,$  $|\Psi^{1d} _{\vec{ \bf \sigma}_{\vec{\bf k}}}\rangle,|\Psi^{2d} _{\vec{ \bf \sigma}_{\vec{\bf k}}}\rangle,|\Psi^{1h1d} _{\vec{ \bf \sigma}_{\vec{\bf k}}}\rangle$,   which  are shown  in Fig. \ref{modes}.

The states labeled with superscripts $1h$  and $2h2m$ are states which  have one and two atoms populating a mode not belonging to the initially populated  $\vec{\bf n}=\{ {\bf n}_1,{\bf n}_2,\dots, {\bf n}_N \}$ manifold respectively. The states labeled with superscripts  $2h1m$ are states which  have two atoms both populating one  mode not belonging to the initially populated manifold.
The states labeled with superscripts $1d$ and $2d$  have  one and two   modes in the  initially populated  manifold occupied by two atoms (doublon) respectively. Finally states labeled by superscripts $1h1d$ have  one atom populating a mode outside the initially occupied manifold and a doublon. We will consider the various contributions case by case in Appendix 2. Here we only quote the final result,
assuming collective interactions and dropping terms that are constants of motion.

 \begin{eqnarray}
{\hat H}_{\vec {\bf n}}^{S_2}&=&a^{T}_1 {\hat S}^{z}+a^{T}_2 ({\hat S}^{z})^2+  a^{T}_3({\hat S}^{z})^3, \label{efespi}
 \end{eqnarray} The parameters $a^{T}_{1,2,3}$ are explicitly computed in Appendix 2.

 Among those corrections, linear and quadratic contributions   can be absorbed into the spin Hamiltonian, and only the processes proportional to $({\hat S}^{z})^3$ give rise to an additional term.   Multi-body interactions arising from virtual scattering processes have been  shown  to introduce measurable corrections in
quantum phase revivals  in the Bose-Hubbard model \cite{Will2010,Johnson2009}. The cubic terms can give rise to visible corrections to  the  Ramsey-fringe contrast as recently measured in the JILA Sr clock and  reported in Ref. \cite{Martin2012}. Here we also show, in Sec.\ref{secyb}, that they play an important  role in the Yb  Ramsey-fringe contrast decay.

 In principle, terms proportional to  $({\hat S}^{z})^4$ could appear in the effective Hamiltonian.  We find, however, that they vanish in our system due to the fact that when an atom in mode ${\bf n}_j$ scatters to an already occupied mode, ${\bf n}_{j'}$, forming a doublon, the reverse process (an atom in mode ${\bf n}_{j'}$ scatters to mode ${\bf n}_{j}$ forming a doublon), contributes with the opposite sign to the effective model, and thus, both cancel.

\section{ Ramsey interrogation: Beyond the collective regime} \label{noncoll}
Negligible excitation inhomogeneity is not necessarily a good approximation in all optical lattice clock experiments. For example, in the Yb optical lattice clock at NIST, the current operating  temperature is typically around $10 \mu K$ and $\Delta \Omega/\bar{\Omega}$ can  be in some cases as high a 0.25 (see Fig.~\ref{inte2}). In the following sections we describe various ways to treat inhomogenous excitation and interaction effects that cause population of non-collective states. We first consider the case of collective interactions, where an analytic solution to the full quantum problem in the presence of an excitation inhomogeneity  can be derived based on perturbation theory in $\Delta \Omega/\bar{\Omega}$. Next, we discuss a  solution based on the TWA that incorporates in the model, not only excitation inhomogeneity but also inhomogeneity in
the spin coupling constants and losses.

\subsection{Analytic solution}\label{anal}

In the  presence of excitation inhomogeneity,  an analytic treatment based on perturbation theory can be performed under the collective Hamiltonian approximation, {\it i.e.} neglecting two body losses and assuming collective two-body interactions.

To accomplish that  one writes $\Omega_{{\bf n}_j}=\bar{\Omega}_{\vec{\bf n}} +\delta \Omega_{{\bf n}_j}$, with $\bar{\Omega}_{\vec{\bf n}} =\sum_{j} \Omega_{{\bf n}_j}/N$ the mean Rabi frequency  and treats $\delta \Omega_{{\bf n}_j}/\bar{\Omega}$ as a perturbation parameter. Keeping only the leading order correction,  which can be shown to be quadratic in  $\Delta{\Omega}_{\vec{\bf n}} =\sqrt{\sum_{j} \Omega_{\bf n_j}^2/N-\bar{\Omega}_{\vec{\bf n}}^2}$, {\it i.e.} the first-order corrections vanish,   and after  a lengthly but straightforward calculation  described in Appendix 3, one can compute  the number of excited atoms, $N_{e,\vec {\bf n}}(t_1,t_2)$, in Ramsey spectroscopy. The expression is quite complicated and we do not explicitly show  it in the main text.  Instead  here we discuss its behavior in the weakly interacting regime, $\chi_{\vec{\bf n}}\tau\ll 1$ and ${J}^\perp_{\vec{{\bf n}}}\tau\ll 1$, where  one can show that the density shift  becomes:

\begin{small}
\begin{eqnarray}
&&2 \pi  \Delta\nu_{\vec{{\bf n}}}^{{\rm Inh}}=(N-1)\Big[{C}_{\vec{{\bf n}}} -  {\chi}^{*}_{\vec{{\bf n}}} \cos({\bar\theta}_1^{\vec{\bf n}} )\Big] -N \Delta \theta_2 ^{\vec{\bf n}}\Delta \theta_1^{\vec{\bf n}}  \left[  {J}^\perp_{\vec{{\bf n}}}   + \frac{ \chi_{\vec{{\bf n}}}}{N} \sin^2({\bar\theta}_1^{\vec{\bf n}} ) \right] \left (\frac{\cot({\bar\theta}_2^{\vec{\bf n}})}{\sin({\bar\theta}_1^{\vec{\bf n}} )}\right),\label{spfn2}\\
&&{\chi}^{*}_{\vec{{\bf n}}}=\chi_{\vec{{\bf n}}} \left [1+ \frac{(N-3)}{2(N-1)}(\Delta \theta_1^{\vec{\bf n}})^2 \right],
 \end{eqnarray}\end{small}
   with $\Delta \theta_{1,2}^{\vec{\bf n}} =\Delta{\Omega}_{\vec{\bf n}} t_{1,2}$ and  $\bar{ \theta}_{1,2} =\bar{\Omega}_{\vec{\bf n}} t_{1,2}$.

 The expression of the shift in the weakly interacting limit is very illuminating.  Note that while to leading order ($\Delta\theta=0$) the density shift is only determined by the excitation fraction,  $\cos({\bar\theta}_1 )= (N_g-N_e)/2$,  and is independent of the second pulse area, ${\bar\theta}_2$, excitation inhomogeneity does introduce a dependence of the density shift on the second pulse area. An intuitive explanation for that can be obtained by noticing that the density shift is extracted from a collective measurement.  Therefore, if there are atoms excited outside the symmetric Dicke manifold due to the first pulse, they only contribute to the measured signal if they are brought back to the symmetric  Dicke manifold by the second pulse.
The mean-field expression, Eq.~(\ref{spfn2}), also  tells us that  terms proportional to $(\Delta \theta_{1}^{\vec{\bf n}})^2$ just slightly renormalize the overall magnitude of the density shift, $\chi_{\vec{\bf n}}\to \chi^*_{\vec{\bf n}}$,  without affecting its general dependence on the excitation fraction. On the other hand, the term proportional to  $\Delta \theta_{1}^{\vec{\bf n}}\Delta \theta_{2}^{\vec{\bf n}}$ does modify the dependence of the density shift on pulse area and  vanishes  at ${\bar\theta}_2^{\vec{\bf n}}=\pi/2$.

\begin{figure*}
  \begin{center}
   \includegraphics[width=140mm]{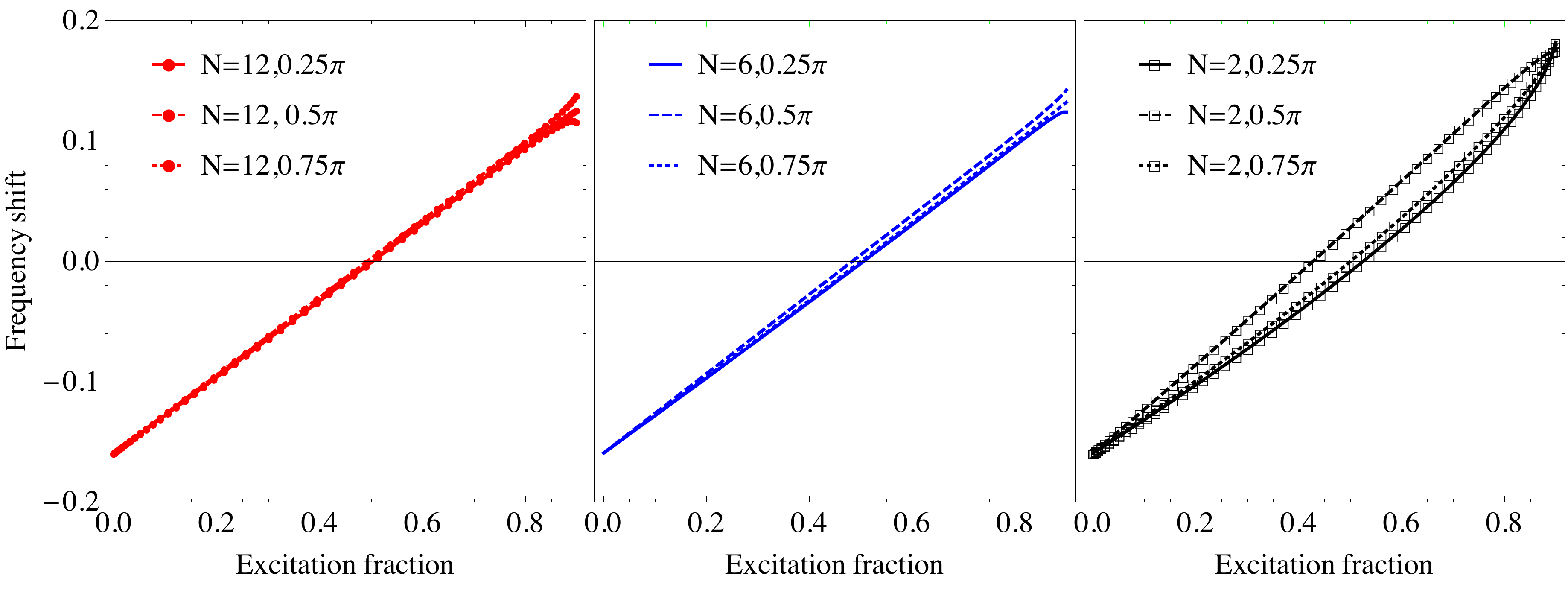}\\
     \includegraphics[width=140mm]{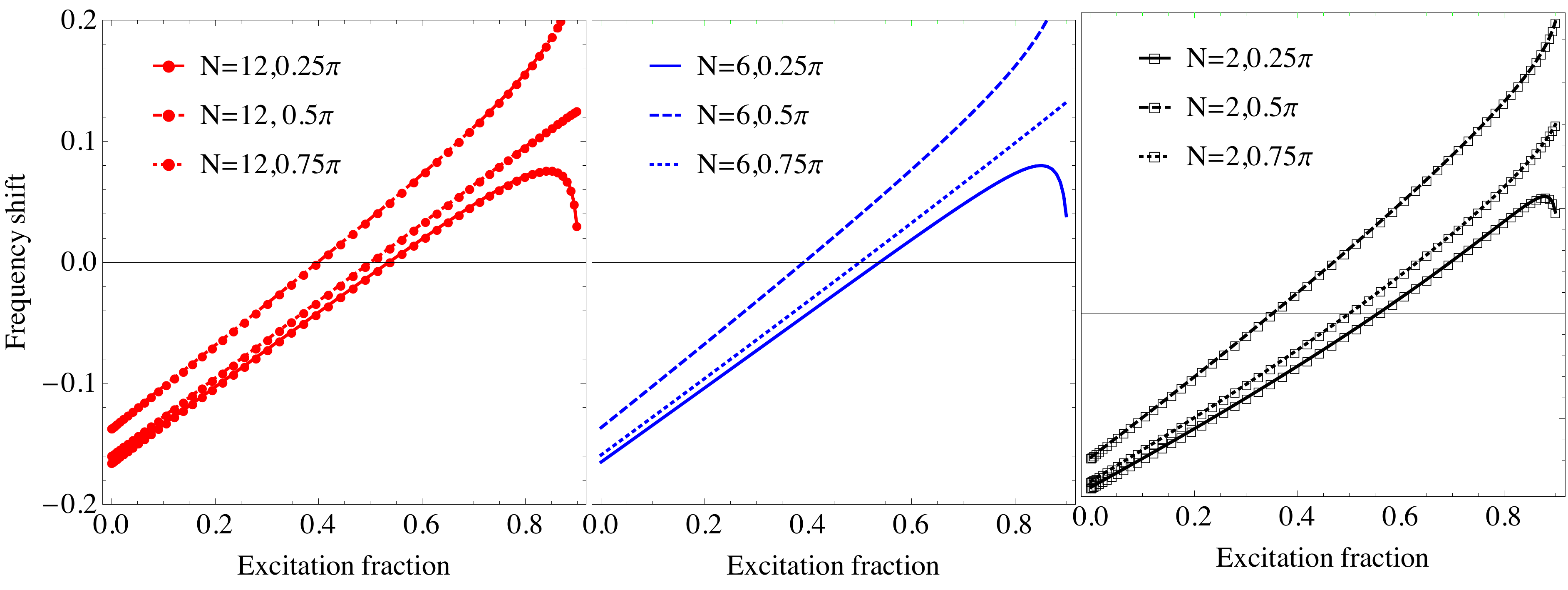}
     \end{center}  \caption{(Color Online)  Shift vs excitation fraction for different atom numbers: $N=2$ (Black open squares), $6$ (no symbol), and $12$ (red circles). The convention for the second pulse area is: $\bar{\theta}_2^{\vec{\bf n}}$:  $0.25\pi$ (solid), $0.5\pi$ (dashed line) and $0.75\pi$ (short-dashed line).  We set $C_{\vec{\bf n}}=0$.  The values of the other  interactions were chosen such that $\chi_{\vec{\bf n}} (N-1)$ are all equal   as well as ${J}^{\perp}_{\vec{\bf n}} N$. For $N=2$, we set ${\chi}_{\vec{\bf n}}=1$ and ${{J}^{\perp}_{\vec{\bf n}}}=0$ and $-0.6$, for the top and bottom panels respectively. Note that for $N>2$ and ${J}^{\perp}_{\vec{\bf n}}=0$, the density shift is quite insensitive to the second pulse area as predicted by theory.}\label{inhofi}
\end{figure*}

Outside the weakly interacting regime, the assumption that the density shift  scales mostly linearly with the particle number breaks down. One important consequence of this observation is that  the behavior of the $N=N_T$ particle system  cannot be reproduced by solving the $N=2$ system and then rescaling the interactions, as claimed in Ref. \cite{Gibble2009},
 {\it i.e.} replacing $\chi_{\vec{\bf n}} $ and ${J}^{\perp}_{\vec{\bf n}}$ in the $N=2$ solutions by ${\chi}_{\vec{\bf n}}^{\rm eff} \to \chi_{\vec{\bf n}} (N_T-1)$ and ${{J}^{\perp}_{\vec{\bf n}}}^{\rm eff} \to {{J}^{\perp}_{\vec{\bf n}}} N_T/2$.  The relevant role of non-linear effects is illustrated in Fig.~\ref{inhofi}. In the top panel we compare the behavior of three systems with different $N=2,6,12$. We fix the interactions such that ${\chi}_{\vec{\bf n}}  (N-1)$ are all equal   as well as all ${{J}^{\perp}_{\vec{\bf n}} } N$.  While the curves $N=6$ and $12$ are very close to each other, they do not agree with the  $N=2$ solution.

 Note also that for the many-body system,  $N\gg 2$, the part of the shift sensitive to the second pulse area
 is  mainly proportional to  $N({J}^\perp_{\vec{\bf n}} )$ (up to $1/N$ corrections) and not  to  $ 2{J}^\perp_{\vec{\bf n}} + \chi_{\vec{\bf n}}\sin^2({\bar\theta}_1)$, as it is the case for $N=2$.  This phenomenon is visible in the bottom panel of  Fig.~\ref{inhofi}, which  shows the shift  vs excitation fraction for three different second pulse areas: $(0.25,0.5$,  and $0.75 )\pi$.
  At ${{J}^{\perp}}=0 $ and   $\bar{\theta}_1^{\vec{\bf n}}=\pi/2$ (close to zero crossing for the parameters used in the plot) the density shift only exhibits a significant  dependence on the second pulse area when $N=2$.

\subsection{TWA and Mean-field solutions}
We will now  use the TWA to go beyond the collective approximation  that was required  to derive  the equations of motion in prior sections. The TWA  allows us to include different types of mechanisms,  which bring  the system out of the collective Dicke  manifold,  such as  single-particle  excitation inhomogeneities  and
non-collective elastic and inelastic two-body collisions.  We can also easily include one-body losses due to, for example, background gas collisions. The latter are characterized by the jump operators $\sqrt{\Gamma_{j\alpha}} \hat{c}_{{\bf n}_j\alpha}$,  which annihilate the  atom $j$ in spin state $\alpha$.

For the mean-field treatment, we assume the reduced density matrix of the many-body system factorizes as described in Eq.~(\ref{fac}). With the aim of dealing with long-range interactions, it is better to write the equations of motion of the Fourier transformed quantities:

\begin{eqnarray}
\rho_{\alpha \alpha}(k)&=&\sum_{j=1}^N e^{i \frac{2 \pi j k}{N}} \tilde{\rho}^j_{\alpha,\alpha} \\
\rho_{eg}(k)&=&\sum_{j=1}^N e^{-i \frac{2 \pi j k}{N}} \tilde{\rho}^j_{e,g}=\rho_{ge}^* (k)
\end{eqnarray}

The  equations of motion derived in this way are general,  and  in section~\ref{compasec},  will be applied to model the Yb clock dynamics. These equations reduce to   Eqs.~(\ref{eqgp0}-\ref{eqgp}) when the interaction parameters are collective, {\it i.e.} have only $k=k'=0$ components. In their most general form, the equations of motion are  nontrivial, and  we  present them in  Appendix 4. Here instead we proceed to  analyze simpler relevant cases, which will allow us to test the validity of their corresponding  TWA dynamics (obtained after sampling   the solutions of the   mean-field equations with  the Wigner distribution).   We proceed by considering two  cases, one in which  the inhomogeneity is caused by interactions and the other in which  it  is  caused  by single-particle terms. Since analytic solutions are only available when one restricts the dynamics to a pure Hamiltonian evolution, in the reminder of this section we will neglect single-particle and two-body losses.

\subsubsection{Non-collective interactions}

$\bullet$~{\bf Ising Case}

\begin{figure}
 \begin{center}
 \includegraphics[width=70mm]{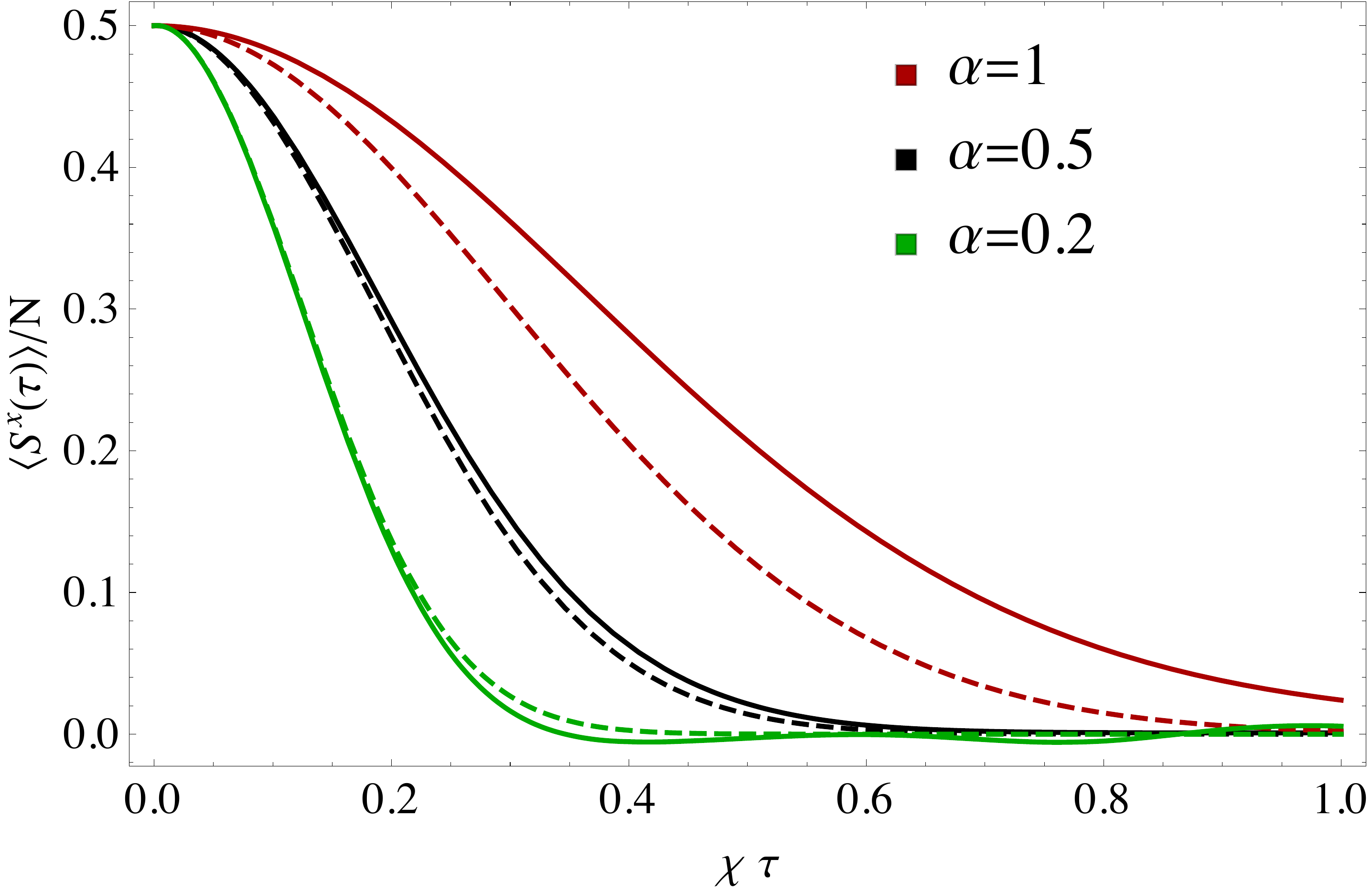}
 \end{center}
 \caption{ (Color Online) Ramsey-fringe-contrast decay for the case of power-law Ising interactions in a 2D square array composed of 121 atoms (See text). The first pulse is set to $\pi/2$. The solid  lines  correspond to  the exact solution obtained from Ref. \cite{FossFeig2013b}. The dashed lines  are the TWA solution. From top to bottom the curves correspond to $\alpha=1,0.5$, and $0.2$. }\label{kats}
\end{figure}

\begin{figure*}
  \begin{center}
   \includegraphics[width=70mm]{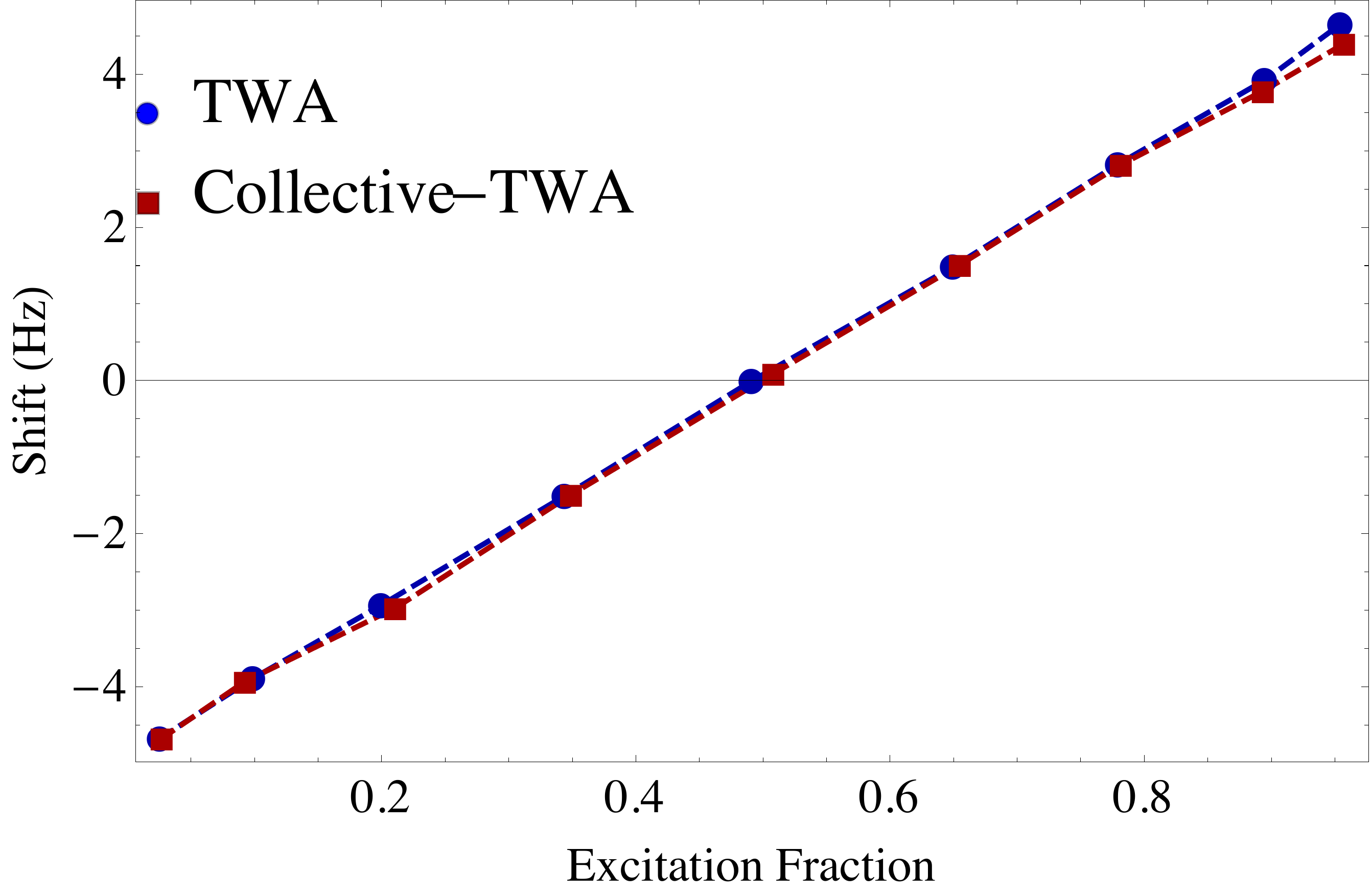}\quad  \includegraphics[width=70mm]{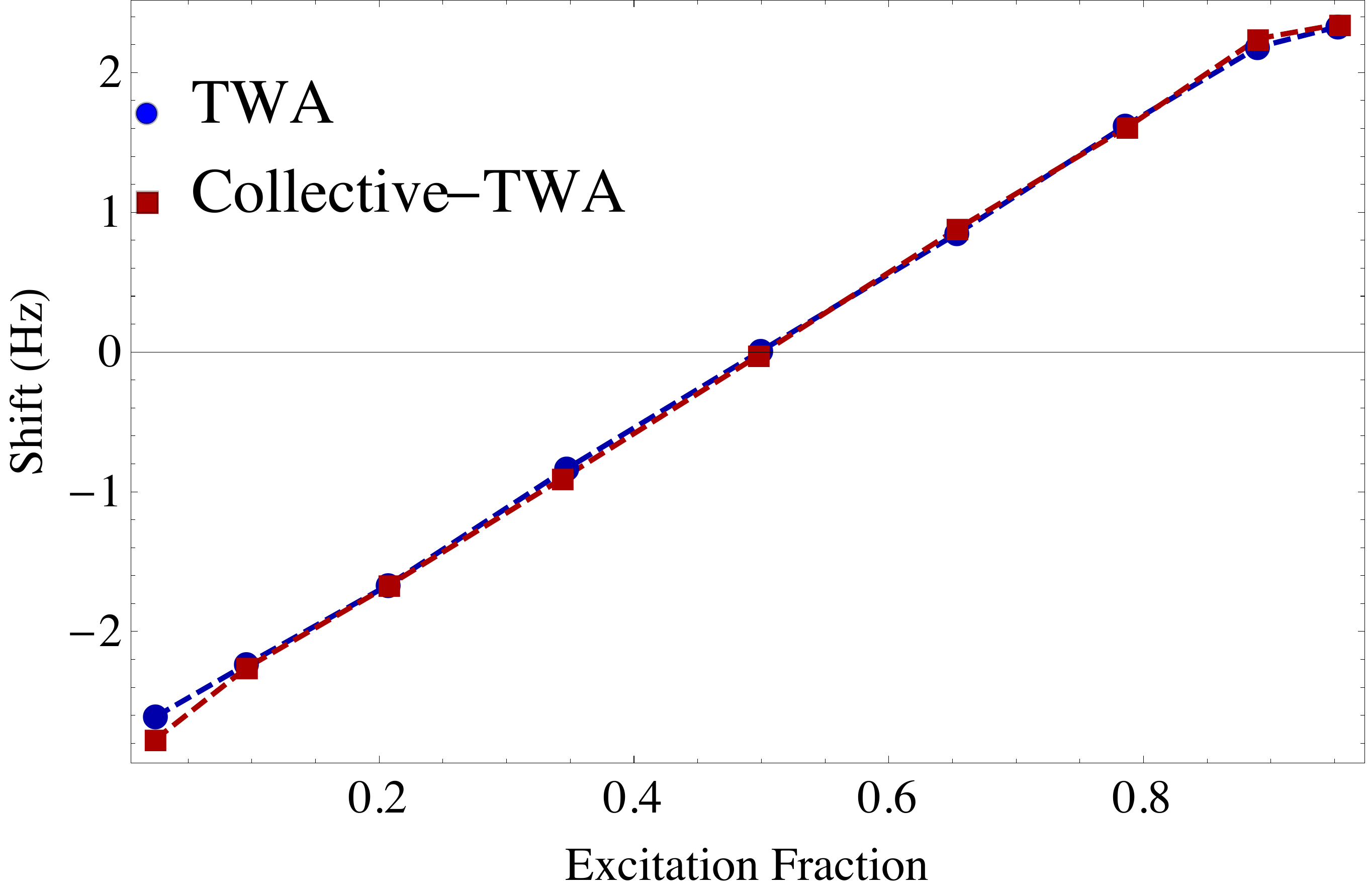}\\
       \includegraphics[width=70mm]{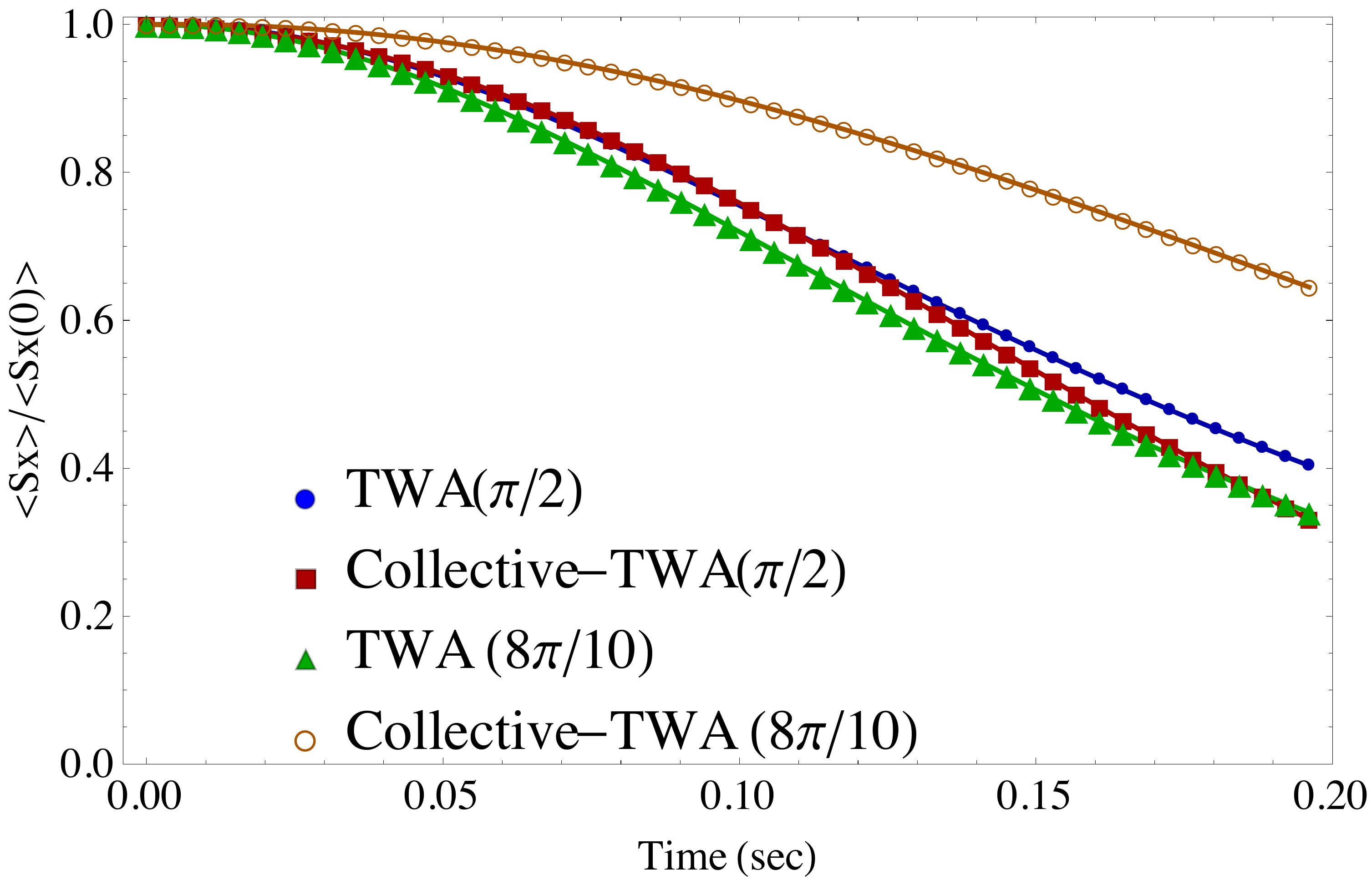}\quad  \includegraphics[width=70mm]{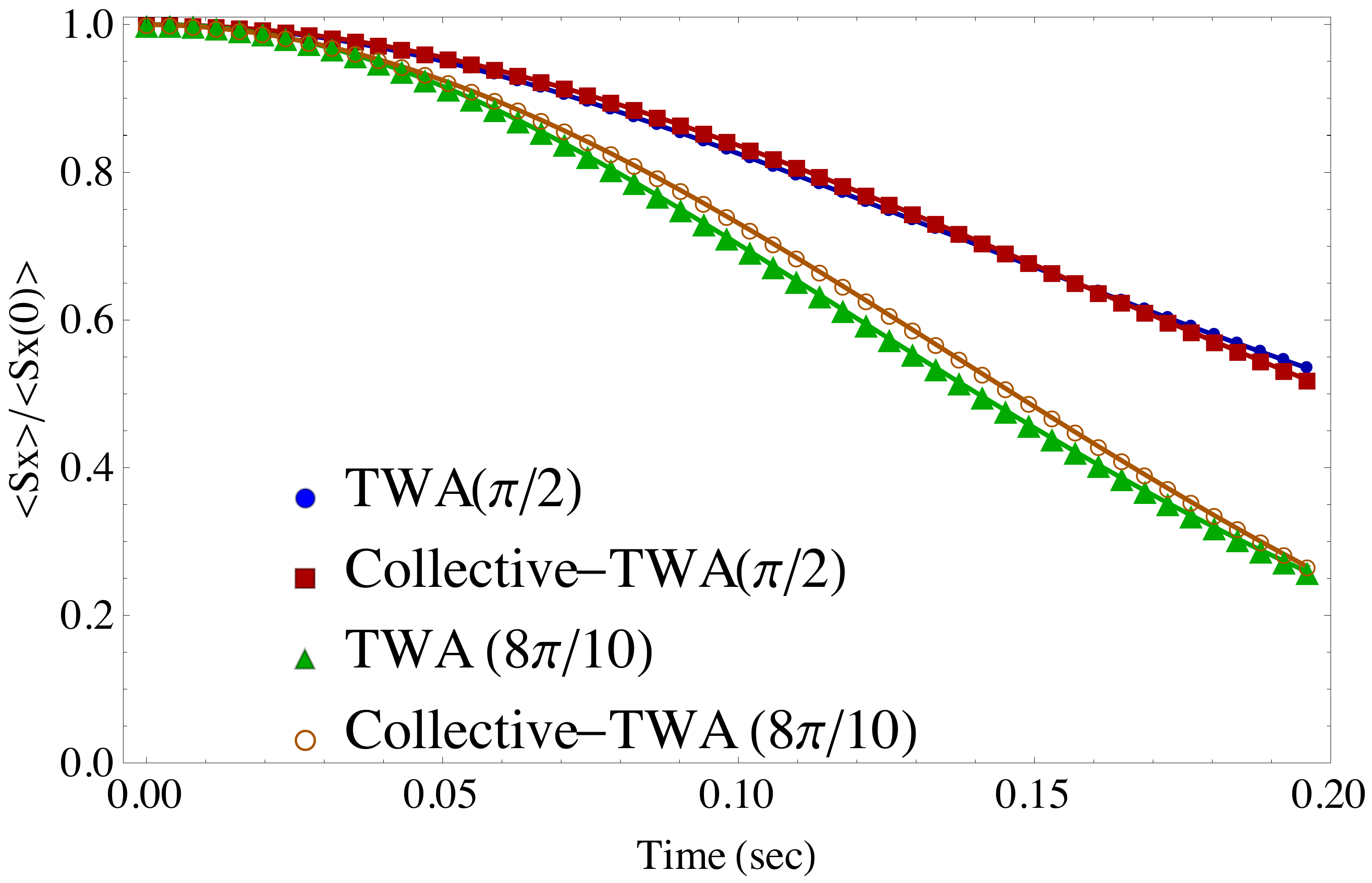}\\
     \end{center}  \caption{(Color online) Top panels: Comparisons of the density shift vs excitation fraction predicted by the collective TWA and the  generic TWA  at $\tau=50$ ms. The generic TWA accounts for the mode dependence of  $p$-wave interactions. Here zero excitation inhomogeneity is assumed ($\Delta\Omega=0$). A thermal average over initially populated modes is performed. The left panel is for a fixed  number of atoms,  $N=20$. In the  right panel,  we have done an average over  the atom number in  the pancakes using  a Poissonian distribution with $\bar{N}\sim 9$ (see Sec. \ref{tem}). Here $v^{ee}=v^{gg}$, thus $C_{{\bf n}_j,{\bf n}_{j'}}=0$. Bottom  panels: Comparison of the contrast decay  vs time predicted by the collective TWA and  generic TWA for different first pulse areas (indicated in the plot). Again, the left panels is for a fixed  number of atoms,  $N=20$,  and  the right panel is computed after performing an atom number average. }\label{coll}
\end{figure*}

The Ising case corresponds to  ${J}^{\perp}_{{\bf n}_j,{\bf n}_{j'}}=0$. An exact solution for the Ising dynamics exists for an initial product  state and for arbitrary spin  coupling constants \cite{FossFeig2013b,FossFeig2012b, Hazzard2012,Emch,Kastner2011}. This analytic solution allows us to benchmark the validity of the TWA beyond the collective regime, assuming   zero transverse interactions ${J}^{\perp}_{{\bf n}_j,{\bf n}_{j'}}=0$ and  no losses. We will also assume, for simplicity, no excitation inhomogeneity during pulses. In this case, at the mean-field level, one obtains:
\begin{eqnarray}
 {\rho_{eg}(k=0,\tau)}=(S^x_0+ {\rm i} S^y_0) \exp[ -{\rm i} \tau \mathcal{M}^o]_{0,0},
 \end{eqnarray}with
\begin{eqnarray}
\mathcal{M}^o_{k,k'}&\equiv& \delta  I_{k,k'} -2 {\hat S}^{z}_0\chi_{k'-k,0}-N C_{k'-k,0},\\
\chi_{k,0}&\equiv&\frac{1}{N^2} \sum_{j,j'=1}^{N}  e^{i \frac{2 \pi j k}{N}} \chi_{\vec {\bf  n}_j,\vec { \bf  n}_{j'}},\\
C_{k,0}&\equiv&\frac{1}{N^2} \sum_{j,j'=1}^{N}  e^{i \frac{2 \pi j k}{N}} C_{\vec {\bf  n}_j,\vec {\bf  n}_{j'}}.
\end{eqnarray} Here $I$ is the identity matrix. $\{S^x_0,S^y_0,S^z_0\}$ are the components of the collective Bloch vector after the first pulse.
In Fig.~\ref{kats},  we show the comparisons for a  system  of $N=121=11\times 11$ atoms assuming that modes  $\vec {\bf  n}$  form a 2D square lattice array.  We set  $\chi_{\vec {\bf  n}_j,\vec { \bf  n}_{j'}}\propto 1/|{\bf  n}_j-{\bf  n}_{j'}|^{\alpha}$ and vary $\alpha=0.2,0.5,1$. Note this choice of $\chi_{\vec {\bf  n}_j,\vec { \bf  n}_{j'}}$ is done just for convenience without any direct link with the  mode dependence of $\chi_{\vec {\bf  n}_j,\vec { \bf  n}_{j'}}$ when it describes actual $s$ or $p$- wave interactions. The physically relevant case for the clock will be described below.  For   $\alpha=0.2,0.5$, the exact and TWA solutions agree  well. However, for  $\alpha=1$, deviations start to become important.  An approximate criterion for the validity of the TWA for a translationally invariant system is that,   if one defines \cite{Hazzard2012} $\Xi_m=\sum_j \chi_{\vec {\bf  n}_0,\vec { \bf  n}_{j}}^m$, then $\Xi_1^2\approx N\Xi_2$. Note that, in this analysis,  we have not added any quantum vacuum noise to the initially empty $k\neq 0$ modes.  However, we have confirmed that its addition only slightly changes the conclusions.

$\bullet$~{\bf Pure $p$-wave interactions}

We now take advantage of the TWA to benchmark the validity of the collective model when the interaction inhomogeneity  comes from the mode dependence of the $p$-wave interactions. In this case, for simplicity we also assume no excitation inhomogeneity during the pulses and assume no single particle or two-body losses. We also  set $s$-wave  interactions to zero and  perform a thermal average over populated modes assuming a Boltzmann distribution (see Sec.~\ref{tem}). In   Fig.~\ref{coll}, we show  the density shift and the Ramsey fringe contrast decay  predicted by the TWA solution, both with and without interaction inhomogeneity.
While the density shift agrees fairly well between the collective and non-collective solutions, the contrast shown for  the high ($\theta_1 = 0.8 \pi$) initial  pulse area has a slower decay for the case of collective interactions. On the contrary, at initial pulse areas close to $\pi/2$,  the agreement between the two solutions is fair, and the collective model exhibits just a slightly faster decay of the contrast.
The disagreement of the contrast at high excitation fraction  is because, at these conditions (a similar effect happens at small tipping angles), the contrast reduction comes primarily from simple dephasing  due to the different precession rates of the Bloch vector for different realizations of the interaction parameters. Recall that the dynamics induced by mean-field interactions correspond to a precession around an effective  magnetic field whose strength depends  on atom number, interaction strength,  and excitation fraction, Eq.~(\ref{beff}). For the collective model that type of  dephasing is much weaker. On the contrary, if $C_{\vec {\bf  n}_j,\vec { \bf  n}_{j'}}=0$, as it is in this example, the effective magnetic field vanishes at $\theta_1=\pi/2$.  Thus,  for $\theta_1$ close to $\pi/2$,  dephasing is minimal and  contrast decay is dominated by many-body effects, which are fairly well-captured by the collective model.

\begin{figure}
  \begin{center}
   \includegraphics[width=70mm,height=50mm]{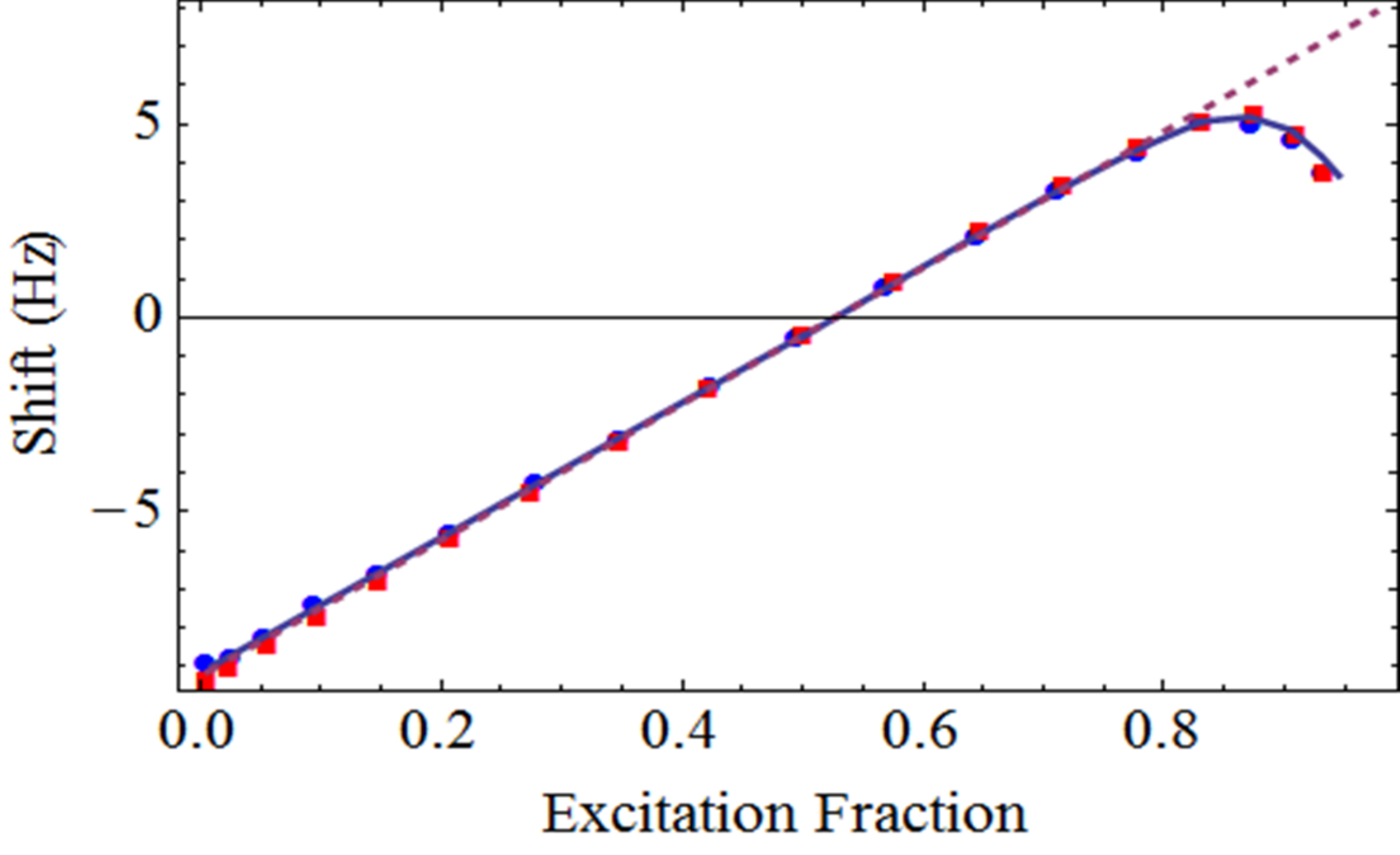}\quad
     \includegraphics[width=70mm,height=50mm]{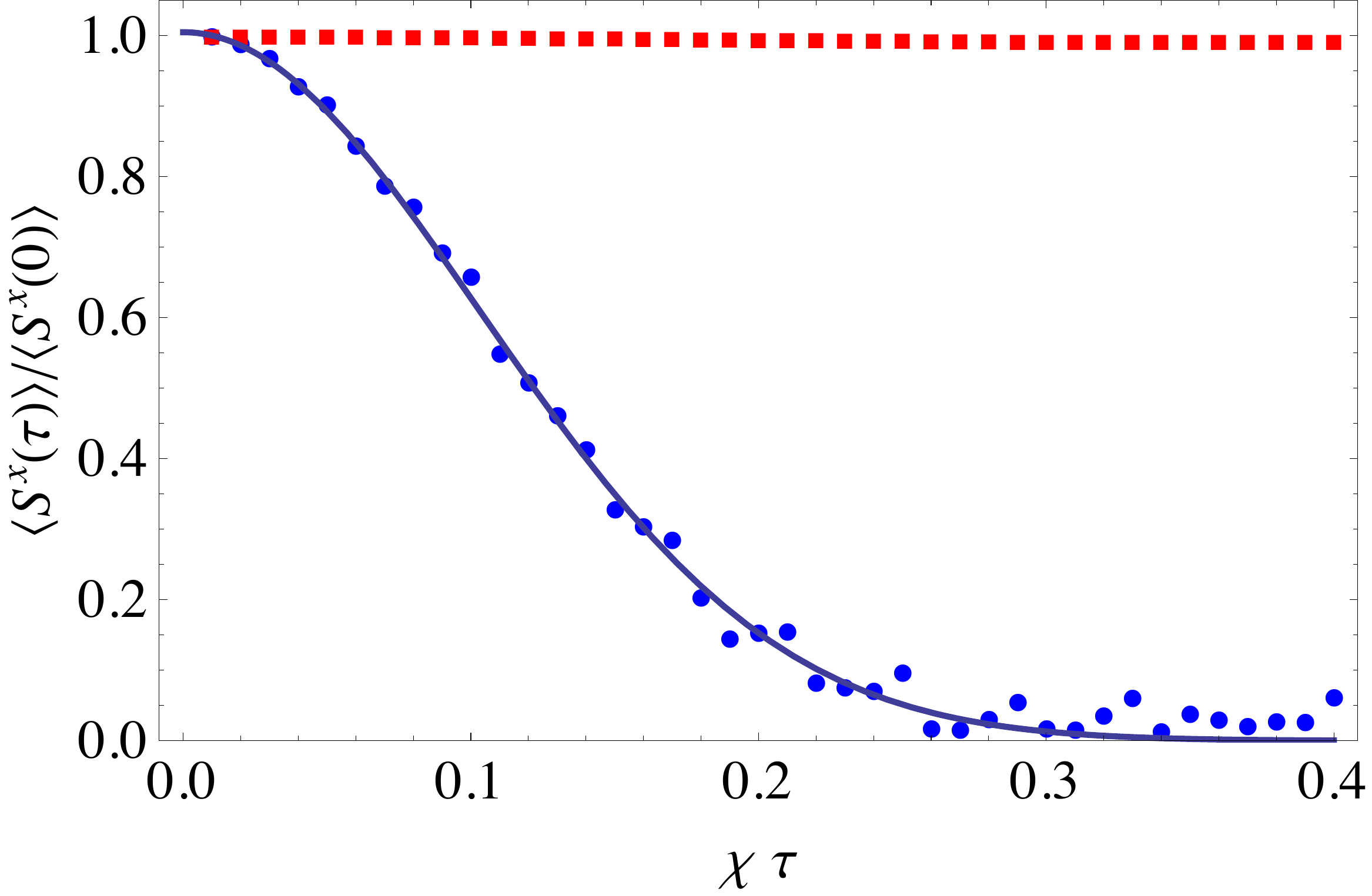}
     \end{center}  \caption{(Color online)  Left panel: Comparison of  the density shift vs excitation fraction predicted by the  analytic perturbative treatment (solid blue line),  the TWA (filled blue dots) and the mean-field (red squares) with excitation inhomogeneity. We assumed collective interactions, set $\Delta \Omega/\bar{\Omega}=0.25$ and fixed $N=30$. The density shift is computed at $\chi_{\vec{\bf n}} \tau=0.04 $. Here $v^{ee}=v^{gg}$, thus $C_{\vec{\bf n}}=0$. Right panel:  Ramsey contrast decay  vs time including excitation inhomogeneity predicted by the analytic perturbative treatment (solid blue line),  the TWA (filled blue dots), and the mean-field (red squares).  The pulse area is set to $\pi/2$.}\label{inh}
\end{figure}

Despite the disagreement of the contrast at low/high excitation fractions and   at zero  $s$-wave interactions, the collective model remains very useful for describing experimental observations during the interrogation of the clock. The reason is that,  instead of a single pancake with fixed atom number, in the experiment, there is an array of pancakes populated with a Poissonian distribution (as explained below). If one takes into account the Poissonian average, one recovers a  good agreement between the collective and non-collective models even at low/high excitation fractions, as shown in Fig.~\ref{coll}.   In this case, dephasing due to the different atom number in each lattice pancake dominates the contrast decay at low/high excitation fractions. When the Poissonian average is included,  the cases with low/high pulse areas are the ones that exhibit the fastest Ramsey fringe  contrast decay. Note that, for Fig.~\ref{coll},
the mean-atom number per  the pancake used for the plots with a Poissonian distribution was   $\bar{N}\sim 9$. On the contrary, for the fixed-atom-number plots  we set $N=20$. The difference in atom number  is the only reason why the overall timescale for the
contrast decay in the Poissonian-averaged plots is slower than in the fixed-atom-number ones.

We also expect that finite $|J^\perp_{\vec {\bf  n}_j,\vec { \bf  n}_{j'}}|>0$ interactions (induced by $s$-wave collisions) will  enhance the gap  between the collective Dicke manifold, $S=N/2$, and the other manifolds. A larger gap  will further suppress population leakage outside the former (gap protection)\cite{Rey2008a} and
 will improve the collective approximation. The validity of the gap protection idea  is explicitly shown in Appendix 5, where we use a box potential instead of a harmonic potential to make the mode dependence of the $p$-wave interactions even stronger. There we show that the dynamics converge to the collective dynamics as the mean value of  $|J^\perp_{\vec {\bf  n}_j,\vec { \bf  n}_{j'}}|$  is increased.

\subsubsection{Excitation  induced inhomogeneity}

In the presence of finite $\varphi$, Eq.~(\ref{inhoeq}), the gap protection does not help to prevent leakage of  population  outside the Dicke manifold. This is because the transfer is made during  short pulses where interactions play no role
and single-particle excitation inhomogeneity is the leading transfer mechanism. In Fig.~\ref{inh}, we compare the analytic solution found in Sec.~\ref{anal}  using  perturbation theory in $\Delta\Omega/\bar{\Omega}$ (solid blue line), the TWA (filled blue circles), and a pure  mean-field with excitation inhomogeneity (red squares). Fig.~\ref{inh} assumes collective interactions  and equal first and second pulse areas. For comparison purposes, we also show the analytic solution for the density shift in the case of zero inhomogeneity (dashed purple line). Note that excitation inhomogeneity manifests itself in the density shift  mainly at high  excitation fractions where it  gives rise to some curvature. The figure also shows  that the density shift is well captured by  both the TWA and the mean-field. On the contrary, the  Ramsey fringe contrast decay  is  only  captured by the TWA and  analytic solutions, which agree well with each other. The mean-field does not predict any contrast decay. The pulse area for the Ramsey fringe contrast shown in Fig.~\ref{inh}  was set to be $\pi/2$ .

\section{Comparisons with optical lattice clock experiments}\label{compasec}

After developing  and benchmarking in great detail the theoretical formalism, we now proceed to apply it to describe experimental measurements performed in  the ${}^{87}$Sr JILA optical lattice clock \cite{Campbell2009,Martin2012}  and  in the NIST ${}^{171}$Yb \cite{Lemke2011,Ludlow2011}  optical lattice clock.  One of the  main differences between the two clocks  is their temperature. While the JILA clock operates typically at temperatures between $(1-4) \mu $K  and with a typical excitation inhomogeneity $\Delta \Omega/\bar{\Omega}\lesssim 0.1$, the ${}^{171}$Yb  clock typically operates at temperatures  at or somewhat above $10 \mu $K with corresponding excitation inhomogeneity  $\Delta \Omega/\bar{\Omega}\sim 0.25-0.3$.  These differences have significant consequences for the proper theory model required to describe the experiment.  While the pure collective model is sufficient to describe the many-body dynamics in the ${}^{87}$Sr clock, as explicitly shown below, the inclusion of excitation inhomogeneity is needed to reproduce the  ${}^{171}$Yb measurements. We first start by describing in  more detail the experimental set-up, which is common for the two clocks, and then  present the experimental measurements and comparisons with theory.

\subsection{Spatial averaging  and finite temperature  } \label{tem}

\begin{figure}
  \begin{center}
   \includegraphics[width=70mm]{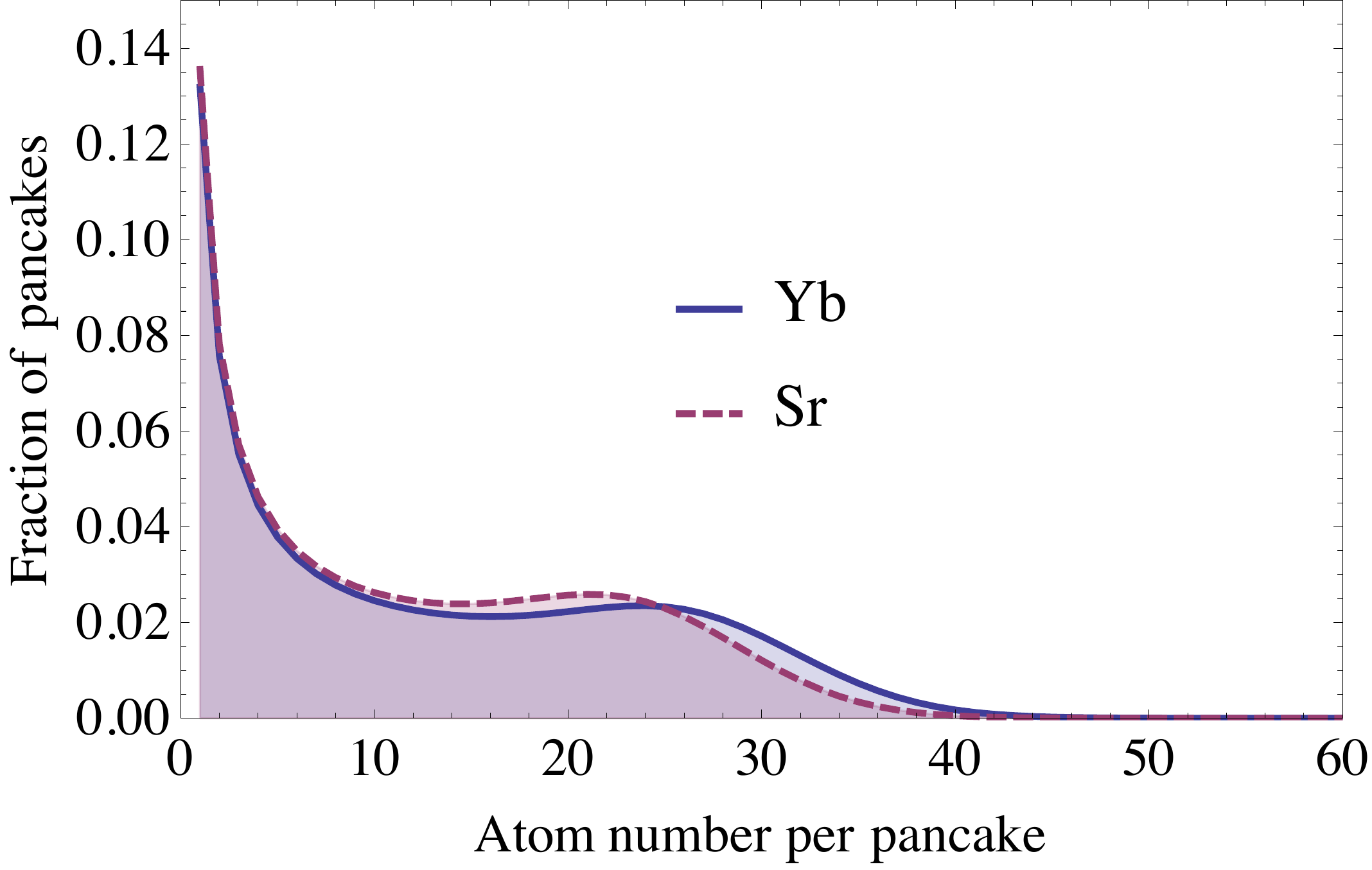}
     \end{center}  \caption{(Color Online)  Atom number distribution: The solid  blue distribution  is for the Yb 1D optical lattice clock at a typical atom number  $N_T=26000$; the dashed purple distribution is for the  Sr 1D optical lattice clock at typical atom number  $N_T=5000$.  }\label{atom}
\end{figure}

In the experiments, a one dimensional lattice is used to confine the atoms creating  an array of 2D  disk-shaped arrays  or pancakes (see Fig.~\ref{schma}b).  The atom  number distribution along the array is determined  by assuming the atoms are loaded in the MOT (magneto-optical trap)
with a Gaussian density profile, and that this profile is kept the same but simply partitioned into lattice sites when the 1D lattice is turned on. In this case   $\rho(l)=\frac{N_T a}{\sigma_M \sqrt{2\pi}} e^{-(  l a)^2/(2 \sigma_M^2)}$ is taken as the density at  pancake $l$ after integrating along the other two directions. Here $\sigma_M$  is the $1/e$ MOT radius, which is   $\sim 30 \mu$m for ${}^{87}$Sr clock and  $\sim 130 \mu$m for the ${}^{171}$Yb clock, $a\sim 0.4$ $\mu$m is the lattice spacing (approximately the same for both experiments) and $N_T$ the total atom number. The atom number fluctuates from realization to realization according to a Poissonian distribution, and the probability of having $N>0$ atoms at  lattice site $l$ is
${\mathcal{R}}(l,N) =A e^{-\rho(l)} \frac{\rho(l)^N}{N!}$  with $A$ a normalization constant. Any experimental observable should be then calculated as $\mathcal{O}_{\rm{Exp}}=\sum_{N} \mathcal{O}(N) {\mathcal{R}}(N)$ with ${\mathcal{R}}(N)=\sum_{l} {\mathcal{R}}(l,N)$. In Fig.~\ref{atom}, we show an atom number distribution for a typical number of $N_T\sim 26000$ atoms in the Yb clock. The Sr clock operates with a smaller  atom number, $N_T=5000$, but, due to  different trapping conditions, the average distribution of atoms in both experiments is similar, as can be seen in Fig.~\ref{atom}.

In most of the calculations performed above,  we  fixed the set of vibrational modes to be $\vec{\bf {n}}$. However, to compare with  experiments, we need to perform a thermal average. For  $\mu$K temperatures, at which the clocks operate, we can assume the modes are initially populated according to a Maxwell-Boltzmann  distribution,  computed using  single-particle energies.  Although so far we have assumed that atoms are frozen along  the axial direction, this assumption is not fully correct for the Yb lattice clock operated at $\sim 10 \mu$K with a typical axial trapping frequency of $\nu_Z=75-80$ KHz.  In this case,  the mean axial mode occupation number $\bar{n}_Z\sim 2.5$, and to compare with the Yb experiment, we  perform a thermal average over excited-optical-lattice bands.

\begin{figure}\begin{center}
   \includegraphics[width=80mm]{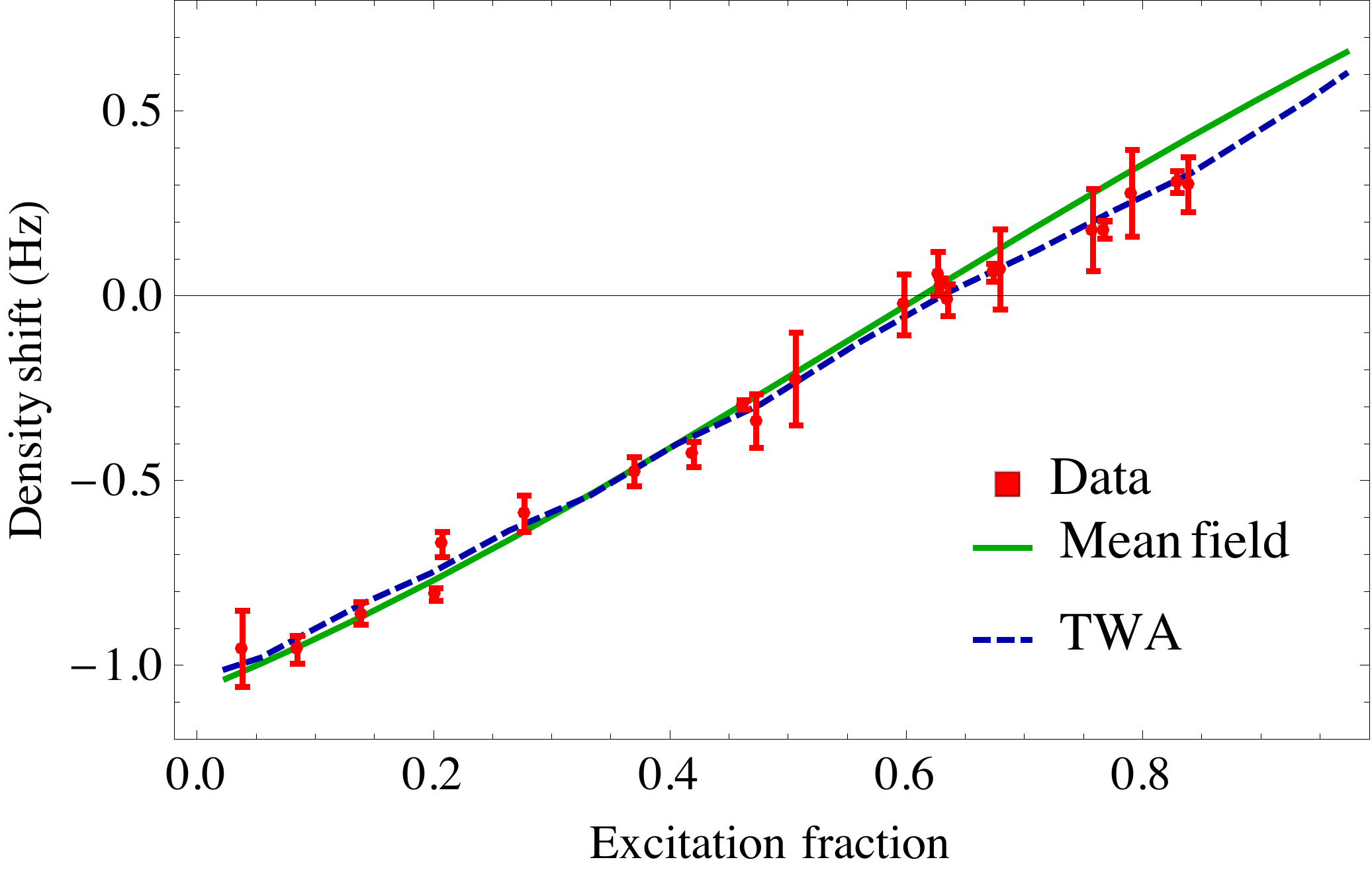}
     \end{center}  \caption{(Color Online)  Density shift in Ramsey spectroscopy vs excitation fractions measured in the Sr clock: The filled symbols show experimental data. The dashed(blue) line is the  prediction from the  collective  TWA model and the solid (green) line  the collective mean-field model.
     The experimental data was normalized  to a typical  time-averaged total atom  number of 1000 atoms. The excitation fraction was also time-averaged. See Ref.~\cite{Martin2012} and text for more information.
      }\label{srsh}
\end{figure}

\begin{figure*}
  \begin{center}
   \includegraphics[width=75mm]{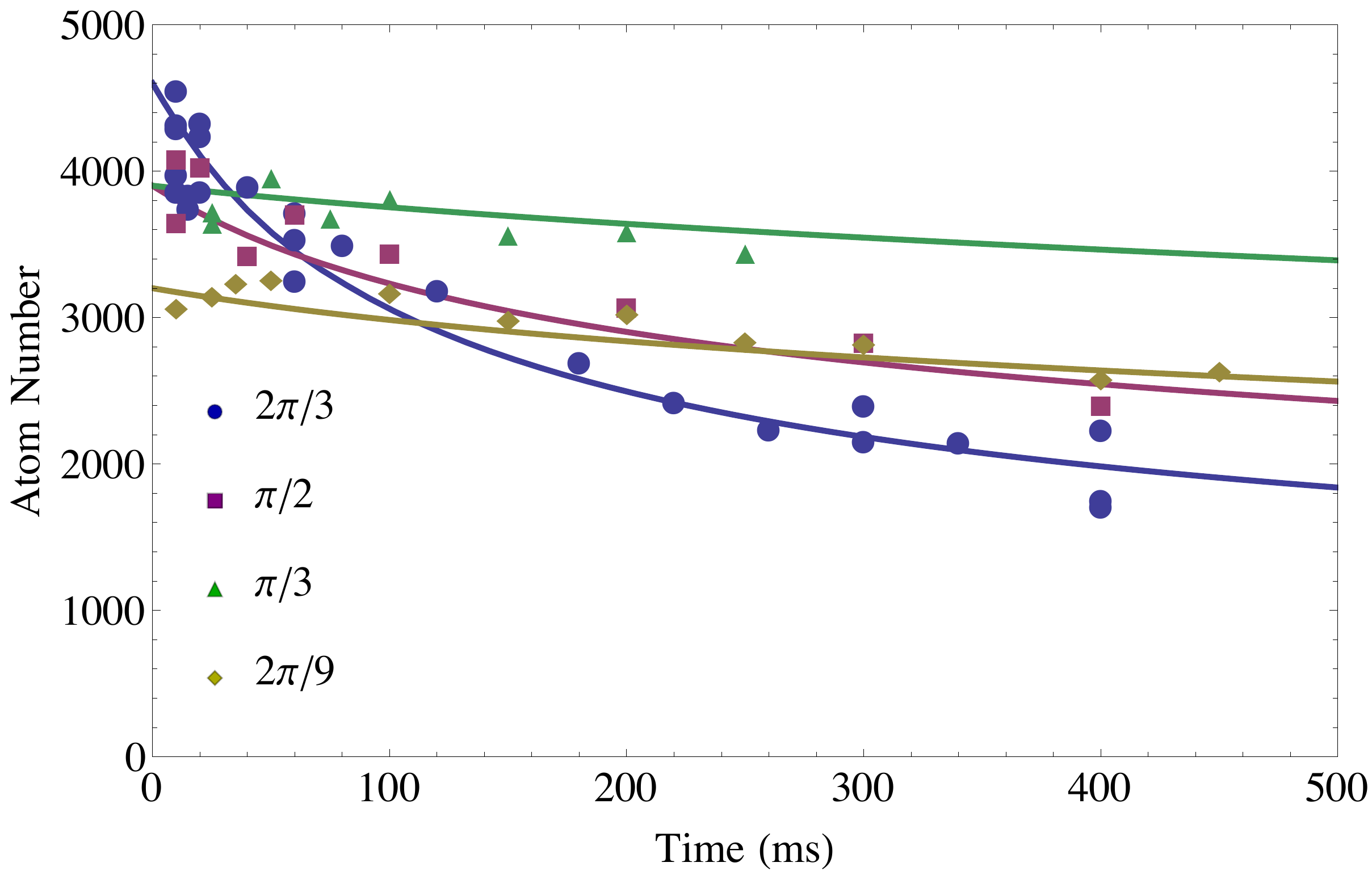}\quad    \includegraphics[width=75mm]{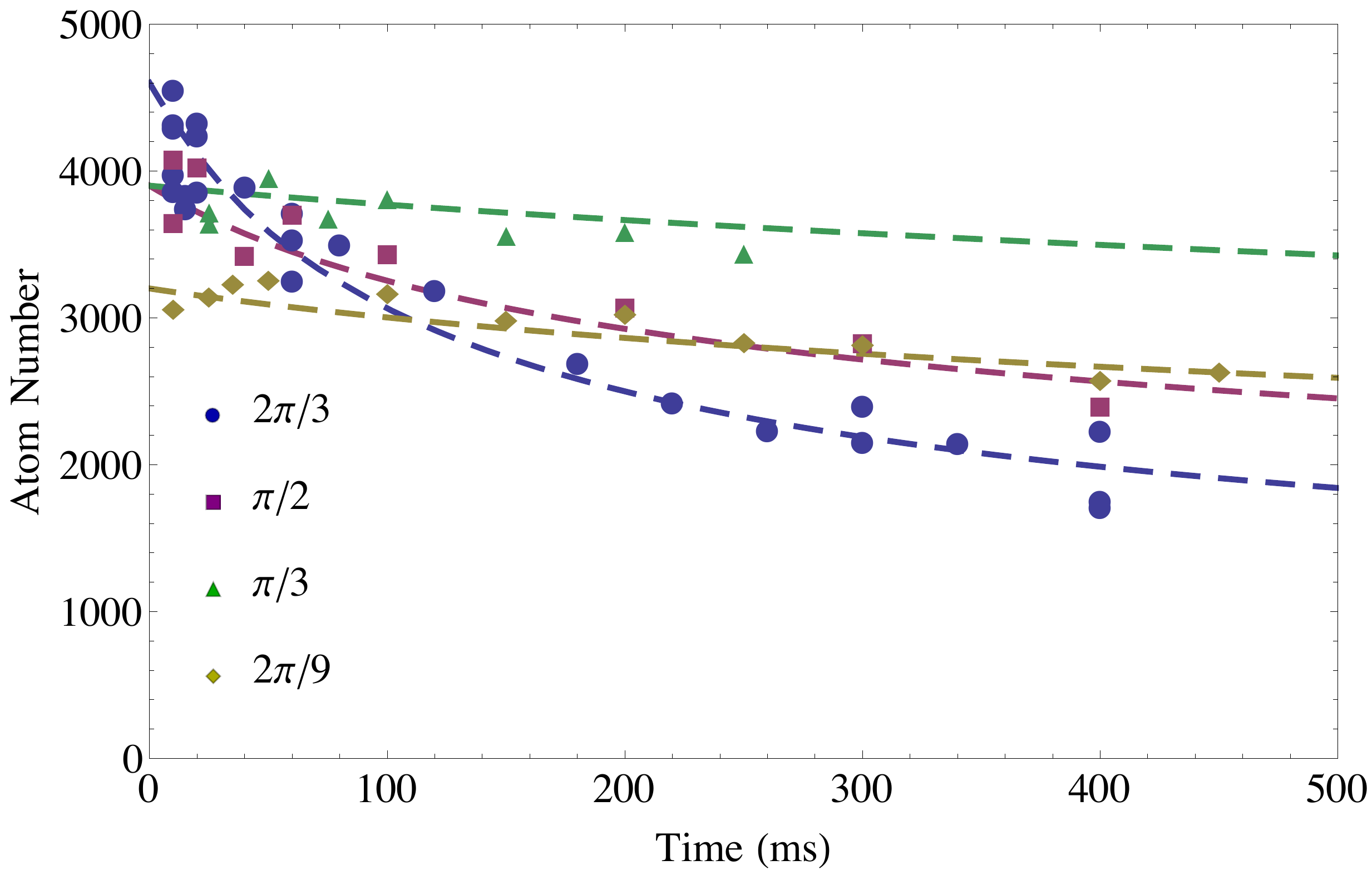}\\
     \includegraphics[width=75mm]{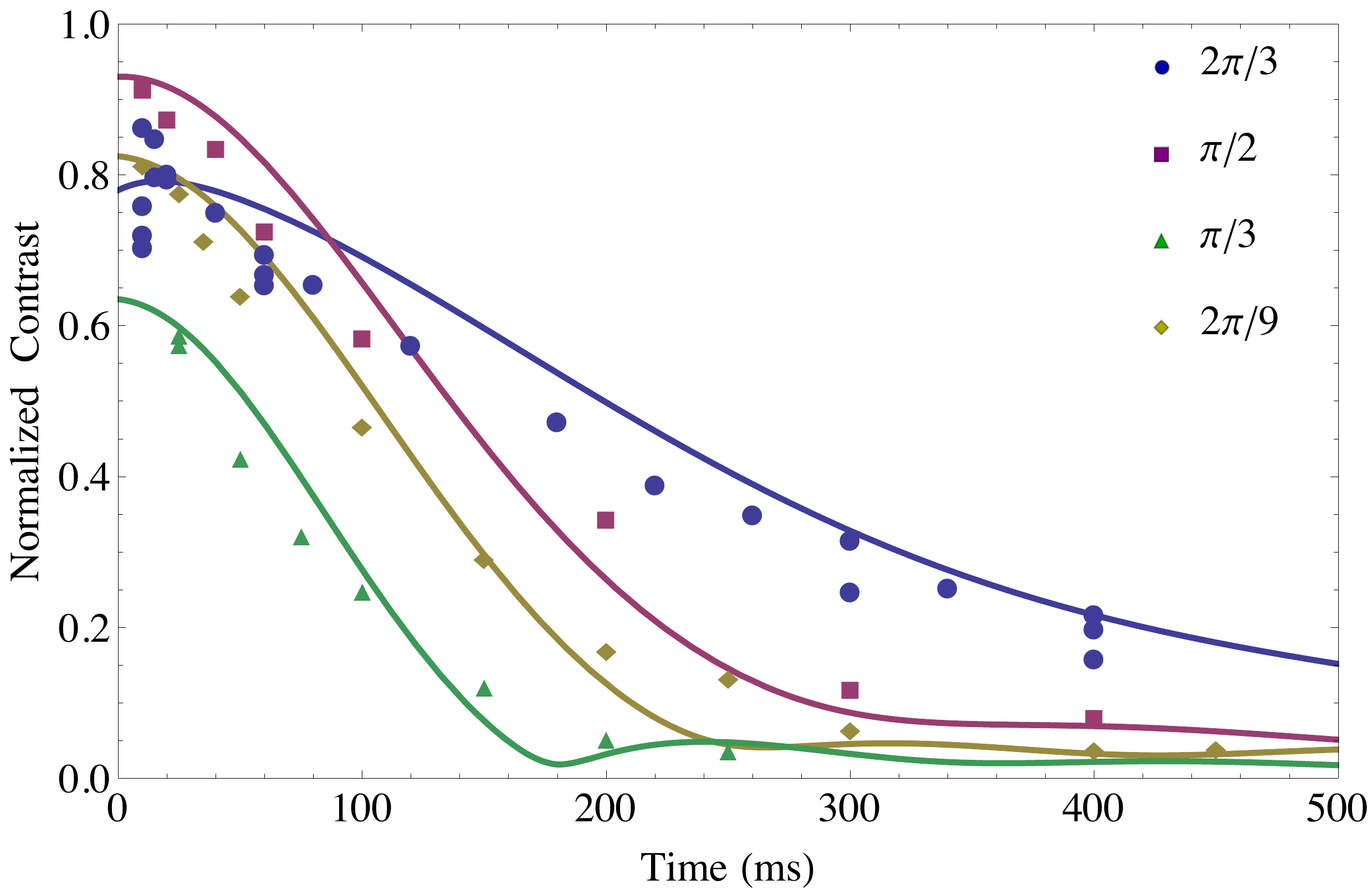}\quad    \includegraphics[width=75mm]{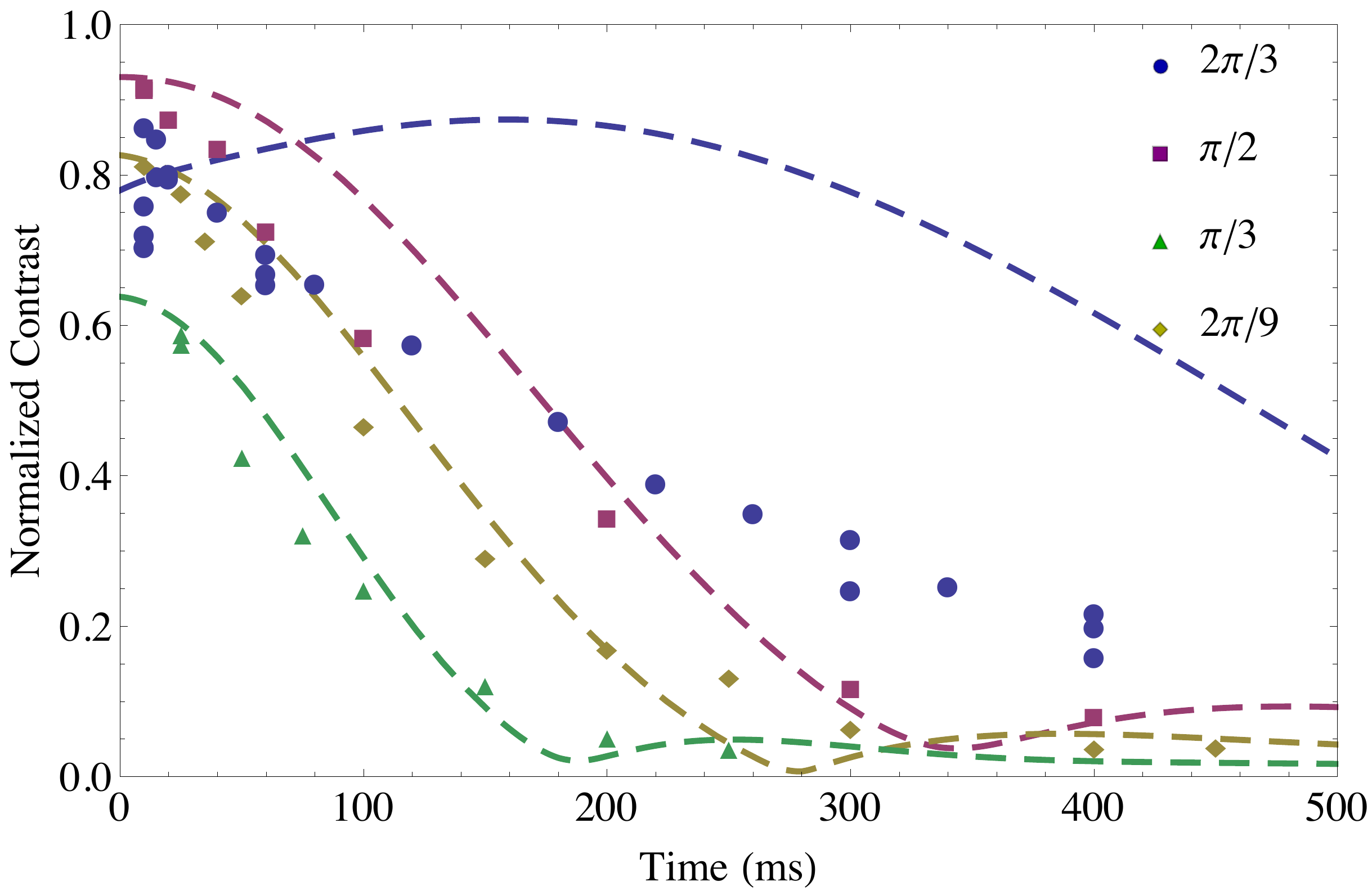}
     \end{center}  \caption{ (Color Online) Atom number decay and Ramsey fringe contrast vs dark time  measured in the Sr clock: The filled symbols show experimental data,  the solid lines are the  predictions from the collective  TWA model, and the dashed lines are the  predictions from the collective mean-field  model. See text for more information. An overall scaling factor on the order of $\sim 0.9-1$ in the contrast plots, that accounted  for the experimental dead-time between the end of the second pulse and the population measurements  was used in the theory curves to match  the contrast at zero dark time. Statistical error bars for experimental data are   comparable to symbol's size. See also Ref.~\cite{Martin2012}.}\label{srco}
\end{figure*}

\subsection{Ramsey Spectroscopy }

\begin{figure*}
  \begin{center}
   \includegraphics[width=150mm]{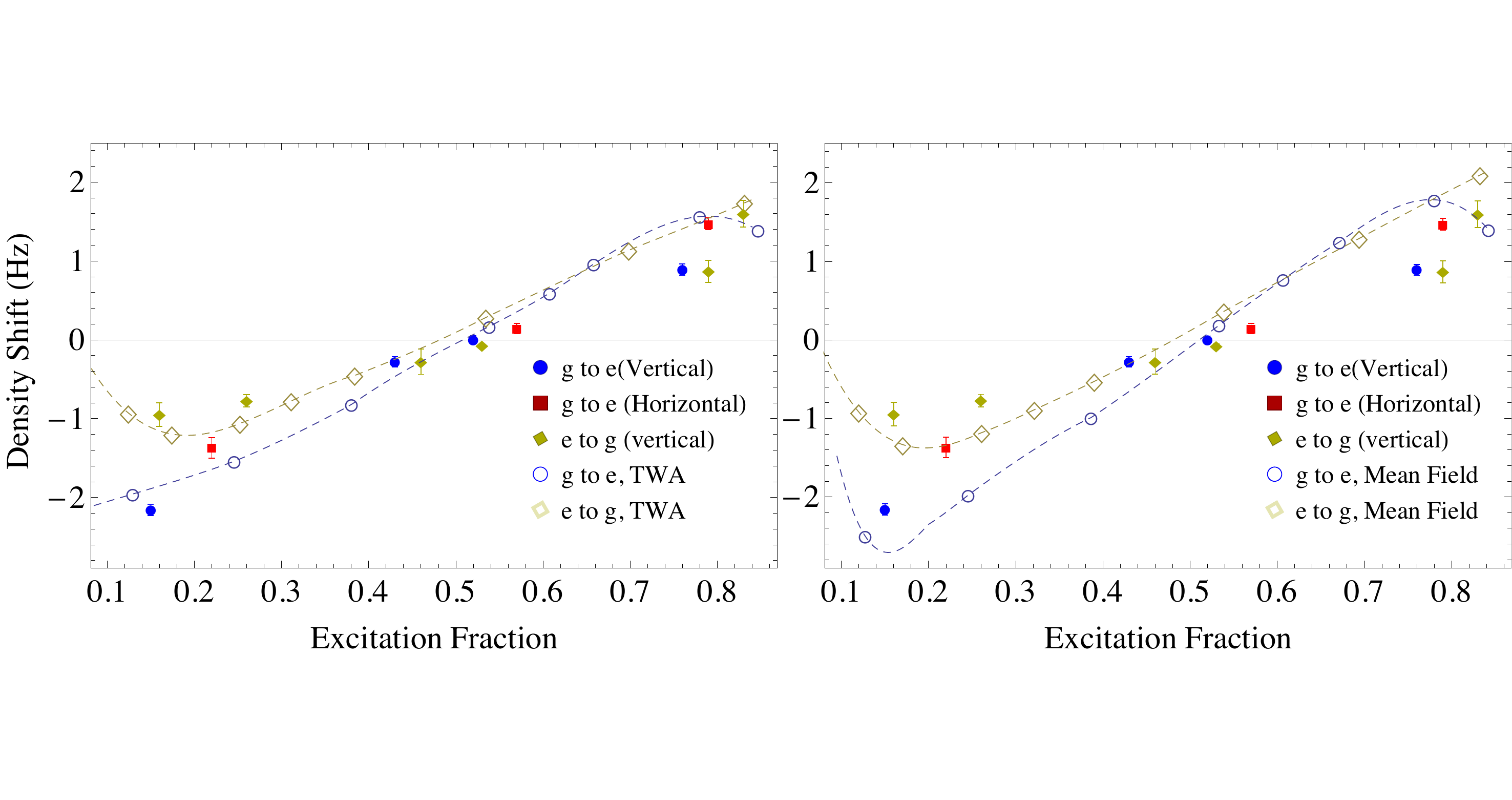}
     \end{center}  \caption{(Color Online)  Density Shift vs excitation fractions  measured in the Yb clock: The filled symbols show experimental data and the open symbols joined by dashed lines show predictions from the non-collective  TWA model (left) and the non-collective mean- field model (right). The blue circles and red squares were taken using a vertical and horizontal 1D lattice, respectively, and used a $g\to e$ interrogation. The yellow diamonds used a vertical lattice but interrogated the $e\to g$ transition.
     The experimental data measured the shift by normalizing it to a typical  time-averaged total atom  number of $\langle N_T\rangle_\tau=25000$ atoms. The excitation fraction was also time-averaged. See text for more information.
      }\label{shiex}
\end{figure*}

\begin{figure*}
  \begin{center}
   \includegraphics[width=150mm]{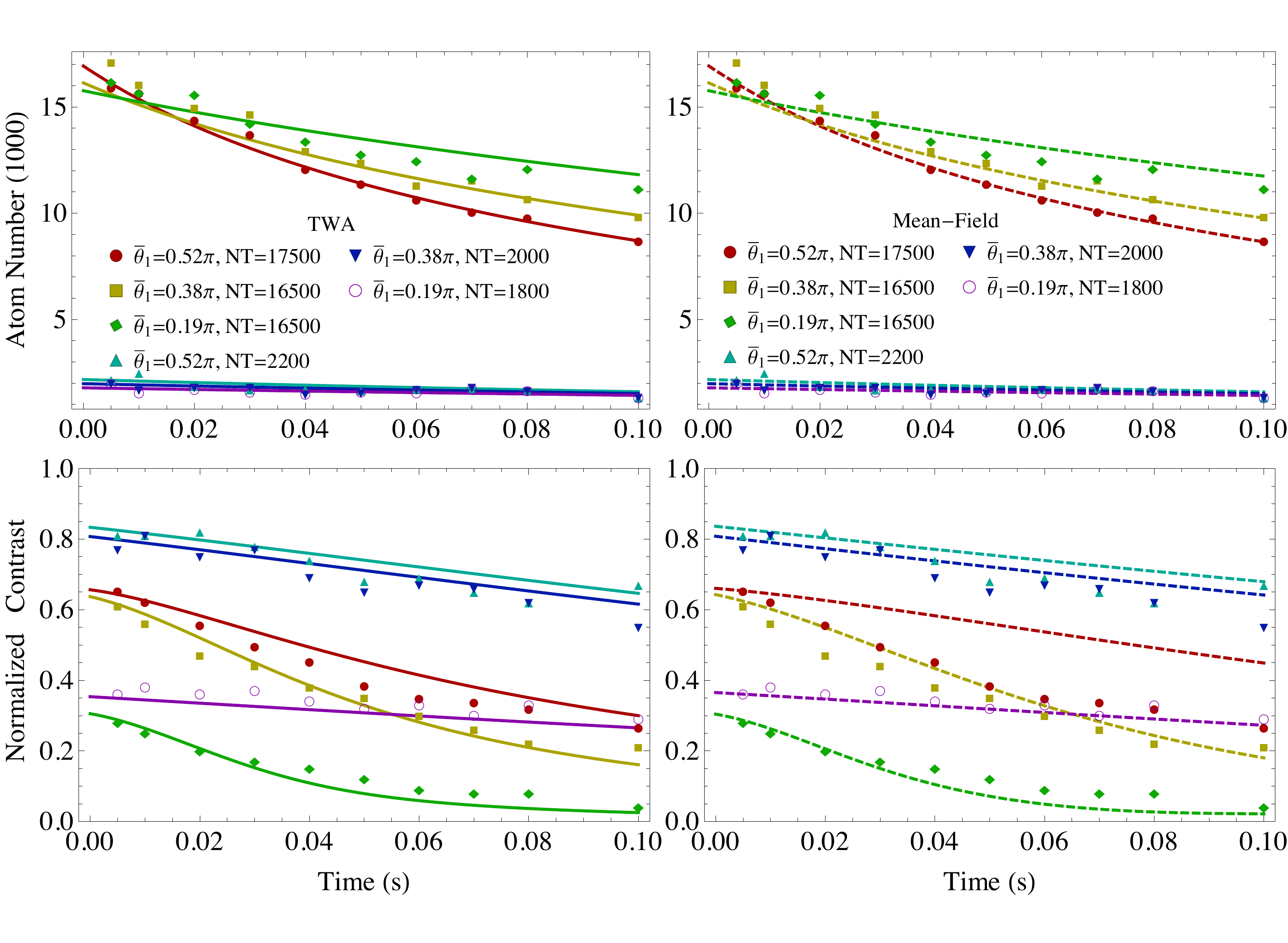}
     \end{center}  \caption{ (Color Online) Atom number  and Ramsey fringe contrast  vs dark time  measured in the Yb clock: The filled symbols show experimental data,  the solid lines are the  predictions from the non-collective  TWA model, and the dashed lines are the  predictions from the non-collective  mean-field  model. See text for more information. To account for the  additional  25 ms dead-time between the end of the second pulse and  the population measurements,  we introduced  an overall scaling factor $\sim 0.8-0.9$ to match the experimentally measured contrast at zero dark time. Experimental error bars are  comparable to symbol's size.}\label{conex}
\end{figure*}

The experiments performed Ramsey spectroscopy  measurements and varied  the atom number, pulse area, and  the initial state of the atoms.
The results are summarized in  Figs. \ref{srsh}-\ref{srcono} for the Sr clock and  Figs.~\ref{shiex}-\ref{conex}  for the Yb clock. In those plots, the   theoretical predictions are also shown.

\subsubsection{Sr optical lattice clock }
Let us first discuss the Sr clock, which admits a simpler theoretical treatment based on a pure collective model.  Figure \ref{srsh} shows the measured density shift vs excitation fraction. The excitation fraction was changed  by varying the first pulse area. In the presence of two-body losses, the excitation fraction is not constant during the dark time, and, to simplify the comparisons with theory, a time-averaged excitation fraction was used for the density shift plot. The latter was extracted in the experiment  by performing a set of independent measurements periodically inserted into the clock sequence \cite{Martin2012}. The duration of the dark time, $\tau$, was set to  80~ms, and the final pulse area  to $\pi/2$. The typical axial and radial sample temperature, measured by sideband spectroscopy and Doppler spectroscopy respectively, were between, $T_Z\sim 1-2 \mu$K and $T_R\sim 2-4 \mu$K; the  trapping frequencies  were set to $\nu_Z=80$ kHz and  $\nu_R\sim 450$ Hz;
and the typical misalignment angles were small enough  that  excitation inhomogeneity  was negligible.   Under these operating conditions,  the collective model, after performing an average over the atom number distribution,   is expected to work fairly well. This expectation is confirmed by the linear dependence of the shift at high excitation fraction and the good agrement between  the  collective TWA predictions and the experimental data.  Since only the $g$ interaction parameters are known, $b_{gg}\sim 76$ $a_B$ ($a_B$ the Bohr radius)\cite{Campbell2009}, we used the density shift data to determine the parameters  $V^{e,g}-V^{g,g}=-2\pi \times 0.27$ Hz and   $V^{e,e}-V^{g,g}=0.4 V^{e,g}$. Note that these parameters are almost temperature independent.

Figure \ref{srco} shows the atom number decay and the normalized Ramsey fringe contrast. By normalized we mean that the contrast was rescaled by the corresponding atom number at a given time. In the experiment, the Ramsey fringe contrast was extracted in a manner that is insensitive to the frequency noise of the local oscillator. For details, see Ref.~\cite{Martin2012,Martinthesis}. Four different first pulse areas were used (the second pulse area was always $\pi/2$) and  are indicated with different symbols in the figure. The Ramsey pulses were always shorter than  $6$~ms. Atom number decay data was used to determine $\Gamma^{e,e}=2\pi \times 0.1$ Hz. $\Gamma^{e,g}$ was consistent with zero in agreement with measurements reported  in Ref.~\cite{Bishof2011b}. There  the single atom decay rates were also  measured to be $\Gamma^e\sim \Gamma^g=2\pi \times 0.01$ Hz, and we used those values. These   parameters together with the values of  $V^{e,g}-V^{g,g}$ and   $V^{e,e}-V^{g,g}$ extracted from  the density shift were used to compute  the Ramsey contrast decay curves.

\begin{figure}
  \begin{center}
   \includegraphics[width=80mm]{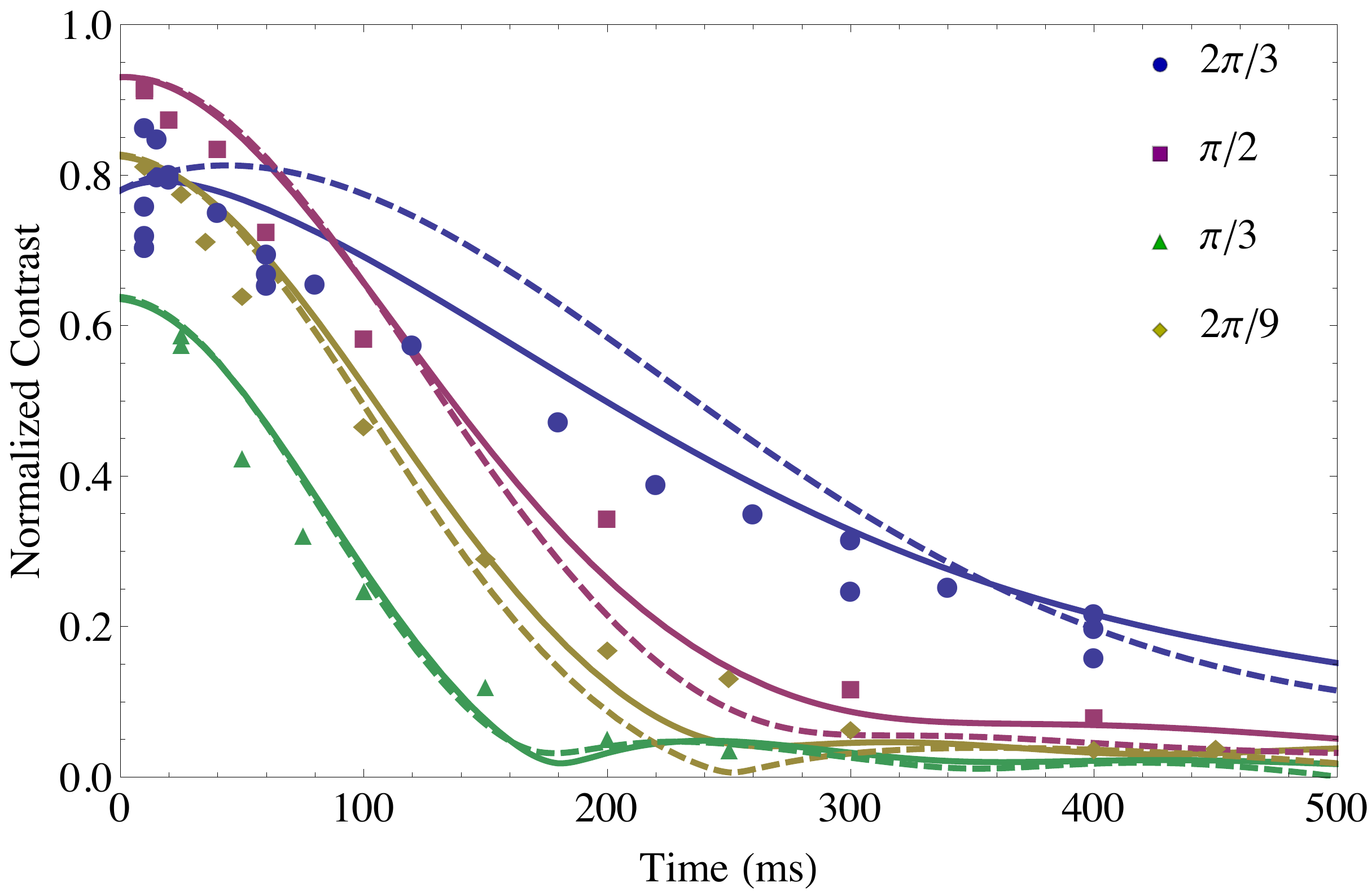}
     \end{center}  \caption{(Color Online)   Ramsey contrast decay obtained  from the the collective TWA with cubic terms (solid lines) and without them (dashed lines). The symbols are experimental data taken at different first pulse areas. Statistical error bars for experimental data are  comparable to symbol's size. See also Ref.~\cite{Martin2012}.
      }\label{srcono}
\end{figure}

We added a cubic term to account for the virtual populations of the off-resonant modes as explained in Sec.\ref{effes}.
Since the coupling constants $a_{1,2,3}^T$  cannot be trivially computed, we  used instead the quantities $
\chi^{(3,6)}_{PP}$ -- which give rise to  the  relevant cubic contributions --  as  fitting parameters (see Appendix 2) and set $a_{eg}^-=0$. The values of $\chi^{(3,6)}_{PP}$  utilized  generated an additional  term $ \sum_{r=0}^3 A_r (\hat{N}_{e})^{3-r} (\hat{N}_{g})^r$  with   $ A_0=2\pi \times 0.0016$ Hz,  $ A_1=- 2\pi \times 0.0058$ Hz, $ A_2=2\pi \times 0.003$ Hz, $ A_3=2\pi \times 0.0011$ Hz. Although the cubic corrections are small and played no role in the density shift,  they improved the  agreement between theory and experiment. To emphasize this point in Fig.~\ref{srcono}, we explicitly compare the theory predictions of the contrast with and without cubic terms.

In addition to the collective TWA solutions, we also show the collective mean-field predictions. Both were computed after performing an average over the atom number distribution. Excellent agreement is observed between the collective TWA and the experimental data \footnote{ In   Figs.~\ref{srsh}, in contrast to the theory presented in Ref.~\cite{Martin2012} where losses were neglected,  we do include losses and perform a time average.}. The mean-field model reproduces the density shift measurements and  is able to capture the fast contrast decay at low excitation fractions due to the dephasing   induced by  the different precession rates  of the Bloch vector across different pancakes. Remember that the effective magnetic field induced by interactions is directly proportional to the density shift and  depends on the pulse area and atom number. At low excitation fraction the effective magnetic field (density shift) is maximum as  shown in Fig.~\ref{srsh}. On the contrary, for  the intermediate excitation fraction data points, the mean-field solution fails to reproduce the contrast decay observed experimentally. At those excitation fractions, the effective magnetic field (density shift) is almost zero --see Fig.~\ref{srsh} -- and the contrast decay  is dominated by many-body correlations not included in the mean-field model.

\subsubsection{Yb optical lattice clock }\label{secyb}
Now we proceed to discuss the Yb clock experiment. Figure \ref{shiex} shows the measured density shift vs excitation fraction seen in the Yb clock \cite{Lemke2011,Ludlow2011}. Two types of initial conditions were considered. For the red and blue points, initially all atoms were prepared in the $g$ state and then optically excited to the $e$ state. The blue data (solid circles) was taken using a vertical optical lattice and the red data (solid squares) using an horizontal optical lattice. In the vertical lattice, tunneling between neighboring lattice sites is expected to be  substantially suppressed by gravity \cite{Lemonde2005,Lemke2011}. That suppression is not present in the horizontal setup. Based on that and  the fact that there is no substantial difference between the two sets of data, we conclude that, at least for the density shift, tunneling can be neglected. The yellow solid diamonds show the density shift measured for the initial condition, in which all the atoms were prepared in the $e$ state and then optically deexcited to the $g$ state. For all the cases, the Ramsey pulse time was $t_1 = t_2 = 1$ ms,  the dark time was $\tau= 80 $ ms, the temperature  was $T=10\mu$ K, the  trapping frequencies were $\nu_Z=75$ kHz and  $\nu_R\sim 500$ Hz and $\Delta\Omega/\bar{\Omega }\sim 0.25$. In  ${}^{171}$Yb, both $e-e$ and $e-g$ losses,  as well as single particle $e$ and $g$ losses, play a non-negligible  role in the quantum dynamics \cite{Ludlow2011}. Consequently, both the excitation fraction and atom number vary with time during the spectroscopy. The data presented here accounts for those effects by using time-averaged excitation fraction and a density shift normalized to a time-averaged atom  number. The different behavior observed between the $e-g$ and $g-e$ interrogation measurements is mainly due to excitation inhomogeneity, which depends on pulse area.

Figure \ref{conex} shows the atom number decay and normalized Ramsey fringe contrast. Three different pulse areas were used (both the first and second pulse areas were the same) as  indicated with different symbols in the figure. For each pulse area, two measurements were done, one at a high and the other at a low atom number. These measurements were carried out at similar but slightly different trapping conditions and temperature as compared to the ones used for the density shift: $\nu_Z=80$ kHz, $\nu_R\sim 550$ Hz,  $T\gtrsim 10\mu$K, and  $t_1 = t_2 = 5$ ms.

The most general TWA, which accounts for excitation inhomogeneity, non-collective interactions, two-body and single-particle losses, as well as interaction effects and virtual population of motional states (cubic terms in the spin Hamiltonian), was used to model the experiment. Both a thermal average and an average over the atom number distribution in the pancakes were  performed.   We also compared the predictions of the pure mean-field model,  which accounts for similar types of effects but without performing the sampling  over the Wigner distribution.  Only the $g$ interaction parameters are known in Yb as well, $b_{gg}\approx 0$, and thus we used  $v^{e,e},v^{e,g},\gamma^{e,e},u_{eg},\chi^{(3,6)}_{PP}$ (see Appendix 2) as fitting parameters.
 Note that, since we are not using the collective model, we quote the $v^{\alpha,\beta}$ values and not the $V^{\alpha,\beta}$ values.  The cubic Hamiltonian was the only term projected onto the collective manifold and included as a term $ \bar{P} \sum_{r=0}^3 a_r [\hat{N}_{e}^{3-r}(k=0) \hat{N}_{g}^r(k=0)]$. Here $\bar{P}=1/N^2 \sum_{j,j'}P_{{\bf n}_j ,{\bf n}_{j'}}$. $a_r$ are functions of  $\chi^{(6)}_{PP},\chi^{(3)}_{PP}$.
 From Ref.~\cite{Ludlow2011},  we set $\gamma^{e,g}=3\gamma^{e,e}/5$ and  $\Gamma^e\sim \Gamma^g=2\pi \times 0.3$ Hz. The theoretical curves shown in Figs.  \ref{shiex} and  \ref{conex}  used the following set of parameters: $v^{e,g}=-2\pi \times 3.8$ Hz,  $v^{e,e}=-0.2 v^{e,g}$,$\gamma^{e,e}=2\pi \times 2.8$ Hz,  $u_{eg}=0$,  $ a_0=a_3=2\pi \times 0.3$ Hz  and  $a_1=a_3/2=-2 a_0/3$.
For the Ramsey contrast measurements, we found that   the $5$ ms pulses used in the Yb experiment were not short enough to make interaction effects negligible  during the  pulses, and, because of that, we included interactions during those in the numerical simulations.  We added in addition a single-particle dephasing term acting  on $\rho_{eg}$  in all cases, $e^{-\gamma_{dp} \tau }$ with $\gamma_{dp}=2$ Hz to capture the contrast decay observed for the low density data since, at those densities, interactions effects could not be the cause of contrast decay.

While there is a substantial number of fitting parameters, there is also a variety of experimental data curves. Note also that $\gamma^{e,e}$ is uniquely determined by the atom loss curves and the ratio $v^{e,e}/v^{e,g}$ by the density shift measurements.   In the Sr experiment, the collective  model based on the TWA was able to fairly reproduce   the data taken under different conditions using the same set of two-body interaction parameters and decay rates. For the Yb experiment, the theory  could reproduce the experimental  measurements of the contrast and density shift using one  set of interactions and loss parameters, up to a  rescaling  by a factor of 2.5  in total atom number when computing  the density shift curves. The factor of 2.5 difference  could in principle come from the  different experimental conditions  used for the contrast decay and density shift measurements, although most likely it has its roots in  additional decoherence mechanisms not accounted for by the model, such as  tunneling, higher vibrational bands, and motional decoherence. Those are likely to  affect  the Ramsey fringe contrast decay more than the density shift and could be responsible for the  overestimation of the interaction parameters when we tried  to fit the contrast decay. We  did find  that, if we set all $a_r=0$ (no cubic terms), we needed a  re-scaling factor much larger than 2.5  to fit all the  experimental data with the same interaction parameters.

While both the TWA and the mean-field model reproduced fairly well the density shift, the mean-field model did a poorer job  capturing the contrast decay than the TWA.  For Yb,  the  effective magnetic field vanishes at pulse areas close to $\pi/2$, and notably, it is  at this pulse area  when the deviation between the mean-field and the TWA is more significant. Similar behavior was seen when modeling the Sr clock and is consistent with the expectation that when the effective magnetic field induced by interactions approaches zero, quantum correlations are the dominant cause of Ramsey fringe contrast decay. It suggests that even at $T=10\mu$K,  quantum correlations can be playing a non-negligible  role in the  dynamics of the interrogated atoms.

\section{Conclusions}
Here we have derived a powerful theoretical formalism capable of dealing with the non-equilibrium many-body dynamics  of open spin models with long-range interactions. The formalism is based on the TWA, which we have applied to  open quantum systems. We have benchmarked the accuracy  of the TWA  by taking advantage of existing exact solutions  as well as by comparisons with analytic perturbative treatments and by numerically  solving  the  master equation.  We have applied the developed formalism to describe the many-body dynamics of  optical lattice clocks during Ramsey spectroscopy and demonstrated that a full treatment of quantum correlations reproduces best the observed dynamics, better than a pure mean-field treatment.

The formalism developed here should be useful  for the description of  a broad range of important
modern quantum systems, including for example, trapped ions, neutral atoms in optical cavities,  quantum dots, and
nitrogen vacancy centers.

Going beyond the spin model formulation, which relies on the assumption of frozen motional single particle states, an alternative approach is needed to deal with colder samples where interactions can dominate over the single-particle energy.  In this parameter regime the spin model assumption is not necessarily valid.  One possible alternative approach includes a kinetic-theory treatment based on solving a Boltzmann equation. That type of treatment has been shown to be useful for describing \cite{Williams2002} the spin segregation observed at JILA  during  Ramsey spectroscopy   with a two-component ultra-cold bosonic gas \cite{Lewandowski2002} and for describing \cite{Natu2009} similar segregation effects observed  at Duke in a two-component cold trapped Fermi gas \cite{Du2008}.

\section{Acknowledgement} The authors  thank K.~R.~A. Hazzard,  M. Foss-Feig, A. Koller, M. Beverland, J. Bollinger, J. von Stecher, A. Polkovnikov  and A.~J. Daley  for numerous conversations and feedback. This work has been  supported
by  AFOSR,  NIST, NSF (JILA-PFC-1125844, JQI-PFC-1125565, IQIM-PFC, and PIF), ARO (individual investigator award),  ARO with funding for the DARPA-OLE, DARPA QuASAR, and the NDSEG, Lee A. DuBridge and Gordon and Betty Moore foundations.

\section*{Appendix 1: Collective Master Equation}

In this Appendix, we present the details behind the master-equation treatment of Sec IV A. Let  $\rho_\mathcal{N}$ be the density matrix for a single sector of $\mathcal{N}$ particles. As explained in the main text, since there is no coherence between the sectors with different particle-number, the master equation can be solved in a ``block-diagonal way''. Specifically, if the system starts with $N$ particles, we need to solve a series of differential equations for each of the subspaces with cascading atom numbers: first for $N$ particles, next for $N-2$ particles, then for  $N-4$ particles, etc.   There are ${N \choose n}\equiv \frac{N!}{n!(N-n)!}$ different sectors with $ N-n$ particles, $n=0,2\dots N$. For mode independent elastic and inelastic interaction parameters, in each of the ${N \choose n}$ sectors with $\mathcal{N}=N-n$ particles, the dynamics are restricted to the collective Dicke states $|S=\mathcal{N}/2,M_\mathcal{N}\rangle\equiv|M_\mathcal{N}\rangle$. Moreover, each of the  ${N \choose n}$ sectors behaves identically.  If we  assume that particles in $\rho_\mathcal{N}$ are numbered from 1 to $\mathcal{N}$ in such a way that atoms $\mathcal{N}+2$ and $\mathcal{N}+1$ are the ones that decay as one goes from $\rho_{\mathcal{N}+2}$ to $\rho_\mathcal{N}$, the resulting equations are
{\small
\begin{eqnarray}
\frac{d}{dt} \rho_{\mathcal{N} } &=& - \frac{i}{\hbar} [H_{\mathcal{N} },\rho_{\mathcal{N} }] - \frac{ \Gamma}{2} \sum^{\mathcal{N} }_{i < j,\alpha=ee,eg} [(\hat{A}^{\alpha}_{{\bf n}_i,{\bf n}_{j}})^\dagger  (\hat{A}^{\alpha}_{{\bf n}_i,{\bf n}_{j}})\rho_{\mathcal{N} } + \rho_{\mathcal{N} } (\hat{A}^{\alpha}_{{\bf n}_i,{\bf n}_{j}})^\dagger  (\hat{A}^{\alpha}_{{\bf n}_i,{\bf n}_{j}})] + \notag \\\label{efe}&& \sum_{\alpha=ee,eg} \Gamma^\alpha {N-\mathcal{N}  \choose 2}(\hat{A}^{\alpha}_{{\bf n}_{\mathcal{N} +2},{\bf n}_{\mathcal{N} +1}}
 \rho_{\mathcal{N} +2}  (\hat{A}^{\alpha}_{{\bf n}_{\mathcal{N} +2},{\bf n}_{\mathcal{N} +1}})^\dagger),\label{efe2}
\end{eqnarray}}for $0\leq\mathcal{N}\leq N-2$. In terms of spin operators,
\begin{eqnarray}
&&\sum_{i < j}^\mathcal{N} (\hat{A}^{ee}_{{\bf n}_i,{\bf n}_{j}})^\dagger  (\hat{A}^{ee}_{{\bf n}_i,{\bf n}_{j}})
= \frac{\mathcal{N} (\mathcal{N}-2)}{8} + \frac{\mathcal{N}-1}{2} \hat S^z + \frac{1}{2}(\hat S^z)^2,\\
&&\sum_{i < j}^\mathcal{N} (\hat{A}^{eg}_{{\bf n}_i,{\bf n}_{j}})^\dagger  (\hat{A}^{eg}_{{\bf n}_i,{\bf n}_{j}})
= \frac{\mathcal{N} (\mathcal{N}-2)}{8} + \frac{1}{2}\Big[(\hat S^x)^2 + (\hat S^y)^2- (\hat S^z)^2 \Big],
\end{eqnarray}
{\small
\begin{eqnarray}
&&\langle M_\mathcal{N} |( \hat{A}^{ee}_{{\bf n}_{\mathcal{N} +2},{\bf n}_{\mathcal{N} +1}}
 \rho_{\mathcal{N} +2} (\hat{A}^{ee}_{{\bf n}_{\mathcal{N} +2},{\bf n}_{\mathcal{N} +1}})^\dagger)|M_\mathcal{N} '\rangle =
\langle  M_{\mathcal{N}+2}+2|\rho_{\mathcal{N}+2}| M_{\mathcal{N}'+2}+2\rangle \sqrt{\frac{{\mathcal{N}\choose M_\mathcal{N}+\mathcal{N}/2} {\mathcal{N}\choose M'_\mathcal{N}+\mathcal{N}/2}}{{\mathcal{N}+2 \choose M_\mathcal{N} +\mathcal{N}/2+2} {\mathcal{N}+2 \choose M'_\mathcal{N} +\mathcal{N}/2+2}}}. \notag\\
&&\langle M_\mathcal{N} |( \hat{A}^{eg}_{{\bf n}_{\mathcal{N} +2},{\bf n}_{\mathcal{N} +1}}
 \rho_{\mathcal{N} +2}  (\hat{A}^{eg}_{{\bf n}_{\mathcal{N} +2},{\bf n}_{\mathcal{N} +1}})^\dagger)|M_\mathcal{N} '\rangle =
\langle  M_{\mathcal{N}+2}+1|\rho_{\mathcal{N}+2}| M_{\mathcal{N}'+2}+1\rangle \sqrt{\frac{{\mathcal{N}\choose M_\mathcal{N}+\mathcal{N}/2} {\mathcal{N}\choose M'_\mathcal{N}+\mathcal{N}/2}}{{\mathcal{N}+2 \choose M_\mathcal{N} +\mathcal{N}/2+1} {\mathcal{N}+2 \choose M'_\mathcal{N} +\mathcal{N}/2+1}}}.\notag
\end{eqnarray}}

After the dynamics of the various density matrix elements $\langle  M'_\mathcal{N} |\rho_\mathcal{N}| M_\mathcal{N} \rangle$ are  known, one can compute any observable. For example, the number of $e$ atoms is  given by
\begin{eqnarray}
N_{e} =\sum_{\mathcal{N}} {N \choose \mathcal{N}} \sum_{ M_\mathcal{N} =-\mathcal{N}/2+1}^{\mathcal{N}/2}  (M_\mathcal{N} +\mathcal{N}/2)\langle  M_\mathcal{N} |\rho_\mathcal{N}| M_\mathcal{N} \rangle,\notag
\end{eqnarray}
where $\mathcal{N}$ runs over $\mathcal{N} = N, N-2, N-4, ...$.

\section*{Appendix 2: Effective Hamiltonian}
In this Appendix,  we explain in detail the steps that  lead to Eq.~(\ref{efespi}) in Sec.~V. As explained there, the states $|\Psi^{\Upsilon}_{\vec \sigma_{\vec {\bf k}}}\rangle$ can be divided into six different categories, according to the intermediate state participating in 2nd-order perturbation theory. Each of these six cases gives rise to a different contribution in the Hamiltonian; we will now present these contributions. Let define $\varsigma$ the set of initially unoccupied modes.

\textbf{1. Case $| \Psi^{1h} _{\vec{ \bf \sigma}_{\vec{\bf k}}}\rangle $}\\
We start with process I, where in the intermediate state, one atom is populating a mode that does not belong to the initially populated manifold. We obtain terms with one, two, and three spin operators, {\it i.e.}
\begin{align}\label{eq:process1}
H_{\vec {\bf n}}^{1h} &= \sum_{j \notin \varsigma ,k \in \varsigma} \frac{1}{E_{{\bf n}_j}-E_{{\bf n}_k}}H^{1h}_{j,k},\\
H^{1h}_{j,k} &= H^{1h,S}_{j,k}  +H^{1h,SS}_{j,k}+H^{1h,SSS}_{j,k}.
\end{align}
{\small
\begin{align}
&H^{1h,S}_{j,k} =\sum_{m\neq j}P_{{{\bf n}_j}{{\bf n}_m}{{\bf n}_m}{{\bf n}_k}}(v^{ee}-v^{gg})\Bigg\{  {\hat S}^{z}_{{\bf n}_j} \bigg[(v^{ee}+v^{gg})\sum_{p\leq m} P_{{{\bf n}_j}{{\bf n}_p}{{\bf n}_p}{{\bf n}_k}} + v^{eg}\sum_{p<m}P_{{{\bf n}_j}{{\bf n}_p}{{\bf n}_p}{{\bf n}_k}} + \frac{u_{eg}}{2}\sum_{p\neq m}S_{{{\bf n}_j}{{\bf n}_p}{{\bf n}_p}{{\bf n}_k}}\bigg]\nonumber\\
&+ {\hat S}^{z}_{{\bf n}_m} \bigg[(v^{ee}+v^{gg})\sum_{p\neq m}P_{{{\bf n}_j}{{\bf n}_p}{{\bf n}_p}{{\bf n}_k}}+ \frac{u_{eg}}{2}\sum_{p\neq m}S_{{{\bf n}_j}{{\bf n}_p}{{\bf n}_p}{{\bf n}_k}}\bigg]\Bigg\} + \frac{u_{eg}}{2}\sum_{m\neq j}S_{{{\bf n}_j}{{\bf n}_m}{{\bf n}_m}{{\bf n}_k}}{\hat S}^{z}_{{\bf n}_m}\sum_{p\neq m}(v^{ee}-v^{gg})P_{{{\bf n}_j}{{\bf n}_p}{{\bf n}_p}{{\bf n}_k}},
\end{align}}

{\small\begin{align}
&H^{1h,SS}_{j,k} =\sum_{m\neq j}P_{{{\bf n}_j}{{\bf n}_m}{{\bf n}_m}{{\bf n}_k}}\Bigg\{ \nonumber\\
& v^{eg}\left({\hat S}^{x}_{{\bf n}_j}{\hat S}^{x}_{{\bf n}_m} + {\hat S}^{y}_{{\bf n}_j}{\hat S}^{y}_{{\bf n}_m}\right)\left[\sum_{p\neq m}(v^{ee}+v^{gg}+v^{eg})P_{{{\bf n}_j}{{\bf n}_p}{{\bf n}_p}{{\bf n}_k}} + u_{eg}S_{{{\bf n}_j}{{\bf n}_p}{{\bf n}_p}{{\bf n}_k}} + 2v_{eg}P_{{{\bf n}_j}{{\bf n}_m}{{\bf n}_m}{{\bf n}_k}} \right]\nonumber\\
& + v^{eg}\sum_{p<m}\left({\hat S}^{x}_{{\bf n}_p}{\hat S}^{x}_{{\bf n}_m} + {\hat S}^{y}_{{\bf n}_p}{\hat S}^{y}_{{\bf n}_m}\right)(v^{eg}P_{{{\bf n}_j}{{\bf n}_p}{{\bf n}_p}{{\bf n}_k}} - u_{eg}S_{{{\bf n}_j}{{\bf n}_p}{{\bf n}_p}{{\bf n}_k}})\nonumber\\
& + {\hat S}^{z}_{{\bf n}_j} {\hat S}^{z}_{{\bf n}_m} \left[(2(v^{ee})^2 + 2(v^{gg})^2 - (v^{eg})^2)\sum_{p} P_{{{\bf n}_j}{{\bf n}_p}{{\bf n}_p}{{\bf n}_k}} - (v^{eg})^2 P_{{{\bf n}_j}{{\bf n}_m}{{\bf n}_m}{{\bf n}_k}} + u_{eg}(v^{ee}+v^{gg}-v^{eg})\sum_{p\neq m}S_{{{\bf n}_j}{{\bf n}_p}{{\bf n}_p}{{\bf n}_k}} \right]\nonumber\\
& + \sum_{p<m}{\hat S}^{z}_{{\bf n}_p}{\hat S}^{z}_{{\bf n}_m}\left[(2(v^{ee})^2 -2v^{ee}v^{eg} + (v^{eg})^2 - 2v_{eg}v_{gg}+ 2(v^{gg})^2)P_{{{\bf n}_j}{{\bf n}_p}{{\bf n}_p}{{\bf n}_m}} - u_{eg}(v^{ee}+v^{gg}-v^{eg})S_{{{\bf n}_j}{{\bf n}_p}{{\bf n}_p}{{\bf n}_m}}\right]\nonumber\\
& - u_{eg}^2 \sum_{m\neq j}S_{{{\bf n}_j}{{\bf n}_m}{{\bf n}_m}{{\bf n}_k}}^2\Bigg\{ ({\hat S}^{x}_{{\bf n}_j} {\hat S}^{x}_{{\bf n}_m} + {\hat S}^{y}_{{\bf n}_j}{\hat S}^{y}_{{\bf n}_m} + {\hat S}^{z}_{{\bf n}_j} {\hat S}^{z}_{{\bf n}_m})\left[\sum_{p} S_{{{\bf n}_j}{{\bf n}_p}{{\bf n}_p}{{\bf n}_k}}+S_{{{\bf n}_j}{{\bf n}_m}{{\bf n}_m}{{\bf n}_k}}\right] \nonumber\\
&- \sum_{p<m} ({\hat S}^{x}_{{\bf n}_m} {\hat S}^{x}_{{\bf n}_p} + {\hat S}^{y}_{{\bf n}_m}{\hat S}^{y}_{{\bf n}_p} + {\hat S}^{z}_{{\bf n}_m} {\hat S}^{z}_{{\bf n}_p})S_{{{\bf n}_j}{{\bf n}_p}{{\bf n}_p}{{\bf n}_k}}\Bigg\}\nonumber\\
& -u_{eg}\sum_{m \neq j}S_{{{\bf n}_j}{{\bf n}_m}{{\bf n}_m}{{\bf n}_k}}\Bigg\{ ({\hat S}^{x}_{{\bf n}_j} {\hat S}^{x}_{{\bf n}_m} + {\hat S}^{y}_{{\bf n}_j}{\hat S}^{y}_{{\bf n}_m} + {\hat S}^{z}_{{\bf n}_j}{\hat S}^{z}_{{\bf n}_m})(v^{ee}+v^{gg}+ v^{eg})\sum_{p\neq m}P_{{{\bf n}_j}{{\bf n}_p}{{\bf n}_p}{{\bf n}_k}} \nonumber\\
&+ v^{eg}\sum_{p<m}({\hat S}^{x}_{{\bf n}_m} {\hat S}^{x}_{{\bf n}_p} + {\hat S}^{y}_{{\bf n}_m}{\hat S}^{y}_{{\bf n}_p})P_{{{\bf n}_j}{{\bf n}_p}{{\bf n}_p}{{\bf n}_k}}+(v^{ee}+v^{gg}-v^{eg})\sum_{p<m}P_{{{\bf n}_j}{{\bf n}_p}{{\bf n}_p}{{\bf n}_k}}{\hat S}^{z}_{{\bf n}_m}{\hat S}^{z}_{{\bf n}_p}\Bigg\},
\end{align}}

{\small\begin{align}
&H^{1h,SSS}_{j,k} =\sum_{m\neq j}\left(v^{eg}P_{{{\bf n}_j}{{\bf n}_m}{{\bf n}_m}{{\bf n}_k}}-u_{eg}S_{{{\bf n}_j}{{\bf n}_m}{{\bf n}_m}{{\bf n}_k}}\right)\Bigg\{\nonumber\\
&\sum_{p\leq m}\left({\hat S}^{x}_{{\bf n}_j}{\hat S}^{x}_{{\bf n}_m}{\hat S}^{z}_{{\bf n}_p} + {\hat S}^{x}_{{\bf n}_j} {\hat S}^{x}_{{\bf n}_p} {\hat S}^{z}_{{\bf n}_m}+{\hat S}^{y}_{{\bf n}_j}{\hat S}^{y}_{{\bf n}_m}{\hat S}^{z}_{{\bf n}_p} + {\hat S}^{y}_{{\bf n}_j} {\hat S}^{y}_{{\bf n}_p} {\hat S}^{z}_{{\bf n}_m}\right)2P_{{{\bf n}_j}{{\bf n}_p}{{\bf n}_p}{{\bf n}_k}}(v^{ee}-v^{gg})\Bigg\}\nonumber\\
& + (v^{ee}-v^{gg})\sum_{m\neq j}P_{{{\bf n}_j}{{\bf n}_m}{{\bf n}_m}{{\bf n}_k}}\sum_{p\leq m}{\hat S}^{z}_{{\bf n}_j}{\hat S}^{z}_{{\bf n}_m}{\hat S}^{z}_{{\bf n}_p}\left[4P_{{{\bf n}_j}{{\bf n}_p}{{\bf n}_p}{{\bf n}_k}}(v^{ee}-v^{eg}+v^{gg})- 2u_{eg}S_{{{\bf n}_j}{{\bf n}_p}{{\bf n}_p}{{\bf n}_k}}\right]\nonumber\\
&-2u_{eg}(v^{ee}-v^{gg})\sum_{m\neq j}S_{{{\bf n}_j}{{\bf n}_m}{{\bf n}_m}{{\bf n}_k}}\sum_{p<m}P_{{{\bf n}_j}{{\bf n}_p}{{\bf n}_p}{{\bf n}_k}}{\hat S}^{z}_{{\bf n}_j}{\hat S}^{z}_{{\bf n}_m}{\hat S}^{z}_{{\bf n}_p}.
\end{align}}

\textbf{2. Case $| \Psi^{2h2m} _{\vec{ \bf \sigma}_{\vec{\bf k}}}\rangle $}\\
Next, we investigate the case where two atoms occupy two different states outside the initially populated modes. We arrive at terms with one and two spin operators:
\begin{align}\label{eq:process2}
H_{\vec {\bf n}}^{2h2m} &= \sum_{(j\neq i) \notin \varsigma ;(k \neq q) \in \varsigma} \frac{1}{E_{{\bf n}_i} + E_{{\bf n}_j}-E_{{\bf n}_k}-E_{{\bf n}_q}}H^{2h2m}_{j,i; k,q},
\end{align}
\begin{align}
&H^{2h2m}_{j,i; k,q} = - 2u_{eg}^2 S_{{{\bf n}_j}i{{\bf n}_q}{{\bf n}_k}}^2 \left[{\hat S}^{x}_{{\bf n}_j} {\hat S}^{x}_{{\bf n}_i} + {\hat S}^{y}_{{\bf n}_j} {\hat S}^{y}_{{\bf n}_i} + {\hat S}^{z}_{{\bf n}_j} {\hat S}^{z}_{{\bf n}_i}\right]\nonumber\\
&+P_{{{\bf n}_j}{\bf n}_i{{\bf n}_q}{{\bf n}_k}}^2\bigg\{
((v^{ee})^2 - (v^{gg})^2) \left[{\hat S}^{z}_{{\bf n}_j} + {\hat S}^{z}_{{\bf n}_i}\right]+ 2(v_{eg})^2 \left[{\hat S}^{x}_{{\bf n}_j} {\hat S}^{x}_{{\bf n}_i} + {\hat S}^{y}_{{\bf n}_j} {\hat S}^{y}_{{\bf n}_i}\right] + 2((v^{ee})^2 -(v^{eg})^2 + (v^{gg})^2)\left[{\hat S}^{z}_{{\bf n}_j} {\hat S}^{z}_{{\bf n}_i}\right]\bigg\}.
\end{align}

\textbf{3. Case $| \Psi^{2h1m} _{\vec{ \bf \sigma}_{\vec{\bf k}}}\rangle $}\\
Now we consider processes of type 2h1m where two atoms occupy the same mode outside the initial configuration space:
\begin{align}\label{eq:process3}
H_{\vec {\bf n}}^{2h1m} &= \sum_{(j\neq i) \notin  \varsigma ;k \in  \varsigma} \frac{1}{E_{{\bf n}_i} + E_{{\bf n}_j}-2E_{{\bf n}_k}}H^{2h1m}_{j,i; k},
\end{align}
\begin{align}
H^{2h1m}_{j,i; k} = -u_{eg}^2 S_{{\bf n}_i{\bf n}_j{\bf n}_k{\bf n}_k}^2({\hat S}^{x}_{{\bf n}_j} {\hat S}^{x}_{{\bf n}_i} + {\hat S}^{y}_{{\bf n}_j} {\hat S}^{y}_{{\bf n}_i} + {\hat S}^{z}_{{\bf n}_j} {\hat S}^{z}_{{\bf n}_i}).
\end{align}

\textbf{4. Case $| \Psi^{1d} _{\vec{ \bf \sigma}_{\vec{\bf k}}}\rangle $}\\
In the case of processes of type 1d, two atoms occupy the same mode \emph{within} the initial configuration space. We obtain contributions with one, two, and three spin operators:
\begin{align}\label{eq:process4}
H_{\vec {\bf n}}^{1d} &= \sum_{(j<h) \notin  \varsigma} \frac{1}{E_{{\bf n}_j}-E_{{\bf n}_h}}H^{1d}_{j,h},\\
H^{1d}_{j,h} & = H^{1dS}_{j,h} + H^{1dSS}_{j,h} + H^{1dSSS}_{j,h},
\end{align}

{\small\begin{align}
 H^{1dS}_{j,h} =& \sum_{m\neq j,h}P_{{\bf n}_j{{\bf n}_m}{{\bf n}_m}{{\bf n}_h}}(v^{ee}-v^{gg})\Bigg\{- \frac{u_{eg}}{2}(S_{{{\bf n}_j}{{\bf n}_j}{{\bf n}_j}{{\bf n}_h}}-S_{{{\bf n}_j}{{\bf n}_h}{{\bf n}_h}{{\bf n}_h}}){\hat S}^{z}_{{\bf n}_m}\nonumber\\
&+({\hat S}^{z}_{{\bf n}_j} - {\hat S}^{z}_{{\bf n}_j})
\left[(v^{ee} + v^{gg})\sum_{p\leq m}P_{{{\bf n}_j}{{\bf n}_p}{{\bf n}_p}{{\bf n}_h}} + v^{eg}\sum_{p<m}P_{{{\bf n}_j}{{\bf n}_p}{{\bf n}_p}{{\bf n}_h}} + \frac{u_{eg}}{2}\sum_{p\neq m}S_{{{\bf n}_j}{{\bf n}_p}{{\bf n}_p}{{\bf n}_h}}\right]\Bigg\},
\end{align}}
{\small\begin{align}
& H^{1dSS}_{j,h} = \sum_{m\neq j,h}P_{{{\bf n}_j}{\bf{n}_m}{{\bf n}_m}{{\bf n}_h}}\Bigg\{({\hat S}^{x}_{{\bf n}_j}{\hat S}^{x}_{{\bf n}_m} + {\hat S}^{y}_{{\bf n}_j} {\hat S}^{y}_{{\bf n}_m} - {\hat S}^{x}_{{\bf n}_h} {\hat S}^{x}_{{\bf n}_m} - {\hat S}^{y}_{{\bf n}_h} {\hat S}^{y}_{{\bf n}_m})\times\nonumber\\
&\left[(v^{ee}+v^{eg}+v^{gg})\sum_{p\neq h,j}(v^{eg} P_{{{\bf n}_j}{{\bf n}_p}{{\bf n}_p}{{\bf n}_h}} - u_{eg}S_{{{\bf n}_j}{{\bf n}_p}{{\bf n}_p}{{\bf n}_h}}) + v^{eg}u_{eg}\sum_{p} S_{{{\bf n}_j}{{\bf n}_p}{{\bf n}_p}{{\bf n}_h}}\right]\nonumber\\
& + ({\hat S}^{z}_{{\bf n}_j}{\hat S}^{z}_{{\bf n}_m}  -{\hat S}^{z}_{{\bf n}_h}{\hat S}^{z}_{{\bf n}_m})\times\nonumber\\
&\left[(2 (v^{ee})^2 + 2(v^{gg})^2 - (v^{eg})^2)\sum_{p}P_{{{\bf n}_j}{{\bf n}_p}{{\bf n}_p}{{\bf n}_h}} + u_{eg}(v^{ee}+v^{gg}+v^{eg})\sum_{p\neq m}S_{{{\bf n}_j}{{\bf n}_p}{{\bf n}_p}{{\bf n}_h}} - 2u_{eg}v^{eg}\sum_{p} S_{{{\bf n}_j}{{\bf n}_p}{{\bf n}_p}{{\bf n}_h}}\right]\nonumber\\
&-u_{eg}^2 \sum_{m\neq j,h}S_{{{\bf n}_j}{{\bf n}_m}{{\bf n}_m}{{\bf n}_h}}\left({\hat S}^{x}_{{\bf n}_j}{\hat S}^{x}_{{\bf n}_m} + {\hat S}^{y}_{{\bf n}_j} {\hat S}^{y}_{{\bf n}_m} - {\hat S}^{x}_{{\bf n}_h} {\hat S}^{x}_{{\bf n}_m} - {\hat S}^{y}_{{\bf n}_h} {\hat S}^{y}_{{\bf n}_m}+{\hat S}^{z}_{{\bf n}_j}{\hat S}^{z}_{{\bf n}_m}  -{\hat S}^{z}_{{\bf n}_h}{\hat S}^{z}_{{\bf n}_m}\right)\sum_{p} S_{{{\bf n}_j}{{\bf n}_p}{{\bf n}_p}{{\bf n}_h}}\nonumber\\
& -u_{eg}(v^{ee}+v^{gg}+v^{eg}) \sum_{m\neq j} S_{{{\bf n}_j}{{\bf n}_m}{{\bf n}_m}{{\bf n}_h}}\sum_{p\neq m}P_{{{\bf n}_j}{{\bf n}_p}{{\bf n}_p}{{\bf n}_h}}\left({\hat S}^{x}_{{\bf n}_j}{\hat S}^{x}_{{\bf n}_m} + {\hat S}^{y}_{{\bf n}_j} {\hat S}^{y}_{{\bf n}_m} - {\hat S}^{x}_{{\bf n}_h} {\hat S}^{x}_{{\bf n}_m} - {\hat S}^{y}_{{\bf n}_h} {\hat S}^{y}_{{\bf n}_m}+{\hat S}^{z}_{{\bf n}_j}{\hat S}^{z}_{{\bf n}_m}  -{\hat S}^{z}_{{\bf n}_j}{\hat S}^{z}_{{\bf n}_m}\right)\nonumber\\
&+ u_{eg}(S_{{{\bf n}_j}{{\bf n}_j}{{\bf n}_j}{{\bf n}_h}}-S_{{{\bf n}_h}{{\bf n}_h}{{\bf n}_h}{{\bf n}_j}})\left({\hat S}^{x}_{{\bf n}_j} {\hat S}^{x}_{{\bf n}_h} + {\hat S}^{y}_{{\bf n}_j} {\hat S}^{y}_{{\bf n}_h} + {\hat S}^{z}_{{\bf n}_j} {\hat S}^{z}_{{\bf n}_h}\right)\left[u_{eg}\sum_{p} S_{{{\bf n}_j}{{\bf n}_p}{{\bf n}_p}{{\bf n}_h}} +(v^{ee}+v^{gg}+v^{eg})\sum_{p \neq j,h}P_{{{\bf n}_j}{{\bf n}_p}{{\bf n}_p}{{\bf n}_h}} \right],\end{align}}
{\small
\begin{align}
& H^{1dSSS}_{j,h} = \sum_{m\neq j,h}P_{{\bf n}_j{{\bf n}_m}{{\bf n}_m}{{\bf n}_h}}\left({\hat S}^{x}_{{\bf n}_j}{\hat S}^{x}_{{\bf n}_m}{\hat S}^{z}_{{\bf n}_h} -{\hat S}^{x}_{{\bf n}_h}{\hat S}^{x}_{{\bf n}_m} {\hat S}^{z}_{{\bf n}_j} + {\hat S}^{y}_{{\bf n}_j}{\hat S}^{y}_{{\bf n}_m}{\hat S}^{z}_{{\bf n}_h} -{\hat S}^{y}_{{\bf n}_h}{\hat S}^{y}_{{\bf n}_m} {\hat S}^{z}_{{\bf n}_j}\right)\times\nonumber\\
&(v^{eg}P_{{{\bf n}_j}{{\bf n}_m}{{\bf n}_m}{{\bf n}_h}}-u_{eg}S_{{\bf {n}_{{\bf n}_j}}{{\bf n}_m}{{\bf n}_m}{{\bf n}_h}})2v^{eg}(v^{ee}-v^{gg})\nonumber\\
&+ \sum_{m \neq j,h}P_{{{\bf n}_j}{{\bf n}_m}{{\bf n}_m}{{\bf n}_h}}2(v^{ee}-v^{gg})\sum_{{{\bf n}_p}<{{\bf n}_m}}(P_{{{\bf n}_j}{{\bf n}_p}{{\bf n}_p}{{\bf n}_h}}v^{eg}-u_{eg}S_{{{\bf n}_j}{{\bf n}_p}{{\bf n}_p}{{\bf n}_h}})\times\nonumber\\
&({\hat S}^{x}_{{{\bf n}_j}}{\hat S}^{x}_{{\bf n}_p}{\hat S}^{z}_{{\bf n}_m} -{\hat S}^{x}_{{\bf n}_h}{\hat S}^{x}_{{\bf n}_p} {\hat S}^{z}_{{\bf n}_m} + {\hat S}^{y}_{{{\bf n}_j}}{\hat S}^{y}_{{\bf n}_p}{\hat S}^{z}_{{\bf n}_m} -{\hat S}^{y}_{{\bf n}_h}{\hat S}^{y}_{{\bf n}_p} {\hat S}^{z}_{{\bf n}_m})\nonumber\\
&+ \sum_{m \neq j,h}(v^{eg}P_{{{\bf n}_j}{{\bf n}_m}{{\bf n}_m}{{\bf n}_h}}-u_{eg}S_{{{\bf n}_j}{{\bf n}_m}{{\bf n}_m}{{\bf n}_h}})\sum_{p< m}P_{{{\bf n}_j}{{\bf n}_p}{{\bf n}_p}{{\bf n}_h}}({\hat S}^{x}_{{{\bf n}_j}}{\hat S}^{x}_{{\bf n}_m}{\hat S}^{x}_{{\bf n}_p} -{\hat S}^{x}_{{\bf n}_h}{\hat S}^{x}_{{\bf n}_m} {\hat S}^{z}_{{\bf n}_p} + {\hat S}^{y}_{{{\bf n}_j}}{\hat S}^{y}_{{\bf n}_m}{\hat S}^{z}_{{\bf n}_p} -{\hat S}^{y}_{{\bf n}_h}{\hat S}^{y}_{{\bf n}_m} {\hat S}^{z}_{{\bf n}_p})\nonumber\\
& +2u_{eg}(v^{ee}-v^{gg})\sum_{m\neq j,h}P_{{{\bf n}_j}{{\bf n}_m}{{\bf n}_m}{{\bf n}_h}}(S_{{{\bf n}_j}{{\bf n}_j}{{\bf n}_j}{{\bf n}_h}}-S_{{{\bf n}_j}{{\bf n}_h}{{\bf n}_h}{{\bf n}_h}})({\hat S}^{x}_{{{\bf n}_j}}{\hat S}^{x}_{{\bf n}_h}{\hat S}^{z}_{{\bf n}_m}+{\hat S}^{y}_{{{\bf n}_j}}{\hat S}^{y}_{{\bf n}_h}{\hat S}^{y}_{{\bf n}_m}+{\hat S}^{z}_{{{\bf n}_j}}{\hat S}^{z}_{{\bf n}_h}{\hat S}^{z}_{{\bf n}_m})\nonumber\\
& + \sum_{m\neq j,h}(v^{ee}-v^{gg})\sum_{p<m}({\hat S}^{z}_{{{\bf n}_j}}{\hat S}^{z}_{{\bf n}_m}{\hat S}^{z}_{{\bf n}_p} - {\hat S}^{z}_{{\bf n}_h}{\hat S}^{z}_{{\bf n}_m}{\hat S}^{z}_{{\bf n}_p})
\nonumber\\
&\left[4P_{{{\bf n}_j}{{\bf n}_m}{{\bf n}_m}{{\bf n}_h}}P_{{{\bf n}_j}{{\bf n}_p}{{\bf n}_p}{{\bf n}_h}}(v^{ee}-v^{eg}+v^{gg}) - 2u_{eg}(P_{{{\bf n}_j}{{\bf n}_m}{{\bf n}_m}{{\bf n}_h}}S_{{{\bf n}_j}{{\bf n}_p}{{\bf n}_p}{{\bf n}_h}}+P_{{{\bf n}_j}{{\bf n}_p}{{\bf n}_p}{{\bf n}_h}}S_{{{\bf n}_j}{{\bf n}_m}{{\bf n}_m}{{\bf n}_h}})\right ].\end{align}}

\textbf{5. Case $| \Psi^{2d} _{\vec{ \bf \sigma}_{\vec{\bf k}}}\rangle $}\\
Next, we consider the case when two modes inside the initial configuration space are doubly occupied. We obtain contributions with one, two and three spin operators:
\begin{align}\label{eq:process5}
H_{\vec {\bf  n}}^{2d} &= \sum_{j>h, (i\neq j)> (r \neq h),(i,j,h,r)\notin  \varsigma} \frac{1}{E_{{\bf n}_j}-E_{{\bf n}_h}+E_{{\bf n}_i}-E_{{\bf n}_r}}H^{2d}_{j,h,i,r},
\end{align}
{\small
\begin{align}
&H^{2d}_{j,h,i,r} = \frac{1}{2}P_{{{\bf n}_j}{{\bf n}_i}{{\bf n}_r}{{\bf n}_h}}^2 ((v^{ee})^2 - (v^{gg})^2)({\hat S}^{z}_{{\bf n}_j} + {\hat S}^{z}_{{\bf n}_i} - {\hat S}^{z}_{{\bf n}_h} - {\hat S}^{z}_{{\bf n}_r}) \nonumber\\
&-2((v^{ee})^2 - (v^{gg})^2)P_{{{\bf n}_j}{{\bf n}_i}{{\bf n}_r}{{\bf n}_h}}^2({\hat S}^{z}_{{\bf n}_j}{\hat S}^{z}_{{\bf n}_i}{\hat S}^{z}_{{\bf n}_r} + {\hat S}^{z}_{{\bf n}_j}{\hat S}^{z}_{{\bf n}_i}{\hat S}^{z}_{{\bf n}_h} - {\hat S}^{z}_{{\bf n}_j} {\hat S}^{z}_{{\bf n}_r}{\hat S}^{z}_{{\bf n}_h} - {\hat S}^{z}_{{\bf n}_i} {\hat S}^{z}_{{\bf n}_r}{\hat S}^{z}_{{\bf n}_h})\nonumber\\
&-2(v^{ee}-v^{gg})P_{{{\bf n}_j}{{\bf n}_i}{{\bf n}_r}{{\bf n}_h}}(v^{eg}P_{{{\bf n}_j}{{\bf n}_i}{{\bf n}_r}{{\bf n}_h}}-u_{eg}S_{{{\bf n}_j}{{\bf n}_i}{{\bf n}_r}{{\bf n}_h}}) \times\nonumber\\
&\Big[{\hat S}^{x}_{{\bf n}_j}{\hat S}^{x}_{{\bf n}_r}{\hat S}^{z}_{{\bf n}_i} + {\hat S}^{y}_{{\bf n}_j}{\hat S}^{y}_{{\bf n}_r}{\hat S}^{z}_{{\bf n}_i} + {\hat S}^{x}_{{\bf n}_i} {\hat S}^{x}_{{\bf n}_h} {\hat S}^{z}_{{\bf n}_j} + {\hat S}^{y}_{{\bf n}_i}{\hat S}^{y}_{{\bf n}_h}{\hat S}^{z}_{{\bf n}_j}- {\hat S}^{x}_{{\bf n}_j}{\hat S}^{x}_{{\bf n}_r}{\hat S}^{z}_{{\bf n}_h} - {\hat S}^{y}_{{\bf n}_j}{\hat S}^{y}_{{\bf n}_r}{\hat S}^{z}_{{\bf n}_h} - {\hat S}^{x}_{{\bf n}_i} {\hat S}^{x}_{{\bf n}_h} {\hat S}^{z}_{{\bf n}_r} - {\hat S}^{y}_{{\bf n}_i}{\hat S}^{y}_{{\bf n}_h}{\hat S}^{z}_{{\bf n}_r}\Big]\nonumber\\
&-2(v^{ee}-v^{gg})P_{{{\bf n}_j}{{\bf n}_i}{{\bf n}_r}{{\bf n}_h}}(v^{eg}P_{{{\bf n}_j}{{\bf n}_i}{{\bf n}_r}{{\bf n}_h}}+u_{eg}S_{{{\bf n}_j}{{\bf n}_i}{{\bf n}_r}{{\bf n}_h}})\times\nonumber\\
&\Big[{\hat S}^{x}_{{\bf n}_j}{\hat S}^{x}_{{\bf n}_h}{\hat S}^{z}_{{\bf n}_i} + {\hat S}^{y}_{{\bf n}_j}{\hat S}^{y}_{{\bf n}_h}{\hat S}^{z}_{{\bf n}_i} + {\hat S}^{x}_{{\bf n}_i} {\hat S}^{x}_{{\bf n}_r} {\hat S}^{z}_{{\bf n}_j} + {\hat S}^{y}_{{\bf n}_i}{\hat S}^{y}_{{\bf n}_r}{\hat S}^{z}_{{\bf n}_j} - {\hat S}^{x}_{{\bf n}_j}{\hat S}^{x}_{{\bf n}_h}{\hat S}^{z}_{{\bf n}_r} - {\hat S}^{y}_{{\bf n}_j}{\hat S}^{y}_{{\bf n}_h}{\hat S}^{z}_{{\bf n}_r} - {\hat S}^{x}_{{\bf n}_i} {\hat S}^{x}_{{\bf n}_r} {\hat S}^{z}_{{\bf n}_h} - {\hat S}^{y}_{{\bf n}_i}{\hat S}^{y}_{{\bf n}_r}{\hat S}^{z}_{{\bf n}_h}\Big].
\end{align}}

\textbf{6. Case $| \Psi^{1h1d} _{\vec{ \bf \sigma}_{\vec{\bf k}}}\rangle $}\\
Finally, we investigate the case of  double occupancy of a mode  $ \in \varsigma$ and one mode  $ \notin \varsigma$:
\begin{align}\label{eq:process6}
H_{\vec {\bf n}}^{1h1d} &= \sum_{(i\neq j\neq h)\notin  \varsigma, q\in  \varsigma} \frac{1}{E_{{\bf n}_j}-E_{{\bf n}_h}+E_{{\bf n}_i}-E_{{\bf n}_q}}H^{(1h1d)}_{j,h,i,q},
\end{align}
{\small
\begin{align}
&H_{j,h,i,q}^{1h1d} = \frac{1}{2}P_{{{\bf n}_j}{{\bf n}_i}{{\bf n}_h}{{\bf n}_q}}^2 ((v^{ee})^2 - (v^{gg})^2)({\hat S}^{z}_{{\bf n}_j} + {\hat S}^{z}_{{\bf n}_i} - {\hat S}^{z}_{{\bf n}_h} )\nonumber\\
& - u_{eg}^2 S_{{{\bf n}_j}{{\bf n}_i}{{\bf n}_h}{{\bf n}_q}}^2 ({\hat S}^{x}_{{\bf n}_j}{\hat S}^{x}_{{\bf n}_i} + {\hat S}^{y}_{{\bf n}_j}{\hat S}^{y}_{{\bf n}_i} +{\hat S}^{z}_{{\bf n}_j}{\hat S}^{z}_{{\bf n}_i} )\nonumber\\
& + (v^{eg})^2 P_{{{\bf n}_j}{{\bf n}_i}{{\bf n}_h}{{\bf n}_q}}^2  ({\hat S}^{x}_{{\bf n}_j}{\hat S}^{x}_{{\bf n}_i} + {\hat S}^{y}_{{\bf n}_j}{\hat S}^{y}_{{\bf n}_i}) + ((v^{ee})^2 -(v^{eg})^2 + (v^{gg})^2 )P_{{{\bf n}_j}{{\bf n}_i}{{\bf n}_h}{{\bf n}_q}}^2{\hat S}^{z}_{{\bf n}_j}{\hat S}^{z}_{{\bf n}_i}\nonumber\\
& + P_{{{\bf n}_j}{{\bf n}_i}{{\bf n}_h}{{\bf n}_q}}^2 \left[ -v^{eg}(v^{ee}+v^{gg})({\hat S}^{x}_{{\bf n}_j}{\hat S}^{x}_{{\bf n}_h} + {\hat S}^{y}_{{\bf n}_j}{\hat S}^{y}_{{\bf n}_h}+{\hat S}^{x}_{{\bf n}_i}{\hat S}^{x}_{{\bf n}_h} + {\hat S}^{y}_{{\bf n}_i}{\hat S}^{y}_{{\bf n}_h}) - ((v^{ee})^2 + (v^{gg})^2)({\hat S}^{z}_{{\bf n}_i}{\hat S}^{z}_{{\bf n}_h}+ {\hat S}^{z}_{{\bf n}_j}{\hat S}^{z}_{{\bf n}_h})\right]\nonumber\\
& + u_{eg}S_{{{\bf n}_j}{{\bf n}_i}{{\bf n}_h}{{\bf n}_q}}P_{{{\bf n}_j}{{\bf n}_i}{{\bf n}_h}{{\bf n}_q}}\left[(v^{ee}+v^{gg})({\hat S}^{x}_{{\bf n}_j}{\hat S}^{x}_{{\bf n}_h} + {\hat S}^{y}_{{\bf n}_j}{\hat S}^{y}_{{\bf n}_h} - {\hat S}^{x}_{{\bf n}_i}{\hat S}^{x}_{{\bf n}_h} - {\hat S}^{y}_{{\bf n}_i}{\hat S}^{y}_{{\bf n}_h}) - 2v^{eg}({\hat S}^{z}_{{\bf n}_i}{\hat S}^{z}_{{\bf n}_h} - {\hat S}^{z}_{{\bf n}_j}{\hat S}^{z}_{{\bf n}_h})\right]\nonumber\\
& - 2(v^{ee}-v^{gg})v^{eg}P_{{{\bf n}_j}{{\bf n}_i}{{\bf n}_h}{{\bf n}_q}}^2({\hat S}^{x}_{{\bf n}_j}{\hat S}^{x}_{{\bf n}_h}{\hat S}^{z}_{{\bf n}_i}+{\hat S}^{y}_{{\bf n}_j}{\hat S}^{y}_{{\bf n}_h}{\hat S}^{z}_{{\bf n}_i} + {\hat S}^{x}_{{\bf n}_i}{\hat S}^{x}_{{\bf n}_h}{\hat S}^{z}_{{\bf n}_j}+{\hat S}^{y}_{{\bf n}_i}{\hat S}^{y}_{{\bf n}_h}{\hat S}^{z}_{{\bf n}_j} )\nonumber\\
&  +2(v^{ee}-v^{gg})u_{eg}P_{{{\bf n}_j}{{\bf n}_i}{{\bf n}_h}{{\bf n}_q}}S_{{{\bf n}_j}{{\bf n}_i}{{\bf n}_h}{{\bf n}_q}}({\hat S}^{x}_{{\bf n}_j}{\hat S}^{x}_{{\bf n}_h}{\hat S}^{z}_{{\bf n}_i}+{\hat S}^{y}_{{\bf n}_j}{\hat S}^{y}_{{\bf n}_h}{\hat S}^{z}_{{\bf n}_i} - {\hat S}^{x}_{{\bf n}_i}{\hat S}^{x}_{{\bf n}_h}{\hat S}^{z}_{{\bf n}_j}-{\hat S}^{y}_{{\bf n}_i}{\hat S}^{y}_{{\bf n}_h}{\hat S}^{z}_{{\bf n}_j} )\nonumber\\
& - 2((v^{ee})^2 - (v^{gg})^2)P_{{{\bf n}_j}{{\bf n}_i}{{\bf n}_h}{\bf {n}_q}}^2{\hat S}^{z}_{{\bf n}_j}{\hat S}^{z}_{{\bf n}_i}{\hat S}^{z}_{{\bf n}_h}.
\end{align}}

The formulas for the contributions of the effective Hamiltonian can be simplified significantly  if we consider the case of collective interactions, {\it i.e.}we use the fact that the coupling constants $P_{{{\bf n}_j}{{\bf n}_i}{{\bf n}_h}{{\bf n}_q}}$ and $S_{{{\bf n}_j}{{\bf n}_i}{{\bf n}_h}{{\bf n}_q}}$ do not depend significantly on $i,j,h,$ and $q$.

Under the assumption of equal coupling constants, it follows immediately that the contributions to the effective Hamiltonian stemming from processes of type 1d and 2d,  Eq.~\eqref{eq:process4} and Eq.~\eqref{eq:process5}, vanish due to symmetry considerations. We then  arrive at the following contributions:

\begin{align}
H^{1h} =& \frac{1}{12} (v^{ee}-v^{gg})\left[\chi^{(3)}_{PP}(3(3N-2)(N-2) (v^{ee}+v^{gg} + 3N(N-2)v^{eg}) +  \chi^{(3)}_{PS}3 u^{eg}(N-1)(N-2)\right] {\hat S}^{z}\nonumber\\
& + \frac{({\hat S}^{x})^2 + ({\hat S}^{y})^2}{4}\bigg [\chi^{(3)}_{PP}(2(N-2)v^{eg}(v^{ee}+v^{gg}) + (3N-2)(v^{eg})^2 ) \nonumber\\
&- \chi^{(3)}_{PS} u^{eg}((N-2)(2v^{ee}+2v^{gg}) + (3N-2)v^{eg})\bigg]\nonumber\\
&+ \frac{({\hat S}^{z})^2}{4}\bigg[ \chi^{(3)}_{PP}((6N-8)((v^{ee })^2+ (v^{gg})^2 )- 2(N-2)v^{eg}(v^{ee}+v^{gg}) - (N+2)(v^{eg})^2 )\nonumber\\
& - \chi^{(3)}_{PS} 2(N-2)u^{eg}(v^{ee}+v^{eg}+v^{gg}) \bigg] \nonumber\\
& -\frac{\vec S \cdot \vec S}{4}(N+2)\chi^{(3)}_{SS}(u^{eg})^2\nonumber\\
& + ({\hat S}^{z})^3 (v^{ee} - v^{gg})\bigg[\chi^{(3)}_{PP}(v^{ee}-v^{eg}+v^{gg}) - \chi^{(3)}_{SP}u^{eg}\bigg]\nonumber\\
& + (({\hat S}^{x})^2 + ({\hat S}^{y})^2){\hat S}^{z} (v^{ee}-v^{gg})\bigg[\chi^{(3)}_{PP}v^{eg}- \chi^{(3)}_{PS}u^{eg}\bigg]\\
H^{2h2m} =& \chi^{(4)}_{PP}\bigg[((v^{ee})^2 - (v^{gg})^2)2(N-1){\hat S}^{z} + 2(v^{eg})^2 (({\hat S}^{x})^2 + ({\hat S}^{y})^2) + 2((v^{ee})^2 - (v^{eg})^2 + (v^{gg})^2 )({\hat S}^{z})^2 \bigg] \nonumber\\
&- 2 (u_{eg})^2 \chi^{(4)}_{SS} \vec S \cdot \vec S\\
H^{2h1m} &= - 2 (u_{eg})^2 \chi^{(5)}_{SS} \vec S \cdot \vec S\\
H^{1h1d} &= \frac{1}{4} {\hat S}^{z} (v^{ee}- v^{gg}) \left[\chi_{PP}^{(6)}((N^2 + 3N -2)(v^{ee}+v^{gg}) +4N v^{eg}) - 4N \chi_{PS}^{(6)}u_{eg}\right]\nonumber\\
& - \chi_{SS}^{(6)} (N-2)(u_{eg})^2 \vec S \cdot \vec S \nonumber\\
& +  (({\hat S}^{x})^2 + ({\hat S}^{y})^2)(N-2) \chi_{PP}^{(6)} v^{eg}(v^{eg}-2v^{ee}-2v^{gg})\nonumber\\
& - ({\hat S}^{z})^2 (N-2)\left[\chi_{PP}^{(6)}  ((v^{ee})^2 + (v^{eg})^2 + (v^{gg})^2) + 4 \chi_{PS}^{(6)} u_{eg}v^{eg} \right]\nonumber\\
& + 2(({\hat S}^{x})^2 + ({\hat S}^{y})^2){\hat S}^{z} (\chi_{PP}^{(6)} v^{eg} - \chi_{PS}^{(6)} u_{eg})(v^{gg}-v^{ee})\nonumber\\
& + 2({\hat S}^{z})^3 \chi_{PP}^{(6)}((v^{gg})^2 - (v^{ee})^2)
\end{align}

\begin{align}
\chi^{(3)}_{AB} = \frac{1}{N(N-1)(N-2)}\sum_{\substack{j,p,m\\ k \in \Upsilon}}\frac{A_{{\bf n}_j,{\bf n}_p,{\bf n}_p,{\bf n}_k}B_{{\bf n}_j,{\bf n}_m,{\bf n}_m,{\bf n}_k}}{E_{{\bf n}_j} - E_{{\bf n}_k}}
\end{align}

\begin{align}
\chi^{(4)}_{AB} = \frac{1}{N(N-1)}\sum_{\substack{j,i\\ k, q \in \Upsilon}}\frac{A_{{\bf n}_j,{\bf n}_i,{\bf n}_k,{\bf n}_q}B_{{\bf n}_j,{\bf n}_i,{\bf n}_k,{\bf n}_q}}{E_{{\bf n}_j} + E_{{\bf n}_i}- E_{{\bf n}_k}-E_{{\bf n}_q}}
\end{align}

\begin{align}
\chi^{(5)}_{AB} = \frac{1}{N(N-1)}\sum_{j,i; k \in \Upsilon}\frac{A_{{\bf n}_j,{\bf n}_i,{\bf n}_k,{\bf n}_k}B_{{\bf n}_j,{\bf n}_i,{\bf n}_k,{\bf n}_k}}{E_{{\bf n}_j} + E_{{\bf n}_i}- 2E_{{\bf n}_k}}
\end{align}

\begin{align}
\chi^{(6)}_{AB} = \frac{1}{N(N-1)(N-2)}\sum_{\substack{j\neq i \neq h\\ q \in \Upsilon}}\frac{A_{{\bf n}_j,{\bf n}_i,{\bf n}_h,{\bf n}_q}B_{{\bf n}_j,{\bf n}_i,{\bf n}_h,{\bf n}_q}}{E_{{\bf n}_j} + E_{{\bf n}_i}- E_{{\bf n}_h}-E_{{\bf n}_q}}
\end{align}
Finally, we make use of the fact that within the collective Dicke manifold the relation $({\hat S}^{x})^2 + ({\hat S}^{y})^2 + (S^z)^2 = \tfrac{N}{2}(\tfrac{N}{2}+1)$ holds.

\begin{eqnarray}
H^{1h} &=& a^{1h}_1 {\hat S}^{z}+a^{1h}_2 ({\hat S}^{z})^2+  a^{1h}_3({\hat S}^{z})^3 \\
H^{2h2m} & =& a^{2h2m}_1 {\hat S}^{z} + a^{2h2m}_2({\hat S}^{z})^2 \\
H^{2h1m} &=& b^{2h1m}_2\vec S \cdot \vec S\\
H^{1d} &=& H^{2d}=0 \\
H^{1h1d} &=& a^{1h1d}_1 {\hat S}^{z}+a^{1h1d}_2 ({\hat S}^{z})^2+  a^{1h1d}_3({\hat S}^{z})^3\\
\end{eqnarray}  with

{\small \begin{eqnarray}
&&a^{1h}_1 = \frac{1}{12}{\hat S}^{z} (v^{ee} - v^{gg})\bigg[\chi^{(3)}_{PP}(3(3N-2)(N-2) (v^{ee}+v^{gg}) + 6N^2 v^{eg}) + \chi^{(3)}_{PS}u^{eg}3(N-1)(N-2)\bigg],\\
&&a^{1h}_2 =  \frac{1}{2}\chi^{(3)}_{PP}\bigg[(3N-4)((v^{ee})^2 + (v^{gg})^2 )-2(N-2)v^{eg}(v^{ee} + v^{gg}) - 2N(v^{eg})^2 \bigg],\\
&&a^{1h}_3 =   (v^{ee} - v^{gg})\bigg[\chi^{(3)}_{PP}(v^{ee}-2v^{eg} + v^{gg})\bigg],\\
&&a^{2h2m}_1 =\chi^{(4)}_{PP}\bigg[((v^{ee})^2 - (v^{gg})^2 )2(N-1)\bigg],\\
&&a^{2h2m}_2 =\chi^{(4)}_{PP}\bigg[ 2((v^{ee})^2- 2 (v^{eg})^2 + (v^{gg})^2 )\bigg],\\
&&b^{2h1m}_2 = - 2 (u_{eg})^2 \chi^{(5)}_{SS},\\
&&a^{1h1d}_1= \frac{1}{4} (v^{ee} - v^{gg})\left[\chi_{PP}^{(6)} (N^2 +3N-2)(v^{ee}+v^{gg}) + 4N(N-4)(\chi_{PS}^{(6)}u_{eg}- \chi_{PP}^{(6)} v^{eg})\right],\\
&&a^{1h1d}_2 =-  (N-2)\left[ \chi_{PP}^{(6)} ((v^{ee})^2 + 2(v^{eg})^2 + (v^{gg})^2 - 2v^{eg}(v^{ee} + v^{gg})) + 2\chi_{PS}^{(6)} u_{eg} (v^{ee} + 2v^{eg} + v^{gg}) \right],\\
&&a^{1h1d}_3= - 2 ({\hat S}^{z})^3 (v^{ee} - v^{gg}) \bigg[ \chi_{PP}^{(6)} (v^{ee} - 2v^{eg} + v^{gg}) + 2\chi_{PS}^{(6)}u_{eg}\bigg].
\end{eqnarray}}

In conclusion,  the virtual processes give rise to an effective Hamiltonian, which  after projection onto the collective
Dicke Manifold and  up to constants of motion ({\it i.e.} $\vec S \cdot \vec S$ terms) is  given by:

\begin{eqnarray}
H_{\vec n}^{S_2}&=&a^{T}_1 {\hat S}^{z}+a^{T}_2 ({\hat S}^{z})^2+  a^{T}_3({\hat S}^{z})^3,\\
a^{T}_1& =&a^{1h}_1+a^{2h2m}_1+a^{1h1d}_1,\\
a^{T}_2& =&a^{1h}_2+a^{2h2m}_2+a^{1h1d}_1,\\
a^{T}_3& =&a^{1h}_2+a^{1h1d}_1.
\end{eqnarray}

\section*{Appendix 3: Analytic solution for the case of weak excitation inhomogeneity}

In the  presence of excitation inhomogeneity,  an analytic treatment based on perturbation theory can be performed when dealing with purely unitary evolution ({\it i.e.} neglecting two-body losses) and assuming collective two-body interactions.

To accomplish that,  one writes $\Omega_{\vec{\bf n}_j}=\bar{\Omega}_{\vec{\bf n}} +\delta \Omega_{\vec{\bf n}_j}$, with $\bar{\Omega}_{\vec{\bf n}} =\sum_{j} \Omega_{\vec{\bf n}_j}/N$ the mean Rabi frequency  and treats $\delta \Omega_{\vec{\bf n}_j}/\bar{\Omega}$ as a perturbation parameter.  Note that in this case the Hamiltonian evolution  during the pulse can be written as

 \begin{eqnarray}
&&e^{i t   \sum_{j=1}^N\Omega_{\vec{\bf n}_j} {\hat S}^{y}_{\vec{\bf n}_j}}=   e^{i t   \bar{\Omega} {\hat S}^{y}}\hat {\Re}(t),\\
&&\hat {\Re}(t)=  (1+ i t \hat {O}  - ( t {\hat  O})^2/2\dots),\\
&&\hat O\equiv \sum_{j=1}^N \delta\Omega_{\vec{\bf n}_j} {\hat S}^{y}_{\vec{\bf n}_j}= \sum_{k=1}^N \delta\Omega_{k} \left(\sum_{j=1}^N e^{i 2\pi k j/N}{\hat S}^{y}_{\vec{\bf n}_j}\right).
\end{eqnarray}Here  $k=1,\dots N-1$, ${\hat S}^{x,y,z}=\sum_j {S}^{x,y,z}_{\vec{\bf n}_j}$, $\delta \Omega_{k}\equiv\frac{1}{N} \sum_{j=1}^{N} e^{i \frac{2 \pi j k}{N}} \delta\Omega_{\vec {\bf  n}_j}$, and ${\hat S}^{\pm}_{\vec{\bf n}_j}={\hat S}^{x}_{\vec{\bf n}_j}\pm i {\hat S}^{y}_{\vec{\bf n}_j}$. Since
 \begin{eqnarray}
&&\sum_{j=1}^{N} e^{i 2\pi k j/N} {\hat S}^{\pm}_{\vec{\bf n}_j}|N/2,M\rangle =A^\pm_{N,M} |N/2-1,M\pm1,k\rangle\notag, \\
&&A^\pm_{N,M} \equiv \pm \left(\frac{(N/2 \mp M)(N/2\mp M-1)}{(N-1)} \right)^{1/2},
\end{eqnarray} then  $\hat O |N/2,M\rangle=\frac{1}{2 i} \sum_{k=1}^N \delta\Omega_{k} (A^+_{N,M}|N/2-1,M+1,k\rangle-A^-_{N,M}|N/2-1,M-1,k\rangle)$. This equality means that the role of inhomogeneity is to  populate the $S=N/2-1$ states,  the so called spin-wave states. These states
during the free evolution acquire  a  phase   $  -\tilde{\delta} M \tau + \chi_{\vec{\bf n}} M^2 \tau- N J^\perp_{\vec{\bf n}}\tau$, which   will give rise to an additional interaction energy shift proportional to $J^\perp_{\vec{\bf n}}$  not present in  the fully collective case. $\tilde{\delta} =\delta-C_{\vec{\bf n}} (N-1)$.

Under this approximation, one  obtains that, after the second pulse,  the number of excited atoms, $N_{e,\vec {\bf n}}(t_1,t_2)$ in Ramsey spectroscopy is:

{\small
\begin{eqnarray}
N_{e,{\vec{\bf n}}}(t_1,t_2)&=&\frac{N}{2}+\langle {\hat S}^{z} \rangle=N/2+ \cos(\theta_2^{\vec{\bf n}})   \langle \psi (0^-)|\hat {\Re}^\dagger(t_1) e^{i \tau \hat H^{S}/\hbar}  \hat {\Re}^\dagger(t_2) {\hat S}^{z}  {\Re}(t_2)e^{-i\tau \hat H^{S}/\hbar} \hat {\Re}(t_1)|\psi (0^-)\rangle \notag \\&& + \sin(\theta_2^{\vec{n}})   \langle \psi (0^-)|\hat {\Re}^\dagger(t_1) e^{i \tau \hat H^{S}/\hbar}  \hat {\Re}^\dagger(t_2) {\hat S}^{x} \hat  {\Re}(t_2)e^{-i\tau \hat H^{S}/\hbar} {\hat \Re}(t_1)|\psi (0^-)\rangle. \label{stateer2}
\end{eqnarray}}Keeping only the leading order correction,  which can be shown to be proportional to $\Delta{\Omega}_{\vec{\bf n}}^2 =\sum_{j} \Omega_{\vec{\bf n}_j}^2/N-\bar{\Omega}_{\vec{\bf n}}^2$, {\it i.e.} the first-order corrections vanish,  one can show that the Ramsey signal is given  by

{\small
\begin{eqnarray}
&&N_{e,{\vec{\bf n}}}(t_1,t_2)= \frac{N}{2} \Delta \theta _1 \Delta \theta _2  W \cos \left({\bar\theta} _1\right) \cos \left({\bar\theta}_2\right) Z^{N-2} \cos [\tau (\tilde{\delta}+(N-2)\zeta +\vartheta )]  \\&& +\frac{N}{2}  \left(1-\frac{\Delta \theta_2^2}{2}\right)\sin({\bar\theta}_1 )\sin({\bar\theta}_2) Z^{N-1} \cos\left[ \tau(\tilde{\delta}+(N-1)\zeta)\right] \notag \\&& -\frac{N }{2} \left(\frac{\Delta \theta_1^2}{4}\right) \sin({\bar\theta}_1 )\sin({\bar\theta}_2) Z^{N-3} \bigg[ (N-1) \tilde{Z}\cos\left[ \tau(\tilde{\delta}+(N-3)\zeta+\phi)\right]  +(N-3) \cos\left[ \tau(\tilde{\delta}+(N-3)\zeta)\right] \bigg] + K,\notag
\end{eqnarray} }  The parameters $\vartheta,\phi,\tilde{Z},W,$ and $K$ are given by:

 {\small\begin{eqnarray}
&& \tan (\vartheta  \tau) \equiv \tan \left [(N J^\perp+\chi) \tau\right]\sec({\bar\theta}_1 ), \label{exp1} \\
&& \tan (\phi \tau) \equiv \tan \left (2\chi \tau\right)\cos({\bar\theta}_1 ), \label{exp2}  \\
&& \tilde{Z}^2\equiv 1-\sin^2({\bar\theta}_1 ) \sin^2 \left (2\chi \tau\right), \\
&& W^2\equiv 1-\tan^2({\bar\theta}_1 ) \sin^2 \left[(N J^\perp+\chi) \tau\right], \\
&&K\equiv \frac{N}{2} -\frac{N}{2} \cos({\bar\theta}_1 )\cos({\bar\theta}_2)\left(1-\frac{\Delta {\bar\theta}_2^2+ \Delta {\bar\theta}_1^2}{2}\right) - \frac{N}{2} \sin({\bar\theta}_1 )\sin({\bar\theta}_2)\Delta \theta_2 \Delta \theta_1 \cos(N J^\perp \tau) . \label{exp4}
\end{eqnarray} }  and $Z,\zeta$  given by Eqs. (\ref{Zn}) and (\ref{zetan}). For simplicity, we have removed the subscript ${\vec{\bf n}}$ but it is understood that all the interactions and Rabi frequencies are for the selected modes  ${\vec{\bf n}}$ under consideration. For $N=1$,   $\Delta{\Omega}=0$.

\section*{Appendix 4: Generic mean-field equations of motion}

Here we derive the most general equations of motion that include excitation inhomogeneity, non-collective elastic and inelastic two-body interactions, and  single-particle losses. The equations are obtained by assuming  that the density matrix can be factorized according to Eq.~(\ref{fac}).

{\small
\begin{eqnarray}
 \frac{d}{dt} {\rho_{ee}(k)}&=&\frac{1}{2} \sum_{k'=0}^{N-1}\bigg( {\Omega}_{k+k'}  \rho_{eg}(k')+ { \Omega}_{k-k'}\rho_{ge}(k')\bigg)-{\rm i} \sum_{k',k''=0}^{N-1}\rho_{eg}(k') \bigg(\rho_{ge}(k'')\Big[ J^\perp_{k'+k,k''}- J^\perp_{k-k'',-k'}\Big]\bigg) \label{mag1}
\notag\\
&&-\Gamma^e \rho_{ee}(k)  -\sum_{k',k''=0}^{N-1}\rho_{ee}(k') \bigg(\Gamma^{ee}_{k-k',k''} \rho_{ee}(k'')+\Big (\frac{\Gamma^{eg}_{k-k',k''}+ \Lambda^{eg}_{k-k',k''}}{2}\Big)\rho_{gg}(k'')\bigg) \notag \\
&&  -\sum_{k',k''=0}^{N-1}\Big (\frac{\Gamma^{eg}_{k'',k'-k}- \Lambda^{eg}_{k'',k'-k}}{4} \Big) \rho_{ge}(k') \rho_{eg}(k'')
  -\sum_{k',k''=0}^{N-1}\Big(\frac{\Gamma^{eg}_{k'+k,k''}- \Lambda^{eg}_{k'+k,k''}}{4} \Big) \rho_{ge}(k'') \rho_{eg}(k'),\\
   \frac{d}{dt} {\rho_{gg}(k)}&=&-\frac{1}{2} \sum_{k'=0}^{N-1}\bigg( {\Omega}_{k+k'}  \rho_{eg}(k')+ { \Omega}_{k-k'}\rho_{ge}(k')\bigg)
+{\rm i} \sum_{k',k''=0}^{N-1}\rho_{eg}(k') \bigg(\rho_{ge}(k'')\Big[ J^\perp_{k'+k,k''}- J^\perp_{k-k'',-k'}\Big]\bigg)\notag\\
&&-\Gamma^g \rho_{gg}(k) -\sum_{k',k''=0}^{N-1} \Big (\frac{\Gamma^{eg}_{k-k',k''}+ \Lambda^{eg}_{k-k',k''}}{2}\Big)\rho_{gg}(k')\rho_{ee}(k'')\\
&&  -\sum_{k',k''=0}^{N-1}\Big (\frac{\Gamma^{eg}_{k'',k'-k}- \Lambda^{eg}_{k'',k'-k}}{4} \Big) \rho_{ge}(k') \rho_{eg}(k'')
  -\sum_{k',k''=0}^{N-1}\Big(\frac{\Gamma^{eg}_{k'+k,k''}- \Lambda^{eg}_{k'+k,k''}}{4} \Big) \rho_{ge}(k'') \rho_{eg}(k'),\\
 \frac{d}{dt} {\rho_{eg}(k)}&=&  -{\rm i} \delta \rho_{eg}(k)-  \frac{1}{2} \sum_{k'=0}^{N-1}\bigg(
  {\Omega}_{-k-k'} \Big[\rho_{ee}(k')-\rho_{gg}(k')\Big]\bigg)\\
   && -{\rm i} \sum_{k',k''=0}^{N-1}\rho_{eg}(k') \bigg(\Big[\rho_{ee}(k'')-\rho_{gg}(k'')\Big]\Big[ J^\perp_{-k-k'',-k'}- J^\perp_{k'-k,k''}- \chi_{k'-k,k''}\Big]-
  \Big[\rho_{ee}(k'')+\rho_{gg}(k'')\Big ]C_{k'-k,k''}\bigg) \notag \\
  && -\frac{1}{2}(\Gamma^e + \Gamma^g )\rho_{eg}(k) -\sum_{k',k''=0}^{N-1}\rho_{eg}(k') \bigg(\frac{\Gamma^{ee}_{k'-k,k''}}{2} \rho_{ee}(k'')+
  \Big (\frac{\Gamma^{eg}_{k'-k,k''}+ \Lambda^{eg}_{k'-k,k''}}{4}\Big)\Big[ \rho_{ee}(k'')+\rho_{gg}(k'') \Big]\bigg)\notag\\
 && -\sum_{k',k''=0}^{N-1}\Big[ \rho_{ee}(k')+\rho_{gg}(k') \Big]\rho_{eg}(k'')
  \Big (\frac{\Gamma^{eg}_{k'',k'-k}- \Lambda^{eg}_{k'',k'-k}}{4}\Big). \notag \label{mag2}
   \end{eqnarray}   }         Here we have introduced the following quantities:

   \begin{eqnarray}
\Omega_{k}&\equiv&\frac{1}{N} \sum_{j=1}^{N} e^{i \frac{2 \pi j k}{N}} \Omega_{\vec {\bf  n}_j},\\
\Gamma^{\alpha,\beta}_{k,k'}&\equiv&\frac{1}{N^2} \sum_{j,j'=1}^{N}  e^{i \frac{2 \pi j k}{N}} e^{-i \frac{2 \pi j' k'}{N}}\Gamma_{ \vec{\bf  n}_j,\vec {\bf  n}_{j'}}^{\alpha,\beta},\\
\Lambda^{eg}_{k,k'}&\equiv&\frac{1}{N^2} \sum_{j,j'=1}^{N}  e^{i \frac{2 \pi j k}{N}} e^{-i \frac{2 \pi j' k'}{N}}\Lambda_{ \vec{\bf  n}_j,\vec {\bf  n}_{j'}}^{eg},\\
J^{\perp}_{k,k'}&\equiv&\frac{1}{N^2} \sum_{j,j'=1}^{N}  e^{i \frac{2 \pi j k}{N}} e^{-i \frac{2 \pi j' k'}{N}}J^{\perp}_{\vec {\bf  n}_j,\vec { \bf  n}_j'},   \\
\chi_{k,k'}&\equiv&\frac{1}{N^2} \sum_{j,j'=1}^{N}  e^{i \frac{2 \pi j k}{N}} e^{-i \frac{2 \pi j' k'}{N}}\chi_{\vec {\bf  n}_j,\vec { \bf  n}_{j'}},   \\
C_{k,k'}&\equiv&\frac{1}{N^2} \sum_{j,j'=1}^{N}  e^{i \frac{2 \pi j k}{N}} e^{-i \frac{2 \pi j' k'}{N}}C_{\vec {\bf  n}_j,\vec {\bf  n}_{j'}}.
\end{eqnarray} We have also assumed translationally invariant single particle decay and detuning since this is the most relevant experimental case. However, those can be made to be inhomogeneous  straightforwardly. $\Lambda_{ \vec{\bf  n}_j,\vec {\bf  n}_{j'}}^{eg}$ accounts for  $s$-wave  $e$-$g$ inelastic  collisions. We have not introduced $s-$ losses before since those  are only possible  when the dynamics are not restricted to take place in the fully symmetric Dicke manifold \cite{FossFeig2012}.

Note that for a coherent state , $\rho_{ee}(k=0)\rho_{gg}(k=0)=\rho_{eg}(k=0)\rho_{ge}(k=0)$. This  is an important  equality to keep in mind to reproduce  the terms proportional to $\Gamma^{eg}$ in Eqs. (\ref{eqgp0}-\ref{eqgp}) and to reproduce the absence of  $s$-wave losses in the collective limit.

\section*{Appendix 5: Gap protection}
\begin{figure}[tbh]
 \begin{center}
 \includegraphics[width=0.6 \columnwidth]{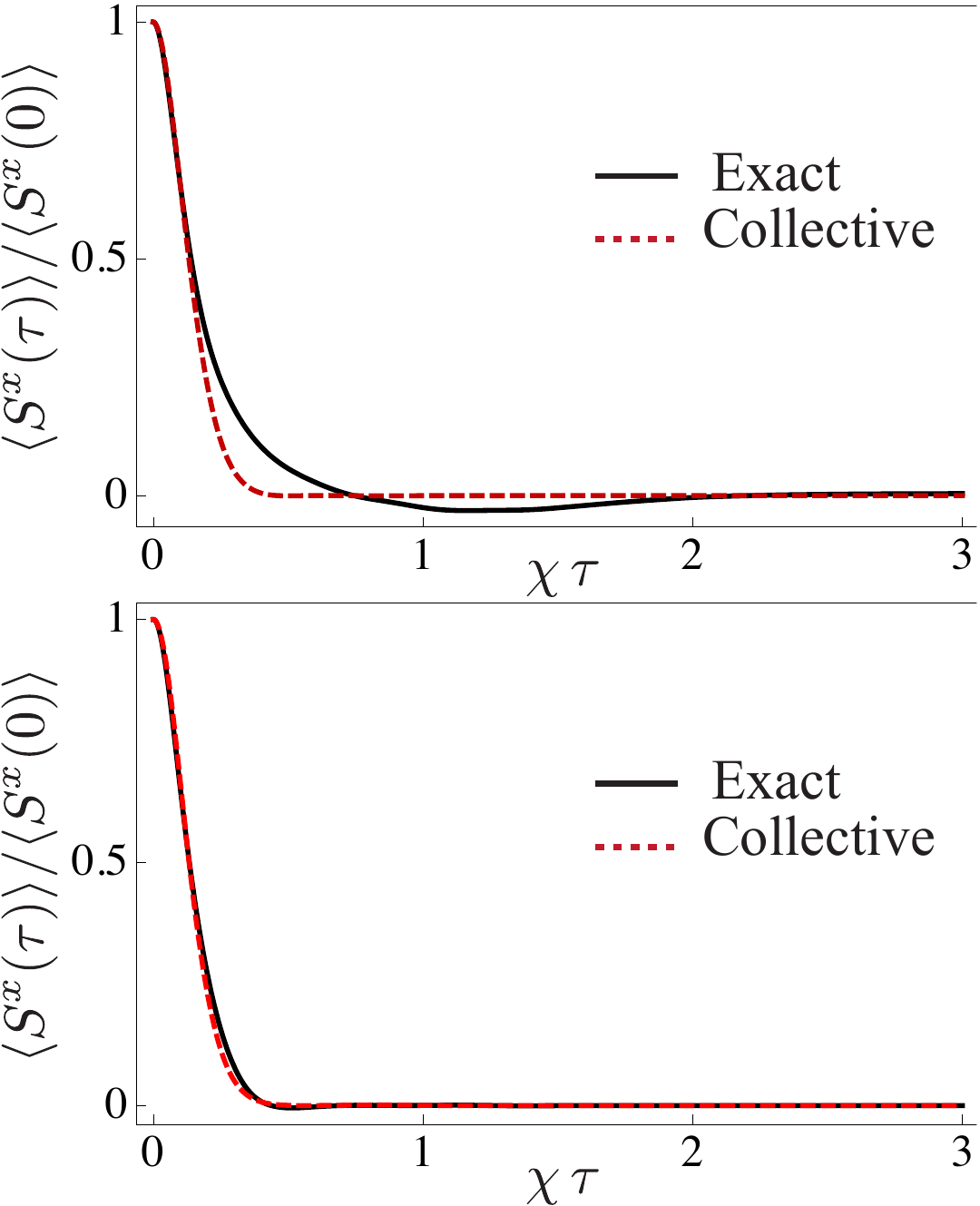}
 \caption{Effect of s-wave interactions on the validity of the collective approximation. Comparison of the exact solution (solid) with the collective approximation (dashed) in the absence of s-wave interactions (top) and for $u/v^{eg} = 15$ (bottom).\label{fig:swave_protction}}
 \end{center}
\end{figure}

The derivation of the collective spin model is based on the assumption that all particles interact collectively with each other and pulses are homogeneous, {\it i.e.} the coupling constants $P_{{\bf {n}_j}{\bf{n}_i}{\bf {n}_h}{\bf {n}_q}}$ and $S_{{\bf {n}_j}{\bf{n}_i}{\bf {n}_h}{\bf {n}_q}}$ are taken to be independent of $j,i,h$ and $q$. This assumption is justified if there exists a large energy gap that prevents transitions between the $S=N/2$ and $S=N/2- 1$ sectors. Such a gap can be generated if the term $J^{\perp}_{\vec{\mathbf{n}}}\vec S \cdot \vec S$ is large compared to the other terms in the Hamiltonian. Recall that $J^{\perp}_{\vec{\mathbf{n}}} = (V_{\vec{\mathbf{n}}}^{eg}  - U_{\vec{\mathbf{n}}}^{eg})/2$, where $V_{\vec{\mathbf{n}}}^{eg}$ and $U_{\vec{\mathbf{n}}}^{eg}$ are functions of the scattering lengths for $p-$and $s$-wave interactions, respectively. We expect that the assumption of a collective interaction approximates well the full dynamics of the system in the presence of a large energy gap. As an illustration, we present in Fig.~\ref{fig:swave_protction} a comparison of the exact solution with the collective approximation in the case of no $s$-wave interaction, $u= 0$, and a ratio of $u/v^{eg} = 15$. In this case, we use a boxed potential to compute the mode-dependence of the $p$-wave interactions. The $s$-wave interactions in the boxed potential become mode independent.

\bibliographystyle{apsrev}
\bibliography{./molref4}

\end{document}